\documentclass[%
 reprint,
 superscriptaddress,
nofootinbib,
 amsmath,amssymb,
 aps,
 prx
]{revtex4-2}
\usepackage[T1]{fontenc}
\usepackage{graphicx}% Include figure files
\usepackage{dcolumn}% Align table columns on decimal point
\usepackage{bm}% bold math
\usepackage[colorlinks=true, allcolors=blue]{hyperref}
\usepackage[capitalise]{cleveref}
\usepackage{braket}
\usepackage[table]{xcolor}
\usepackage{xfrac}
\usepackage{comment}
\usepackage[normalem]{ulem}
\usepackage{mathtools}
\usepackage{layouts}
\usepackage{quantikz}
\usepackage{tabularx}
\usepackage{esint}
\usepackage{soul}% For highlighting
\newcommand{\tr}{\text{tr}}

\newcommand{\vect}[1]{\boldsymbol{\mathbf{#1}}}
\newcommand{\approptoinn}[2]{\mathrel{\vcenter{
  \offinterlineskip\halign{\hfil$##$\cr
    #1\propto\cr\noalign{\kern2pt}#1\sim\cr\noalign{\kern-2pt}}}}}

\begin{document}
\title{{Modeling} Logical Gates and Read-Out of Superconducting Gottesman-Kitaev-Preskill Qubits}
\author{Mackenzie H. Shaw}
\email{M.H.Shaw@tudelft.nl}
\affiliation{ARC Centre of Excellence for Engineered Quantum Systems, School of Physics, The University of Sydney, Sydney, NSW 2006, Australia.}
\affiliation{QuTech, Delft University of Technology, P.O. Box 5046, 2600 GA Delft, The Netherlands}
\affiliation{Faculty of EEMCS, Delft University of Technology, Mekelweg 4, 2628 CD Delft, The Netherlands}
\author{Andrew C. Doherty}
\affiliation{ARC Centre of Excellence for Engineered Quantum Systems, School of Physics, The University of Sydney, Sydney, NSW 2006, Australia.}
\author{Arne L. Grimsmo}
\affiliation{ARC Centre of Excellence for Engineered Quantum Systems, School of Physics, The University of Sydney, Sydney, NSW 2006, Australia.}
\affiliation{AWS Center for Quantum Computing, Pasadena, CA 91125, USA}
\affiliation{California Institute of Technology, Pasadena, CA 91125, USA}
	
\date{March 6, 2024}
\begin{abstract}
The Gottesman-Kitaev-Preskill (GKP) code is an exciting route to fault-tolerant quantum computing since Gaussian resources and GKP Pauli-eigenstate preparation are sufficient to achieve universal quantum computing.
{However, there is a disconnect between the noise model that GKP qubits are in theory designed to correct -- uniform random displacement errors -- and the conditions that affect GKP qubits in superconducting devices in practice: realistic noise channels, logical gates, and inefficient measurements.}
In this work {we bridge this gap in three ways. First, we approximate the effect loss and dephasing on approximate GKP codestates using a random displacement channel, and show that this approximation matches well with numerics. Second,} we analyze the error-spreading properties of GKP Clifford gates and describe how a modification in the decoder following the implementation of each gate can reduce the gate infidelity by multiple orders of magnitude. Finally, we consider the effect of homodyne measurement inefficiencies on logical state read-out and {analyze} a scheme {to improve the measurement efficiency using the theory of quantum trajectories.}
%that implements a measurement with a $0.1\%$ error rate in $630$ ns assuming a two-mode interaction strength of 10 MHz and an efficiency of just~$75\%$.
%we provide a practical proposal to perform Clifford gates and state read-out in GKP codes implemented with active error correction in superconducting circuits. We present a method of performing Clifford circuits without physically implementing any single-qubit gates, reducing the potential for them to spread errors in the system. In superconducting circuits, all the required two-qubit gates can be implemented using the same non-linear element. We analyze the error-spreading properties of GKP Clifford gates and describe how a modification in the decoder following the implementation of each gate can reduce the gate infidelity by multiple orders of magnitude. Moreover, we develop a simple analytical technique to estimate the effect of loss and dephasing on GKP codes that matches well with numerics. Finally, we consider the effect of homodyne measurement inefficiencies on logical state read-out and present a scheme that implements a measurement with a $0.1\%$ error rate in $630$ ns assuming a two-mode interaction strength of 10 MHz and an efficiency of just~$75\%$.
\end{abstract}
\maketitle
%\tableofcontents
\section{Introduction}\label{sec:intro}
To construct a large-scale quantum computer, quantum error correction (QEC) is required to achieve error rates low enough to run useful algorithms. Bosonic QEC codes~\cite{Chuang97,Cochrane99,Gottesman01,Michael16} are a promising approach to QEC because they encode logical information in the formally infinite dimensional Hilbert space of a quantum harmonic oscillator, allowing for robust logical qubits to be constructed from a single physical device. Moreover, bosonic codes can be concatenated with a traditional QEC code such as the surface code~\cite{Vuillot19,Noh20,Noh22,Hopfmueller24} or quantum LDPC codes~\cite{Raveendran22}, using the enhanced error tolerance of the bosonic code to reduce the overhead required to reach a given overall logical error rate.

Actively pursued examples of bosonic codes include the cat code~\cite{Cochrane99}, binomial code~\cite{Michael16}, and Gottesman-Kitaev-Preskill (GKP) code~\cite{Gottesman01}, with a recent GKP code experiment surpassing the break-even point by more than a factor of two~\cite{Sivak23}. GKP codes are also particularly promising since universal quantum computation can be achieved using only Gaussian resources combined with a supply of either GKP Pauli-eigenstates~\cite{Baragiola19} or GKP Hadamard-eigenstates~\cite{Yamasaki20}. Such GKP Pauli-eigenstates have been produced in both superconducting microwave cavities~\cite{Campagne19,Eickbusch22,Sivak23} and the motional states of trapped ions~\cite{Fluehmann19,DeNeeve22}, but currently have error-rates too high to surpass the threshold of the GKP-surface code or toric code~\cite{Vuillot19,Noh22}.

Although GKP codes have been modeled both on the physical single-mode level and in larger multi-mode code concatenations, these studies typically use incompatible noise models, with multi-mode simulations using a Gaussian random displacement channel that is theoretically useful but unrealistic, and single-mode studies using more realistic noise models such as loss and dephasing that are difficult to simulate exactly in multi-mode studies. Moreover, realistic modeling of GKP qubits requires the incorporation of logical operations such as gates and measurements. Strikingly,

due to the low energy of microwave photons, homodyne detection in microwave circuits is severely limited in practice, with state-of-the-art experiments achieving efficiencies on the order of 60\% to 75\%~\cite{Macklin15,Touzard19}.
%Although proposals have been made for realizing GKP one- and two-qubit gates~\cite{Tzitrin20,Grimsmo21,Hastrup21} and practical error-correction schemes~\cite{Hillmann22}, more work is required to develop platform-specific schemes that are fault-tolerant to errors, convenient to implement experimentally, and combat experimentally-relevant sources of error. Indeed, due to the low energy of microwave photons, homodyne detection in microwave circuits is severely limited in practice, with state-of-the-art experiments achieving efficiencies on the order of 60\% to 75\%~\cite{Macklin15,Touzard19}. A separate issue is that multi-mode simulations of GKP codestates typically use an unrealistic noise model -- Gaussian random displacement channels -- that do not accurately capture the performance of GKP codes against loss~\cite{Hastrup23,Shaw24-1}.

To address these gaps, we introduce new theoretical techniques to analyze the effect of realistic noise channels, GKP Clifford gates, and read-out that is theoretically-convenient and a close approximation of the realistic noise channel. First, we analytically estimate the effect of loss and dephasing on GKP codes by taking a lowest-order approximation of the gate infidelity error using the GKP stabilizer subsystem decomposition~\cite{Shaw24-1}. The expressions we obtain coincide with those corresponding to a continuous-variable ``twirling approximation'' of the relevant quantum channel. As such, the technique can also be viewed as both a justification and a generalization of the twirling approximation that has previously been limited to analyzing approximate GKP codestates~\cite{Menicucci14}. Second, we introduce a general method to counteract the spread of errors due to Clifford gates using a modified error correction scheme after each gate. And finally, we analyze the effect of low measurement efficiencies on logical state read-out, and we use the theory of quantum trajectories to analyze a method that improves the effective measurement efficiency of read-out by coupling each high-Q GKP mode to a low-Q read-out ancilla.

%In this work, we introduce three practical proposals to improve the performance of Clifford gates and state read-out for actively-corrected GKP qubits implemented in superconducting devices. The first scheme we present removes the need to physically perform single-qubit Clifford gates, thereby reducing the number of physical gates (and hence the spread of errors) in a given circuit. Second, we introduce a general method to counteract the spread of errors due to Clifford gates using a modified error correction scheme after each gate. Finally, we improve the effective measurement efficiency of logical read-out by coupling each high-Q GKP mode to a low-Q read-out ancilla.

%In this work we present a proposal to perform GKP Clifford gates and state read-out in circuit QED that is hardware-efficient and reduces the impact of logical errors and measurement inefficiency. Our proposal avoids physically performing single-qubit Clifford gates, reducing the number of physical gates (and hence the spread of errors) in a given algorithm. Homodyne detection in microwave circuits is severely limited in practice, with state-of-the-art experiments achieving efficiencies below 90\%~\cite{Macklin15,Touzard19}. To combat this, our proposal improves the effective efficiency of logical read-out by coupling each high-Q GKP mode to a low-Q read-out ancilla.

We utilize two recently developed techniques to analyze the performance of each scheme. We analyze physical noise models and GKP Clifford gates using the \emph{stabilizer subsystem decomposition}~\cite{Shaw24-1}. In particular, the stabilizer subsystem decomposition is designed such that the partial trace over the stabilizer subsystem corresponds exactly to an ideal decoding map, making it suitable to calculate gate fidelities and other quantities of interest.
%We show that the spreading of errors due to the application of any GKP Clifford gate can be exactly counteracted by a modification of the decoder following the gate, under the assumptions of ideal gate execution and error correction.
We analyze our measurement scheme using the methods of Ref.~\cite{Warszawski20}, which allows one to solve for the quantum trajectory evolution of a system with time-independent linear dynamics and Gaussian measurement noise.
%This allows us to present realistic experimental parameters that can be used to execute high fidelity fast logical read-out of GKP states.

%The work in this paper focuses on only two of the steps -- Clifford gates and state read-out -- required to perform fault-tolerant quantum computation using GKP codes. Therefore our proposals need to be combined with other work done on GKP state preparation~\cite{Terhal16,Royer20,Campagne19,Eickbusch22,DeNeeve22,Kolesnikow23} \hl{which typically requires coupling to a qubit such as a transmon}, and non-Clifford gates either implemented directly~\cite{Hastrup21,Royer22,nguyen2025fault} or via magic state preparation~\cite{Baragiola19}. Moreover, in the context of a fault-tolerant algorithm one needs either to concatenate a GKP code with a qubit code~\cite{Fukui18,Fukui23,Vuillot19,Noh20,Noh22} or use a \textquotedblleft genuine\textquotedblright\ multi-mode GKP code~\cite{Gottesman01,Harrington04,Conrad22,Royer22}.

The remainder of this paper is organized as follows. In \cref{sec:GKP}, we provide an overview of GKP codes and the notation we will use throughout the manuscript.
%In \cref{sec:gates}, we describe how to remove single-qubit gates from a quantum circuit, define the \textit{generalized controlled gates} which must be performed instead, and provide superconducting circuits that can implement these gates in a circuit QED experiment.
We begin by considering the effects of GKP Clifford gates in \cref{sec:decoders}, in which we establish the figure of merit we will use to analyze noise in GKP codes and explain how to minimize the spread of logical errors using a modified decoding scheme.
Then in \cref{sec:error_estimates} we introduce our new analytic approximation technique for loss and dephasing acting on GKP codes.
Finally, we analyze the effect of measurement inefficiencies on logical state read-out in \cref{sec:measurements} and analyze a method that uses an additional ancilla mode to perform homodyne detection with an enhanced efficiency. We provide concluding remarks in \cref{sec:conc}.

\section{GKP Codes}\label{sec:GKP}

We now present a brief overview of the properties of GKP codes~\cite{Gottesman01} and the notation that we will use in the remainder of this manuscript. GKP codes are a class of bosonic stabilizer codes in which the codespace is the simultaneous $+1$-eigenspace of operators acting on a continuous variable (CV) Hilbert space. The CV system can be described by ladder operators $[a,a^{\dag}]=1$ or quadrature operators $[q,p]=i$. We denote the number states as $\ket{n}$, and position and momentum eigenstates as $\ket{x}_{q}$ and $\ket{y}_{p}$, respectively. We also introduce the displacement operator
\begin{equation}
    T(\vect{v})=T(v_{1},v_{2})= e^{-iv_{1}p+iv_{2}q}
\end{equation}
for $\vect{v}=(v_{1},v_{2})\in\mathbb{R}^{2}$, which obeys 
\begin{equation}
    T(\vect{v})T(\vect{w})=e^{-i(v_{1}w_{2}-v_{2}w_{1})/2}T(\vect{v}+\vect{w}).
\end{equation}
This definition of the displacement operator also ensures that $T(\lambda,0)\ket{x}_{q}=\ket{x+\lambda}_{q}$ and $T(0,\lambda)\ket{y}_{p}=\ket{y+\lambda}_{p}$.

To define a (single-mode) GKP code, we begin with two real vectors $\vect{\alpha}=(\alpha_{1},\alpha_{2})$ and $\vect{\beta}=(\beta_{1},\beta_{2})$ that satisfy $\alpha_{1}\beta_{2}-\beta_{1}\alpha_{2}=\pi$. The stabilizer generators are then given by $S_{X}=T(2\vect{\alpha})$ and $S_{Z}=T(2\vect{\beta})$, which (together with their inverses) generate the stabilizer group. The logical Pauli operators are given by $\bar{X}=T(\vect{\alpha})$, $\bar{Y}=T(\vect{\alpha}+\vect{\beta})$ and $\bar{Z}=T(\vect{\beta})$, where we use bars to indicate logical operators and states. We define the GKP logical lattice
\begin{equation}\label{eq:GKP_logical_lattice}
    \mathcal{L}=\{m\vect{\alpha}+n\vect{\beta}\>|\>m,n\in\mathbb{Z}\},
\end{equation}
and the corresponding Voronoi cell
\begin{equation}\label{eq:voronoi_cell}
    \mathcal{V}_{\mathcal{L}}=\big\{\vect{v}\in\mathbb{R}^{2}\>\big|\>|\vect{v}|<|\vect{v}-\vect{\ell}|\;\forall\,\vect{\ell}\in\mathcal{L},\vect{\ell}\neq\vect{0}\big\}
\end{equation}
which contains the set of points closer to the origin than any other point in $\mathcal{L}$.

The simplest example of a GKP code is the square GKP code, given by $\vect{\alpha}_{\text{sq}}=(\sqrt{\pi},0)$ and $\vect{\beta}_{\text{sq}}=(0,\sqrt{\pi})$. In this case, $\bar{X}_{\text{sq}}=e^{-i\sqrt{\pi}p}$ and $\bar{Z}_{\text{sq}}=e^{i\sqrt{\pi}q}$. General GKP codes can be conveniently described by introducing the canonically transformed \textit{logical} quadrature operators
\begin{align}\label{eq:logical_quadratures}
    \bar{q}&=\frac{1}{\sqrt{\pi}}(\beta_{2}q-\beta_{1}p),&\bar{p}&=\frac{1}{\sqrt{\pi}}(\alpha_{1}p-\alpha_{2}q),
\end{align}
such that $\bar{X}=e^{-i\sqrt{\pi}\bar{p}}$ and $\bar{Z}=e^{i\sqrt{\pi}\bar{q}}$.
%As such, most of the properties of square GKP codes can be mapped to general GKP codes simply by replacing $q,p\mapsto \bar{q},\bar{p}$.
Of particular interest is the hexagonal GKP code, given by
\begin{align}\label{eq:hex_GKP}
    \vect{\alpha}_{\text{hex}}&=\sqrt{\pi}\bigg(\frac{\sqrt[4]{3}}{\sqrt{2}},-\frac{1}{\sqrt{2}\sqrt[4]{3}}\bigg),&\vect{\beta}_{\text{hex}}&=\sqrt{\pi}\bigg(0,\frac{\sqrt{2}}{\sqrt[4]{3}}\bigg)
\end{align}
which has been shown to have the lowest logical error rate out of all single-mode GKP geometries under a pure loss noise model~\cite{Albert18}. Note that we have chosen a rotated definition of the hexagonal code, such that $\beta_{1}=0$ and $\bar{q}\propto q$, allowing a convenient representation of the GKP codestates in the position basis.

To aid our discussion of logical Pauli operators in \cref{sec:decoders}, we introduce the notation
\begin{align}\label{eq:s_quadratures}
    s_{1}&=-\bar{p},&s_{2}&=\bar{q}-\bar{p},&s_{3}&=\bar{q}.
\end{align}
Using this, we can write the logical Pauli operators as $\bar{\sigma}_{i}=e^{i\sqrt{\pi}s_{i}}$, where $\sigma_{1,2,3}=X,Y,Z$ respectively. Each quadrature $s_{i}$ can be written in polar coordinates given by $s_{i}=r_{i}(q\cos\theta_{i}+p\sin\theta_{i})$.

GKP Clifford operators, which map logical Pauli operators to logical Pauli operators, are given by unitary Gaussian operators $\bar{A}$ acting on the CV space. Concretely, the GKP Clifford group is generated by the operators
\begin{align}
    \bar{H}&=e^{i\pi(\bar{q}^{2}+\bar{p}^{2})/4},& \bar{S}&=e^{i\bar{q}^{2}/2}, &\bar{C}_{Z}&=e^{i\bar{q}\otimes\bar{q}}.
\end{align}

In superconducting GKP qubits a convenient alternative set of generators for the GKP Clifford group is given by the \textit{generalized controlled gates}. For Pauli matrices $\sigma_{i},\sigma_{j}$ ($i,j\in\{1,2,3\}$), we define the $\sigma_{i}$-controlled-$\sigma_{j}$ gate as
\begin{equation}\label{eq:generalised_controlled_gate}
    C_{\sigma_{i}\sigma_{j}}=I\otimes I-\frac{1}{2}(I-\sigma_{i})\otimes(I-\sigma_{j}).
\end{equation}
These gates can be interpreted as applying a $\sigma_{j}$ gate to the target (second) qubit if the control (first) qubit is a $-1$ eigenstate of $\sigma_{i}$, and doing nothing if it a $+1$ eigenstate. For example the $Z$-controlled-$Z$ gate is simply the controlled-$Z$ gate, and the $Z$-controlled-$X$ gate is the controlled-NOT gate. Any universal quantum computing circuit comprising of state preparation in the $\ket{0}$ or $\ket{T}=(\ket{0}+e^{i\pi/4}\ket{1})/\sqrt{2}$ states, Pauli $Z$-measurements, and adaptive Hadamard, phase and controlled-$Z$ gates can be written as a circuit that instead consists of state preparation in the $\ket{0}$ or $\ket{T}$ states, Pauli $X$, $Y$ and $Z$-measurements, and generalized controlled gates (which must be performed adaptively). GKP generalized controlled gates can be implemented by
\begin{equation}
    \bar{C}_{\sigma_{i}\sigma_{j}}=e^{i\bar{s}_{i}\otimes\bar{s}_{j}},
\end{equation}
which is convenient because all of the generalized controlled gates can be implemented using a single piece of hardware in superconducting qubits, as we show in \cref{sec:gates}.

In \cref{sec:decoders} we will use an equivalent representation of Gaussian operators as symplectic matrices. In particular, one can describe an arbitrary $n$-mode Gaussian operator $U$ with a $2n\times2n$ real symplectic matrix $\mathcal{S}(U)\in\mathrm{Sp}(2n,\mathbb{R})$ acting on the vector of quadrature operators $\vect{\xi}=(q_{1},\dots,q_{n},p_{1},\dots,p_{n})$, such that $UT(\vect{v})U^{\dag}=T(S\vect{v})$ and $U\vect{\xi}U^{\dag}=\mathcal{S}(U)^{-1}\vect{\xi}$ (where $U$ and $U^{\dag}$ act component-wise on $\vect{\xi}$). For example, a single-mode rotation operator $R(\theta) = e^{i\theta a^{\dag}a}$ can be described with the symplectic operator
\begin{equation}
    \mathcal{S}\big(R(\theta)\big)=\begin{bmatrix}\cos\theta&-\sin\theta\\\sin\theta&\cos\theta\end{bmatrix}.
\end{equation}
In the square GKP code, we have
\begin{subequations}\label{eq:symplectic_Cliffords}
\begin{align}
    \mathcal{S}\big(\bar{H}_{\text{sq}}\big)&=\begin{bmatrix}0&-1\\1&0\end{bmatrix},\label{eq:symplectic_sq_Hadamard}\\
    \mathcal{S}\big(\bar{S}_{\text{sq}}\big)&=\begin{bmatrix}1&0\\1&1\end{bmatrix},\\
    \mathcal{S}\big(\bar{C}_{Z,\text{sq}}\big)&=\begin{bmatrix}1&0&0&0\\0&1&0&0\\0&1&1&0\\1&0&0&1\end{bmatrix}.
\end{align}
\end{subequations}

To describe Clifford gates in general GKP codes, we first define the change of basis matrix $M_{\mathcal{L}}= [\vect{\alpha},\vect{\beta}]/\!\sqrt{\pi}$ that transforms vectors in the square GKP lattice to the vectors in the lattice $\mathcal{L}$, i.e.~$\mathcal{L}=M_{\mathcal{L}}\mathcal{L}_{\text{sq}}$. Then, the symplectic matrix corresponding to the Clifford gate $\bar{A}$ is given by $\mathcal{S}\big(\bar{A}\big)=M^{\vphantom{-1}}_{\mathcal{L}}\mathcal{S}\big(\bar{A}_{\text{sq}}\big)M_{\mathcal{L}}^{-1}$.
%, where $M_{\mathcal{L}}$ is the change of basis matrix $\bar{\vect{\xi}}=M_{\mathcal{L}}^{-1}\vect{\xi}$ from physical modes $\vect{\xi}$ to logical modes $\vect{\bar{\xi}}=(\bar{q}_{1},\dots,\bar{q}_{n},\bar{p}_{1},\dots,\bar{p}_{n})$.

The ideal codestates of a general GKP code are the simultaneous $+1$-eigenstates of the stabilizers $S_{X}$ and $S_{Z}$. However, these ideal codestates are non-normalizable and hence cannot be realized in any physical system. To construct normalizable codestates, we use the non-unitary envelope operator $e^{-\Delta^{2}a^{\dagger}a}$ to define the approximate GKP codestates $\ket{\bar{\psi}_{\Delta}}\propto e^{-\Delta^{2}a^{\dag}a}\ket{\bar{\psi}}$.
%The envelope operator can be written in the position basis as
%\begin{multline}\label{eq:envelope}
%    {\vphantom{\ket{x}}}_{q}\!\bra{x'}e^{-\Delta^{2}a^{\dagger}a}\ket{x}_{q}=\sqrt{\frac{\mathrm{coth}(\Delta^{2})+1}{2\pi}}\\
%    \times\mathrm{exp}\bigg(\!{-}\frac{1}{2}\coth\big(\Delta^{2}\big)\big({x'}^{2}+x^{2}\big)+\mathrm{csch}\big(\Delta^{2}\big)x'x\!\bigg).
%\end{multline}
This approximation, however, introduces errors as the approximate states are not exact $+1$ eigenstates of the stabilizers. The parameter $\Delta$ characterizes the quality of the approximate GKP codestates, where the limit $\Delta\rightarrow0$ approaches the ideal codestates. We will also commonly quote the average photon number of the approximate GKP states~\cite{Albert18}
\begin{equation}
    \bar{n}=\frac{1}{2}\big(\!\braket{\bar{0}_{\Delta}|a^{\dag}a|\bar{0}_{\Delta}}+\braket{\bar{1}_{\Delta}|a^{\dag}a|\bar{1}_{\Delta}}\!\big)\approx \frac{1}{2\Delta^{2}}-\frac{1}{2}.
\end{equation}
which tends to infinity as $\Delta\rightarrow0$.
In general, the quality of non-ideal codestates can be assessed by evaluating the expectation value of the stabilizers and expressing it as a squeezing parameter~\cite{Weigand20}. For the particular form of the approximate codestates that we consider, this is evaluates to $\Delta_{\mathrm{dB}}=-10\log_{10}(\Delta^{2})$ which also tends to infinity as $\Delta\rightarrow0$.

Approximate GKP codestates have been prepared experimentally in superconducting resonators with an experimentally-determined squeezing of $\Delta_{\mathrm{dB}}=9.1$~\cite{Eickbusch22}. GKP-surface code and toric code studies~\cite{Vuillot19,Noh20,Noh22} have shown that the surface code threshold can be reached using codestates with $\Delta_{\mathrm{dB}}=9.9$ assuming that the dominant source of noise is due solely to the approximate GKP codestates. However, in the presence of circuit noise, a larger squeezing is required to get under the surface code threshold. As such, we use $\Delta_{\mathrm{dB}}=12$ as a rough target for practical quantum computing with GKP codes.
%We will also focus on implementations of GKP qubits in the rotating frame of a microwave resonator with Hamiltonian $H=\hbar\omega a^{\dag}a$.

Since the codestates $\ket{\bar{\mu}_{\Delta}}$ are not orthogonal, we can use the L\"owdin orthonormalization procedure~\cite{Lowdin50} to define orthonormalized GKP codestates $\ket{\bar{\mu}_{\Delta,\text{o}}}$, which form an orthonormal basis of the subspace spanned by $\{\ket{\bar{\mu}_{\Delta}}\}_{\mu=0,1}$.
%The L\"owdin orthonormalization has the advantage over the Gram-Schmidt orthonormalization in that it it symmetric with respect to the basis states that are inputs to the algorithm.
%We will use both orthonormalized and non-orthonormalized GKP codestates at different points throughout this manuscript depending on the application we are using them for.
Note that the difference between $\ket{\bar{\mu}_{\Delta}}$ and $\ket{\bar{\mu}_{\Delta,\text{o}}}$ is negligible for values of $\Delta$ that are small enough to be practical.

\begin{figure}
\includegraphics{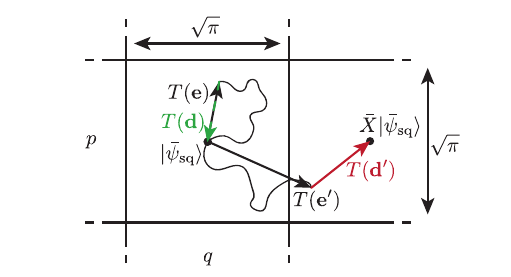}
\caption{The ideal error-correction procedure for the square GKP code, performed over the Voronoi cell of the square GKP lattice. In this case, the Voronoi cell is a square centered at the origin with side length $\sqrt{\pi}$. In green, a random walk of displacement errors results in an overall displacement error $T(\vect{e})$ (up to a geometric phase) acting on the ideal codestate. Since both the components of $\vect{e}$ are less than $\sqrt{\pi}/2$, the shortest displacement that returns the state to the codespace is $T(\vect{d})=T(-\vect{e})$. In red, the displacement error $T(\vect{e}')$ has $e_{1}'>\sqrt{\pi}/2$. This means that the smallest displacement that returns the state to the codespace is $T(\vect{d}')$, which has the net result of applying a logical $\bar{X}$ on the state.}
\label{fig:sq_GKP_decoding}
\end{figure}

Next, we discuss error-correcting the GKP code, beginning with what we refer to as ideal error-correction, which consists of the following steps:
\begin{enumerate}
    \item First, we measure both stabilizers $S_{X}=e^{-2i(\alpha_{1}p-\alpha_{2}q)}$ and $S_{Z}=e^{2i(\beta_{2}q-\beta_{1}p)}$, with measurement outcomes $M_{X}, M_{Z}$. In \textit{ideal} error correction, we assume $M_{X}$ and $M_{Z}$ can be obtained error-free.
    \item We assign each pair of measurement outcomes with a quasi-position and momentum $(k_{q},k_{p})$ such that $M_{X}=e^{-2i(\alpha_{1}k_{p}-\alpha_{2}k_{q})}$ and $M_{Z}=e^{2i(\beta_{2}k_{q}-\beta_{1}k_{p})}$ respectively.
    \item Finally, a displacement $T(-k_{q},-k_{p})$ is applied that returns the state to the ideal codespace of the code.
\end{enumerate}

The periodicity of the complex exponential means that the map $(M_{X},M_{Z})\mapsto(k_{q},k_{p})\in\mathbb{R}^{2}$ is not well-defined; in particular, adding any integer multiple of $\vect{\alpha}$ or $\vect{\beta}$ to the vector $(k_{q},k_{p})$ results in the same pair of measurement outcomes. As such, one must specify a primitive cell of the logical lattice $\mathcal{L}_{\text{log}}$ to serve as the \textit{decoding patch} $\mathcal{P}\subset\mathbb{R}^{2}$. Then, we can choose a unique vector $\vect{k}=(k_{q},k_{p})\in\mathcal{P}$ for any pair of measurement outcomes.

In general, the best choice of decoding patch depends on the noise model being applied to the GKP code. If we consider a noise model defined by a uniform mean-zero Gaussian random walk of displacement errors, the optimal patch to decode over is the Voronoi cell $\mathcal{V}_{\mathcal{L}}$ of the GKP logical lattice $\mathcal{L}$, see \cref{eq:GKP_logical_lattice,eq:voronoi_cell,fig:sq_GKP_decoding}. This choice ensures that the displacement $T(-k_{q},-k_{p})$ is the shortest displacement that returns the state to the codespace, thereby minimizing the chance of a logical error. Note that throughout this paper, by ``optimal patch'' we are really referring to a patch corresponding to a minimum-weight decoder, and further improvements could be made by considering a maximum-likelihood decoder.
%However, we will consider error correction over different patches in our discussion of error-protected logical gates in \cref{sec:decoders}.
The above description also applies to multi-mode GKP codes by increasing the number of stabilizers and the dimension of $\vect{k}$.

In \cref{sec:decoders,sec:error_estimates} we will consider different decoding patches, and for this, we will make use of the following definitions:
\begin{itemize}
    \item The \textit{distance} $d$ of a patch $\mathcal{P}$ is \textit{twice} the length of the shortest vector on the boundary of the patch $\partial\mathcal{P}$, and
    \item The \textit{degeneracy} $a$ of $\mathcal{P}$ is \textit{half} the number of vectors with length $d$ on $\partial\mathcal{P}$.
\end{itemize}
The definition of $d$ is chosen such that any error $T(\vect{e})$ with $|\vect{e}|<d/2$ is correctable. Moreover, the shortest logical Pauli operator $\bar{P}=T(\vect{\ell})$ has $|\vect{\ell}|=d$, while $a$ corresponds to the number of different equal-length logical Pauli operators.

\begin{figure}
    \centering
    \includegraphics{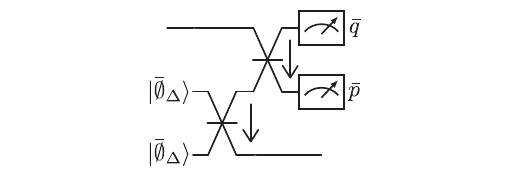}
    \caption{Quantum circuits for teleportation-based QEC. $\ket{\bar{\emptyset}}$ denotes a qunaught GKP codestate, defined as the unique $+1$-eigenstate of the displacements $T(\sqrt{2}\vect{\alpha})$ and $T(\sqrt{2}\vect{\beta})$. Each two-mode gate is a 50-50 beamsplitter given by $e^{i\pi(p_{1}q_{2}-q_{1}p_{2})/4}$. The homodyne measurements are of the logical quadratures from \cref{eq:logical_quadratures}.}\label{fig:tele}
\end{figure}

One concrete way of performing such a QEC cycle is via the teleportation-based circuit~\cite{Walshe20} shown in \cref{fig:tele}. The circuit uses two ancillary modes initialized in the so-called \textit{qunaught} GKP state, which we define for a general GKP lattice as the simultaneous $+1$-eigenstate of the displacement operators $T(\sqrt{2}\vect{\alpha})$ and $T(\sqrt{2}\vect{\beta})$. Note that the factor of $\sqrt{2}$ in the displacements -- instead of the factor of 2 in the standard GKP stabilizer generators -- ensures that their simultaneous +1-eigenspace is one-dimensional and hence defines a unique state. When the input qunaught states are \textit{ideal} GKP codestates, then the teleportation circuit performs a measurement of the stabilizers as well as a correction to the ideal GKP codespace, up to the application of a logical Pauli operator that can be inferred by decoding the measured syndrome. We discuss in \cref{sec:decoders} how to model the effect of approximate qunaught states on the QEC cycle, {and in \cref{sec:measurements} we discuss how one can use an additional physical mode to mitigate the effects of noisy homodyne measurements.}

Teleportation-based QEC is not the only error-correction procedure used for GKP codes; indeed, in superconducting and trapped ions the current favored stabilization procedure makes use of an ancillary two-level system instead~\cite{Royer20,DeNeeve22,Sivak23}. However, a teleportation-based scheme can be used to perform error correction after the GKP states have been prepared, and is particularly natural if GKP codes are concatenated with qubit error correcting codes through foliation methods~\cite{Bolt16}. Moreover, teleportation-based QEC has the advantage that a fresh GKP qubit is prepared in each QEC cycle, and distortions to the envelope and/or GKP lattice are not teleported onto the outgoing qubit. {We also note that any state-preparation method that can be used to prepare an encoded $\ket{\bar{0}_{\Delta}}$ state (for example Refs.~\cite{Rymarz2021,hastrup2021measurement,kolesnikow2025protected,Weigand2016}) can be modified to prepare an encoded qunaught state either by modifying the parameters in the state-preparation method, or by squeezing the $\ket{\bar{0}_{\Delta}}$ state.} We here focus on teleportation-based QEC due to the simple model of approximate error-correction that it provides below.

\section{Error-resistant Clifford gates}
\label{sec:decoders}

We now move our attention to analyzing the effect Clifford gates have on error correction. We begin with the unrealistic case of ideal error correction following the gate, where the only source of noise is from the approximate GKP codewords. Even in this simple case, GKP Clifford gates in general spread the errors and reduce the fidelity of the overall operation. However, we present a modification of the decoding patch that exactly counteracts the spreading of errors due to the gate, thus reducing the average gate infidelity by up to two orders of magnitude. Then, we generalize our approach to consider noisy error correction following the gate. Here, we find that our modification is only able to partly counteract the spreading of errors, but still provides a significant improvement compared to the na\"ive decoding patch.
%These results also justify our proposal from~\cref{sec:gates}, since reducing the number of Clifford gates in a circuit minimizes the spreading of errors.

We will begin by defining the average gate fidelity we will use to quantify the quality of logical gates. Then, we explain how to find the modified decoding patch, which corresponds to the Voronoi cell of a deformed lattice. Our proposed modification is a generalization of the error-protected two-qubit gates described in Ref.~\cite{Noh22} to general Clifford gates and GKP geometries. In the ideal error-correction case, we provide both numerical and analytical results for the quality of the logical gates. Then, we describe how to (approximately) incorporate the effect of non-ideal error correction in a teleportation-based QEC circuit. These results complement recent work which analyzed the performance of two-qubit gates when using dissipative error correction~\cite{Rojkov23}.

Given an $n$-mode GKP logical gate $\bar{U}$ that implements the $n$-qubit gate $U$, we define the \textit{logical} channel $\mathcal{E}_{U,\mathcal{P},\Delta}$ as an $n$-qubit-to-$n$-qubit channel consisting of the following steps:
\begin{enumerate}
    \item First, we encode the $n$-qubits into orthogonalized approximate GKP codestates using the encoding operator $R_{\Delta}=\big(\!\ket{\bar{0}_{\Delta,\mathrm{o}}}\!\bra{0}+\ket{\bar{1}_{\Delta,\mathrm{o}}}\!\bra{1}\!\big)^{\otimes n}$.
    \item Second, we apply the logical operator $\bar{U}$ to the state.
    \item Third, we perform a round of ideal error correction (as described in \cref{sec:GKP}) over a patch $\mathcal{P}$, which returns the state to the ideal codespace, represented by the map $\mathcal{C}_{\mathcal{P}}$.
    \item Finally, the resulting ideal GKP codestate is rewritten as an $n$-qubit state by the map $R^{\dag}=\big(\!\ket{0}\!\bra{\bar{0}}+\ket{1}\!\bra{\bar{1}}\!\big)^{\otimes n}$.
\end{enumerate}
Mathematically, we write the logical channel as
\begin{equation}\label{eq:logical_channel}
    \mathcal{E}_{U,\mathcal{P},\Delta}=\mathcal{J}[R^{\dag}]\circ\mathcal{C}_{\mathcal{P}}\circ\mathcal{J}[\bar{U}R_{\Delta}]
\end{equation}
where $\mathcal{J}[O]\rho=O\rho O^{\dag}$. Note that our definition \cref{eq:logical_channel} assumes ideal error correction; later in this section we will modify this definition to account for the effects of non-ideal error correction.

To quantify how well the map $\mathcal{E}_{U,\mathcal{P},\Delta}$ executes the gate $U$, we use the average gate fidelity. The average gate fidelity $\bar{F}(\mathcal{E},U)$ of a quantum channel $\mathcal{E}$ with a unitary $U$ is defined as~\cite{Nielsen02}
\begin{equation}\label{eq:avg_gate_fid_defn}
    \bar{F}(\mathcal{E},U)=\int d\psi\bra{\psi}U^\dag \mathcal{E}(\ket{\psi}\!\bra{\psi})U\ket{\psi},
\end{equation}
where the integral is over the uniform (Haar) measure $d\psi$ of state space. For notational convenience we write $\bar{F}(\mathcal{E}_{U,\mathcal{P},\Delta},U)=\bar{F}_{U,\mathcal{P},\Delta}$. Specifically, we can calculate the average gate fidelity $\bar{F}_{U,\mathcal{P},\Delta}$ of the logical gate $U$ from the entanglement fidelity of the map $\mathcal{J}[U^{\dag}]\circ\mathcal{E}_{U,\mathcal{P},\Delta}$ as detailed in Ref.~\cite{Nielsen02Simple}. 

\begin{figure}
    \centering
    \includegraphics{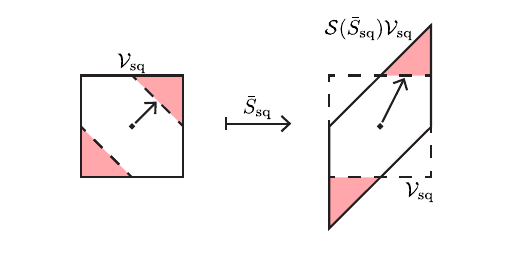}
    \caption{Diagram of the transformation of the square GKP Voronoi cell $\mathcal{V}_{\text{sq}}\mapsto\mathcal{S}(\bar{S}_{\text{sq}})\mathcal{V}_{\text{sq}}$ under a logical phase gate $\bar{S}_{\text{sq}}$ represented by the symplectic matrix $\mathcal{S}(\bar{S}_{\text{sq}})$ (\cref{eq:symplectic_Cliffords}). If error correction is subsequently performed over $\mathcal{V}_{\text{sq}}$, then pre-gate displacement errors lying in the red-shaded regions are mapped to logical errors by $\bar{S}$. The shortest such displacement is represented by the arrow. Alternatively, if error correction is performed over the modified patch $\mathcal{S}(\bar{S}_{\text{sq}})\mathcal{V}_{\text{sq}}$, the pre-image of this patch is $\mathcal{V}_{\text{sq}}$, maximizing the distance of the patch.}
    \label{fig:phase_transformation}
\end{figure}

To compute $\mathcal{E}_{U,\mathcal{P},\Delta}$ we utilize the GKP stabilizer subsystem decomposition (SSSD) that we have recently developed~\cite{Shaw24-1} (see \cref{subsec:SSD} for a summary). The GKP SSSD has the property that the ideal decoding map $\mathcal{J}[R^{\dag}]\circ\mathcal{C}_{\mathcal{P}}$ is equivalent to taking the partial trace over the ``stabilizer subsystem'', which is a natural operation to perform in this formalism. Using the SSSD for our analysis has two key advantages. From an analytical perspective, the SSSD naturally provides a first-order approximation to the gate fidelity that is convenient to use and agrees well with the numerical results (see \cref{sec:gate_error_estimate}); while numerically, calculating the gate fidelity with the SSSD requires fewer computational resources with \textit{increasing} GKP codestate quality ($\Delta\rightarrow0$). This last property contrasts with traditional Fock space simulations which require higher truncation dimensions as $\Delta\rightarrow0$ and thus can only simulate GKP codewords with sufficiently low average photon number.

Now we present our analytical results. In \cref{sec:gate_error_estimate} we estimate the average gate infidelity of the \textit{identity} gate, $1-\bar{F}_{I,\mathcal{P},\Delta}$. Intuitively, this captures idling noise when the only source of errors is due to the use of approximate GKP codestates. The estimate is given by
\begin{equation}\label{eq:avg_fid_estimate}
    1-\bar{F}_{I,\mathcal{P},\Delta}\approx \frac{2^{n}a}{2^{n}+1}\mathrm{erfc}\big(d/(2\Delta)\big),
\end{equation}
where $\mathrm{erfc}(x)$ is the complementary error function, and $d$ and $a$ are the distance and degeneracy (respectively) of the patch $\mathcal{P}$ as defined in \cref{sec:GKP}. Note also that for large $x$ we also have $\mathrm{erfc}(x)\approx e^{-x^{2}}\!/(x\sqrt{\pi})$.

For a given GKP Clifford gate $\bar{A}$, the average gate infidelity will be larger than or equal to that of the identity gate, depending on how $\bar{A}$ propagates displacement errors. Conveniently, the average gate fidelity of a GKP Clifford gate $\bar{A}$ can be expressed in the form of \cref{eq:avg_fid_estimate} by inputting the distance and degeneracy of a modified patch that accounts for this spreading of the errors. Specifically, in \cref{sec:Clifford_gate_estimate} we show that
\begin{equation}\label{eq:Clifford_gate_estimate}
    \bar{F}_{A,\mathcal{P},\Delta\vphantom{)^{-1}}}=\bar{F}_{I,\mathcal{S}(\bar{A})^{-1}\mathcal{P},\Delta}
\end{equation}
for any GKP Clifford gate $\bar{A}$ represented by the symplectic matrix $\mathcal{S}(\bar{A})$. One can then calculate the distance and degeneracy of $\mathcal{S}(\bar{A})^{-1}\mathcal{P}$ and substitute these into \cref{eq:avg_fid_estimate} to estimate the average gate fidelity.
%In words, the average gate fidelity of the logical gate over a patch $\mathcal{P}$ is equal to the average gate fidelity of the identity gate of the modified patch $\mathcal{S}(\bar{A})^{-1}\mathcal{P}$.

If error correction is performed over the Voronoi cell $\mathcal{P}=\mathcal{V}_{\mathcal{L}}$, then typically the modified patch $\mathcal{S}(\bar{A})^{-1}\mathcal{P}$ has a shorter distance than $\mathcal{P}$ and therefore a larger infidelity than the identity gate. Note, however, that any gates implemented by a rotation -- notably the Hadamard gate $\bar{H}_{\text{sq}}=e^{i\frac{\pi}{2}a^\dag a}$ in the square GKP code, and the permutation gate $\bar{H}^{\vphantom{\dag}}_{\text{hex}}\bar{S}^{\dag}_{\text{hex}}=e^{i\frac{\pi}{3}a^\dag a}$ in the hexagonal GKP code -- achieve the same average gate fidelity as the identity gate since they leave Voronoi cells invariant.

\begin{figure*}
\includegraphics{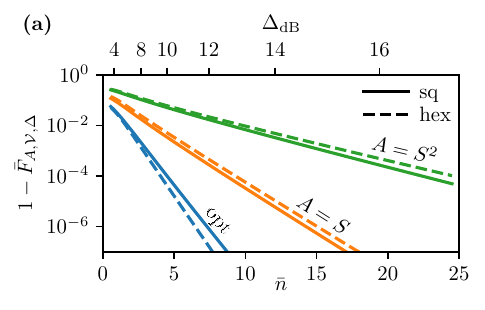}
\hspace*{\fill}
\includegraphics{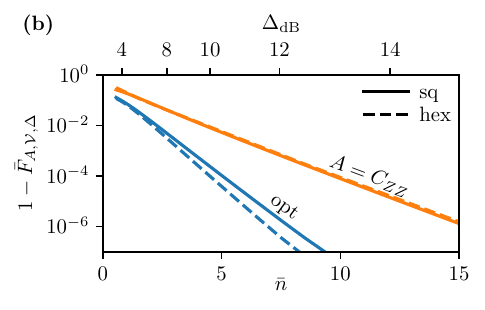}

\caption{\textcolor{black}{Average Gate Infidelity $1-\bar{F}_{A,\mathcal{V}_{\mathcal{L}},\Delta}$ of (a) single-qubit, and (b) two-qubit GKP Clifford gates $A=I,S,S^{2},C_{Z}$ applied to the square and hexagonal GKP codes, decoded over the Voronoi cell $\mathcal{V}_{\mathcal{L}}$ of their respective lattices. The average gate infidelity of the $S, S^{2}$ and $C_{Z}$ gates can be reduced to be the same as that of the identity gate by instead performing ideal error correction over a modified patch $\mathcal{P}=\mathcal{S}(\bar{A})\mathcal{V}_{\mathcal{L}}$ immediately following the application of the gate, where $\mathcal{S}(\bar{A})$ is a symplectic matrix with unitary representation equal to $\bar{A}$. The plots labeled opt represent the average gate fidelity of the identity gate $\bar{F}_{I,\mathcal{V},\Delta}$ and that of Clifford gates with a modified decoding patch $\bar{F}_{\bar{A},\mathcal{S}(\bar{A})\mathcal{V},\Delta}$. Note that the square and hexagonal $C_{Z}$ gates have similar average gate infidelities and are overlapping in (b).}}

\label{fig:gate_error_plots}
\end{figure*}

To counteract the effects of the spreading of errors, one can instead use a modified decoding patch given by
\begin{equation}\label{eq:modified_patch_ideal}
    \mathcal{S}(\bar{A})\mathcal{V}_{\mathcal{L}}=\{\mathcal{S}(\bar{A})\vect{v}\>|\>\vect{v}\in\mathcal{V}_{\mathcal{L}}\}.
\end{equation}
In the case of ideal error correction, the modified patch \cref{eq:modified_patch_ideal} exactly counteracts the spreading of errors due to the Clifford gates, since \cref{eq:Clifford_gate_estimate} guarantees that $\bar{F}_{A,\mathcal{S}(\bar{A})\mathcal{V}_{\mathcal{L}},\Delta}=\bar{F}_{I,\mathcal{V}_{\mathcal{L}},\Delta}$, see \cref{fig:phase_transformation}.

We quantify the effect of the modification~\cref{eq:modified_patch_ideal} both analytically using \cref{eq:Clifford_gate_estimate,eq:avg_fid_estimate} and numerically using the SSSD. In~\cref{tab:estimates}, we present the distance $d$ and degeneracy $a$ of the Voronoi cells of the square and hexagonal GKP codes, as well as the deformed patches $\mathcal{S}(\bar{A})^{-1}\mathcal{V}_{\mathcal{L}}$ due to various single- and two-qubit Clifford gates. The latter deformed patches represent the performance of the gates if the modification~\cref{eq:modified_patch_ideal} is \textit{not} performed, while the Voronoi cells represent the performance with the modification. In \cref{fig:gate_error_plots}, we present the corresponding average gate fidelities calculated numerically. We moreover show in \cref{fig:gate_error_estimate_comparison} in \cref{sec:gate_error_estimate} that the theoretical estimates given by \cref{eq:avg_fid_estimate,tab:estimates} are in agreement with the numerical results from \cref{fig:gate_error_plots} for small enough values of $\Delta$.

Our results show that to achieve a $10^{-2}$ error rate in a $C_{ZZ}$ gate (even in the absence of gate noise), one would need a squeezing of $\Delta_{\mathrm{dB}}\approx 7.5$ (for square GKP) with the modification in \cref{eq:modified_patch_ideal}, compared to a squeezing of $\Delta_{\mathrm{dB}}\approx9.8$ without it. Meanwhile, for fixed $\Delta_{\mathrm{dB}}=12$, the modification improves the average gate infidelity by roughly two orders of magnitude for both the square and hexagonal GKP codes.

Moreover, we find that the hexagonal code outperforms the square code for the identity gate, as expected from the geometry of their respective logical lattices. However, Clifford gates typically decrease the distance of the hexagonal code patch more than for square code, making the use of the modified patch \cref{eq:modified_patch_ideal} more important.
%In the context of a full computation, we envisage that the round of error correction immediately following the implementation of a logical gate is performed over the corresponding modified patch, while all other rounds are performed over the Voronoi cell. This modification is a generalization of ``error-protected'' two-qubit gates, which have already been applied in GKP-surface code simulations~\cite{Noh22}.

%In both the square and hexagonal GKP codes, the patch $\mathcal{S}(\bar{S})\mathcal{V}_{\mathcal{L}}$ corresponding to the logical phase gate $\bar{S}$ has a shorter distance than the original Voronoi cell $\mathcal{V}_{\mathcal{L}}$, and therefore $\bar{S}$ propagates errors due to $\Delta$. Moreover, subsequent applications of the phase gate further decrease the distance of the corresponding patch, even though $S^{2}=Z$ and $S^{4}=I$ at a logical level. A similar reduction in distance occurs for the two-qubit $\bar{C}_{ZZ}$ gate. In the square code, generalized controlled gates involving $Y$ such as $\bar{C}_{ZY}$ and $\bar{C}_{YY}$ produce a further reduction in distance compared to $\bar{C}_{ZZ}$. This is not seen in the hexagonal code, since all generalized controlled gates are related to each other via conjugation by the permutation gate $\bar{H}\bar{S}^{\dag}$.

\begin{table}[]
    \caption{Ideal Clifford gate patches. Summary of the distances $d$ and degeneracies $a$ of the patches $\mathcal{S}(\bar{A})^{-1}\mathcal{V}_{\mathcal{L}}$ corresponding to the GKP Clifford gates $\bar{A}=\bar{I},\bar{S},\bar{S}^{2},\bar{I}{\otimes}\bar{I},\bar{C}_{ZZ},\bar{C}_{YY}$ for the square and hexagonal GKP codes decoded over their Voronoi cells $\mathcal{V}_{\mathcal{L}}$. The average gate fidelity of each gate is estimated by \cref{eq:avg_fid_estimate}, but can be improved to the average gate {fidelity} of the identity gate by modifying the patch over which error correction is performed to $\mathcal{P}=\mathcal{S}(\bar{A})\mathcal{V}_{\mathcal{L}}$. Note that using elementary geometry we can obtain exact expressions for these values, which are displayed in \cref{tab:estimates_full}.}
    \begin{center}
%    \begin{tabular}{c|>{\centering\arraybackslash}m{0.8cm}|c|>{\centering\arraybackslash}m{0.8cm}|c}
%        & \multicolumn{2}{c|}{Square} & \multicolumn{2}{c}{Hexagonal} \\
%        \hline
%        Gate&$a$& $d/\sqrt{\pi}$ & $a$ & $d/\sqrt{\pi}$\rule{0pt}{1em}\rule[-4pt]{0pt}{1em} \\
%        \hline
%        $\bar{I}$ & 2 & 1 & 3 & $1.075$\rule{0pt}{1.6\normalbaselineskip}\\
%        $\bar{S}$ & 1 & $0.707$ & 2 & $0.703$\rule{0pt}{1.6\normalbaselineskip} \\
%        $\bar{S}^{2}$ & 1 & $0.447$ & 1 & $0.427$\rule{0pt}{1.6\normalbaselineskip} \\
%        $\bar{I}\otimes\bar{I}$ & 4 & 1 & 6 & $1.075$\rule{0pt}{1.6\normalbaselineskip} \\
%        $\bar{C}_{ZZ}$ & 2 & $0.707$ & 2 & \rule[-6pt]{0pt}{4pt}$0.703$\rule{0pt}{1.6\normalbaselineskip} \\
%        $\bar{C}_{YY}$ & 4 & $0.577$ & 2 & \rule[-6pt]{0pt}{4pt}$0.703$\rule{0pt}{1.6\normalbaselineskip}
%    \end{tabular}
    \includegraphics{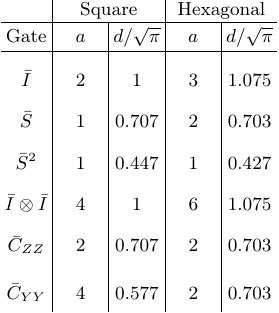}
    \label{tab:estimates}
    \end{center}
\end{table}

Since the modification in \cref{eq:modified_patch_ideal} eliminates the errors caused by the spreading of errors by the gate, the new leading source of error will now be from performing \textit{approximate} error correction. This is a significant consideration because errors can now occur both before \textit{and} after the gate, causing some errors to be spread by the gate, and others not. As a result, neither the original patch $\mathcal{V}_{\mathcal{L}}$ nor the modified patch $\mathcal{S}(\bar{A})^{-1}\mathcal{V}_{\mathcal{L}}$ optimally correct this noise, reducing the performance of the gate.

To tackle this issue we define a third correction patch that accounts for the partially-spread distribution of errors. We derive the required patch by using an analytical approximate expression analogous to \cref{eq:avg_fid_estimate} that incorporates the use of approximate codestates in the QEC cycle. We show that in general the spread of errors due to the gate can only be partly corrected even when the optimal patch is chosen. From this point on, we only provide analytical formulae due to the good agreement between numerical and analytic results in the ideal case, as shown in \cref{fig:gate_error_estimate_comparison}.

\begin{figure}
    \includegraphics{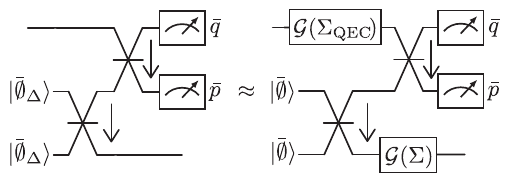}
    
    \caption{\textcolor{black}{Quantum circuits for teleportation-based QEC using approximate GKP codewords. The left circuit is the teleportation-based QEC circuit using approximate GKP qunaught states as in \cref{fig:tele}, while the right circuit consists of an ideal QEC cycle preceded and followed by Gaussian random displacement channels with $\Sigma=\tanh(\Delta^{2}/2)I$ and $\Sigma_{\text{QEC}}$ given in \cref{eq:Sigma_QEC}. We show in \cref{sec:approx_Clifford_gates} that the left and right circuits are approximately equal.}}
    
    \label{fig:tele_general_approx}
\end{figure}

%We begin by redefining the metric that we use to quantify the quality of logical gates.
In order to capture the effects of approximate quantum error correction, we include an additional Gaussian random displacement channel in our model.
%In particular, we include an additional Gaussian random displacement channel to (approximately) account for the use of approximate codestates in the error-correction procedure as we now explain.
The new channel arises from performing a teleportation-based QEC cycle (\cref{fig:tele_general_approx}) with approximate GKP codestates, as we show in \cref{sec:approx_Clifford_gates}. % and its proposed use in optical implementations of the GKP code~\cite{Walshe20}; for an analysis of two-qubit gates with the small-Big-small scheme we refer the reader to Ref.~\cite{Rojkov23}.

Explicitly, instead of \cref{eq:logical_channel}, we define the logical channel as
\begin{equation}\label{eq:logical_channel_approx}
    \mathcal{E}_{U,\mathcal{P},\Delta}=\mathcal{J}[R^{\dag}]\circ\mathcal{C}_{\mathcal{P}}\circ\mathcal{G}(\Sigma_{\text{QEC}})\circ\mathcal{J}[\bar{U}R_{\Delta}],
\end{equation}
where $\mathcal{G}(\Sigma_{\text{QEC}})$ is a Gaussian random displacement (GRD) channel
\begin{multline}\label{eq:GRD}
    \mathcal{G}(\Sigma)(\rho)\\
    =\frac{1}{\sqrt{\det\Sigma}}\!\iint\! \frac{d^{2}\vect{v}}{(2\pi)^{n}}e^{-\vect{v}^{T}\Sigma^{-1}\vect{v}/2}\,T(\vect{v})\rho T(\vect{v})^{\dag},
\end{multline}
and the covariance matrix $\Sigma_{\text{QEC}}$ is given explicitly by
\begin{equation}\label{eq:Sigma_QEC}
    \Sigma_{\text{QEC}}=\tanh(\Delta^{2}/2)M^{}_{\mathcal{L}}P^{}_{1}M^{-1}_{\mathcal{L}}M^{-T}_{\mathcal{L}}\!P^{}_{1}M^{T}_{\mathcal{L}}
\end{equation}
with $M_{\mathcal{L}}=[\vect{\alpha},\vect{\beta}]/\sqrt{\pi}$ and $P_{1}=\mathrm{diag}(-1,1)$. \cref{eq:Sigma_QEC} arises from explicitly calculating the effect of approximate codestates on the ancilla qubits as depicted in \cref{fig:tele_general_approx} (see \cref{sec:approx_Clifford_gates} for a full derivation). Note that for the square GKP code, $\Sigma_{\text{QEC}}=\mathrm{tanh}(\Delta^{2}/2)I$, but in general $\Sigma_{\text{QEC}}$ may not be proportional to the identity matrix. 

Note that we have assumed that the amount of noise introduced by the approximate codestates on the data and ancilla qubits is equal, but there are multiple reasons why this may not be the case in practice. First, additional noise could occur on the data qubit prior to error correction due to loss and/or dephasing. {And second, inefficiency in the homodyne detection can also be modeled under the twirling approximation as Gaussian random displacement noise immediately preceding the measurement.} Such additional noise could, in principle, be accounted for using the same tools as below by modifying the covariance matrix describing the noise. {We also note that it is possible to use the methods described later in \cref{sec:measurements} to reduce the inefficiency of the homodyne measurements (at the cost of using an additional physical mode).}

With the definition of the logical channel in \cref{eq:logical_channel_approx}, we can now calculate the average gate fidelity $\bar{F}_{U,\mathcal{P},\Delta}$. In \cref{sec:approx_Clifford_gates}, we show that for a Clifford gate $A$, the average gate infidelity is approximated by
\begin{equation}\label{eq:avg_fid_estimate_approx}
    1-\bar{F}_{A,\mathcal{P},\Delta}\approx \frac{2^{n}a_{\text{eff}}}{2^{n}+1}\mathrm{erfc}\big(d_{\text{eff}}/(2\sqrt{2}\Delta)\big),
\end{equation}
where $a_{\text{eff}}$ and $d_{\text{eff}}$ are the effective degeneracy and distance of a modified patch $\mathcal{P}'$ that optimally accounts for the additional noise arising from $\Sigma_{\text{QEC}}$.

\begin{figure}
    \includegraphics{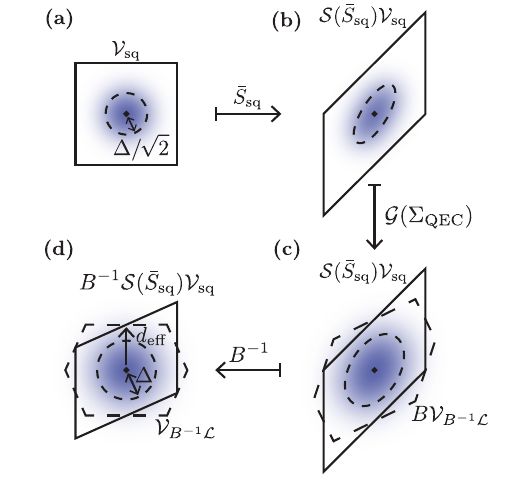}
    \caption{Visual representation of the optimal patch and effective distance of the $\bar{S}_{\text{sq}}$ gate when approximate error-correction is considered. (a) An approximate GKP state, represented (using the twirling approximation) as an ideal state that has incurred Gaussian random displacement error with standard deviation ${\sim} \Delta/\sqrt{2}$. (b) The logical phase gate $\bar{S}_{\text{sq}}$ is applied, deforming the correction patch and the spread of the noise. (c) Finally, an additional Gaussian random displacement channel $\mathcal{G}(\Sigma_{\text{QEC}})$ is applied to incorporate the effects of approximate error correction. The optimal correction patch (dashed) can be found by performing a (non-orthogonal) transformation $B^{-1}$ that reshapes the Gaussian noise to be symmetrical with standard deviation ${\sim} \Delta$. (d) The optimal patch is the Voronoi cell of the transformed lattice, $\mathcal{V}_{B^{-1}\mathcal{L}}$.}\label{fig:phase_approx_transformation}
\end{figure}

It is important to note that there is an additional $\sqrt{2}$ factor in \cref{eq:avg_fid_estimate_approx} compared to \cref{eq:avg_fid_estimate}. This factor arises from the approximate codestates used in the error-correction circuit, and can greatly increase the error rates even in the best-case scenario where the distance remains unchanged, i.e.~$d_{\text{eff}}$ in \cref{eq:avg_fid_estimate_approx} is equal to $d$ in \cref{eq:avg_fid_estimate}. Indeed, multiplying $\Delta$ by a factor of $\sqrt{2}$ is equivalent to subtracting $10\log_{10}(2)\approx3$ dB from the GKP squeezing $\Delta_{\mathrm{dB}}$ -- or, to be concrete, approximate error correction on a 10 dB codestate performs roughly the same as ideal error correction on a 7 dB codestate.

In addition to the unavoidable $\sqrt{2}$ factor introduced by approximate codestates, the effective distance may also alter the error rate, Intuitively, one can understand the origin of $d_{\text{eff}}$ as follows (see \cref{fig:phase_approx_transformation}). First, there is some incoming noise from the approximate GKP codestate, which we approximate as a GRD channel with a covariance matrix $\Sigma=\tanh(\Delta^{2}/2)I$ [\cref{fig:phase_approx_transformation}(a)]. Then, the Clifford gate $\bar{A}$ is applied, spreading the errors and transforming the covariance matrix $\Sigma\mapsto \mathcal{S}(\bar{A})\Sigma\mathcal{S}(\bar{A})^{T}$ [\cref{fig:phase_approx_transformation}(b)].
%At this point, one can modify the patch to exactly counteract the spreading using \cref{eq:modified_patch_ideal}.
Next, we apply the GRD channel $\mathcal{G}(\Sigma_{\text{QEC}})$ that is associated with approximate error-correction, so that the overall noise is given by a GRD channel with covariance matrix $\Sigma_{\text{tot}}=\mathcal{S}(\bar{A})\Sigma\mathcal{S}(\bar{A})^{T}+\Sigma_{\text{QEC}}$ that we now must decode [\cref{fig:phase_approx_transformation}(c)].

To understand how best to decode this noise, we define a (non-symplectic) transformation of phase-space that maps $\Sigma_{\text{tot}}$ to a matrix proportional to the identity [\cref{fig:phase_approx_transformation}(d)]. Labeling the transformation matrix $B^{-1}$, the optimal patch to decode over is then the Voronoi cell $\mathcal{V}_{B^{-1}\mathcal{L}}$ of the transformed lattice $B^{-1}\mathcal{L}$, and the effective distance $d_{\text{eff}}$ is the shortest vector in this lattice. On the physical phase-space [\cref{fig:phase_approx_transformation}(c)] the correction patch is then $B\mathcal{V}_{B^{-1}\mathcal{L}}$.

In \cref{tab:estimates2}, we present the effective degeneracy and distance corresponding to various Clifford gates for the square and hexagonal code, assuming that the modified patch is optimal as described above. Importantly, note that \cref{tab:estimates2} displays the performance of the \textit{optimal} patch against Clifford gates with approximate error-correction, while \cref{tab:estimates} displays the performance of the na\"ive patch in the ideal error-correction case. To make this point explicit, consider the performance of the square GKP $\bar{S}$ gate. In the case of ideal error-correction, the effective distance of the $\bar{S}$ gate can be improved from 0.707$\sqrt{\pi}$ in the na\"ive case to $\sqrt{\pi}$ using \cref{eq:modified_patch_ideal}. In the approximate QEC case, the largest possible effective distance is 0.894$\sqrt{\pi}$ even with an optimally chosen patch.

%\mhs{Updated these numbers with all relevant vectors.} $10^{-2}$ target two-qubit infidelity: square identity needs 10.461 dB, CZZ needs 10.972 (SxS CZZ needs 11.143), CYY needs 11.364. for $10^{-3}$ it's 12.187, 12.757 (12.685), 12.827; for $10^{-4}$ it's 13.440, 14.081 (13.853), 13.935. Crazy how long it takes for the effects of d to catch up with the effects of a! For squeezing $\Delta_{\text{dB}}=10$, you get square error rates of 1.62\%, 2.64\% (3.68\%), 4.87\%; $\Delta_{\text{dB}}=12$ is 0.134\%, 0.292\% (0.304\%), 0.402\%; $\Delta_{\text{dB}}=14$ is $2.86\times10^{-5}$, $11.7\times 10^{-5}$ ($7.18\times10^{-5}$), $8.57\times10^{-5}$. Cross-over between CZZ and SxS CZZ is 12.204 dB. Cross-over between CZZ and CYY is 13.171 dB, and between SxS CZZ and CYY is 15.495.

In general, Clifford gates introduce errors that can only be partly compensated for when approximate QEC is taken into account. Roughly speaking, the two-orders-of-magnitude improvement observed in the ideal case becomes halved to a one-order-of-magnitude improvement in the approximate case. Concretely, to achieve a $10^{-2}$ $C_{ZZ}$ average gate infidelity requires codestates with $\Delta_{\text{dB}}\approx 11.0$ dB of squeezing in the square GKP code. Note that this is roughly $3.5$ dB \textit{more} squeezing than is required to achieve this infidelity with ideal QEC, indicating that there are contributions both from the 3 dB introduced intrinsically by the approximate codestates and from the decreased effective distance of the correction patch.

%10.4 dB for hex 10^-2 HxH CZZ vs 10.6 without the H's

With these general comments made, there are still two interesting features of \cref{tab:estimates2} that we point out now. First, the hexagonal GKP code is outperformed by the square GKP code for the identity gate. This is because just two logical quadratures are measured in \cref{fig:tele_general_approx}, causing an asymmetry in the noise in $\Sigma_{\text{QEC}}$ and breaking the 6-fold rotational symmetry of the hexagonal code. Indeed, the effective degeneracy for the identity gate is 2 for the hexagonal code in \cref{tab:estimates2}, compared to 3 in \cref{tab:estimates}. However, the logical Hadamard gate spreads the noise in such a way that the asymmetry in QEC is \textit{exactly} counteracted, restoring the full effective distance and degeneracy of the hexagonal code (see \cref{sec:approx_Clifford_gates}). Therefore, the hexagonal code only outperforms the square code if a logical Hadamard gate is \textit{physically} implemented before each round of teleportation-based QEC. Note that the hexagonal logical Hadamard gate $\bar{H}_{\text{hex}}$ is \textit{not} a rotation, and single-mode squeezing is required to implement it. Interestingly, the logical $\bar{S}_{\text{hex}}^{\dag}=e^{-i\bar{q}^{2}/2}$ gate also has this property -- meaning that arbitrary single-qubit Clifford gates can be performed while preserving the maximum effective distance during each round of QEC.

%Moreover, we show in \cref{sec:approx_Clifford_gates} that under the model of approximate error-correction we consider, the square GKP code has the best performance against the identity gate out of any single-mode GKP code. Fundamentally,  Note that this result may be sensitive to the specific model of approximate error-correction used here and may change if noise such as loss is also assumed to occur before error correction is performed.

Taking this into account, the hexagonal code also outperforms the square code for the $\bar{C}_{ZZ}$ gate. Specifically, we compare the performance of the between the hexagonal $(\bar{H}\otimes\bar{H})\bar{C}_{ZZ}$ gate and the square $\bar{C}_{ZZ}$ gate. To achieve a target two-qubit gate infidelity of $10^{-2}$, the square GKP code requires codestates with $\Delta_{\text{dB}}=11.0$ while the hexagonal code requires $\Delta_{\text{dB}}=10.8$. Put differently, codestates with $\Delta_{\text{dB}}=12$ achieve a gate infidelity of 0.29\% in the square code and 0.20\% in the hexagonal code. Note that to obtain these values we have used a slightly more accurate formula than \cref{eq:avg_fid_estimate_approx} that takes into account the effects of all Voronoi-relevant vectors of the patch $\mathcal{P}'$ -- as we explain in \cref{sec:approx_Clifford_gates}. To summarize, the hexagonal code modestly outperforms the square code at the cost of requiring single-mode squeezing during every round of QEC.

Finally, remarkably, the effective distance of the square $\bar{C}_{YY}$ gate is exactly $\sqrt{\pi}$ -- the same as that of the identity gate. This is particularly surprising given that this is not true for the $\bar{C}_{ZZ}$ gate, nor for any two-qubit gate of the hexagonal code. In \cref{sec:approx_Clifford_gates} we show that the optimal patch for the $\bar{C}_{YY}$ gate is in fact related to the Voronoi cell of the $D_{4}$ lattice -- the densest lattice-packing in 4D -- giving rise to the very high effective degeneracy of the patch. However, it is worth noting that $\Delta_{\text{dB}}$ must be reasonably large for the improved effective distance to decrease the infidelity enough to make up for the increased effective degeneracy $a_{\text{eff}}=12$ -- indeed, the square $\bar{C}_{YY}$ gate only outperforms the $\bar{C}_{ZZ}$ gate when $\Delta_{\text{dB}}>13.2$. We also briefly comment that the distance of the $\bar{C}_{ZZ}$ gate can be increased by applying single-qubit gates immediately afterward -- however the advantages of this are small and only apply for $\Delta_{\text{dB}}>12.2$ (see \cref{sec:approx_Clifford_gates} for more details). In contrast, each generalized controlled gate performs identically in the hexagonal code due to the 6-fold rotational symmetry of the code. We leave further investigations of the consequences of this to future work.

\begin{table}[]
    \begin{center}
    \caption{Clifford gate patches taking into account approximate error correction. Summary of the effective distances $d_{\text{eff}}$ and degeneracies $a_{\text{eff}}$ of the optimal patches $\mathcal{P}(\bar{A})$ corresponding to various Clifford gates followed by approximate error correction for the square and hexagonal GKP codes. The average gate fidelity of each gate is estimated by \cref{eq:avg_fid_estimate_approx}. Note that these values can all be evaluated analytically, as displayed in \cref{tab:estimates2_full}. Finally, note that Hadamard gates do not affect the performance of the square GKP code, so that the two-qubit gates $(\bar{H}\otimes\bar{H})\bar{C}_{ZZ}$ and $(\bar{H}\otimes\bar{H})\bar{C}_{YY}$ perform identically to $\bar{C}_{ZZ}$ and $\bar{C}_{YY}$ respectively.}
%    \begin{tabular}{c|>{\centering\arraybackslash}m{0.8cm}|c|>{\centering\arraybackslash}m{0.8cm}|c}
%        Code & \multicolumn{2}{c|}{Square} & \multicolumn{2}{c}{Hexagonal} \\
%        \hline
%        Gate&$a_{\text{eff}}$& $d_{\text{eff}}/\sqrt{\pi}$ & $a_{\text{eff}}$ & $d_{\text{eff}}/\sqrt{\pi}$\rule{0pt}{1em}\rule[-4pt]{0pt}{1em} \\
%        \hline
%        $\bar{I}$ & 2 & 1 & 2 & $0.931$\rule{0pt}{1.6\normalbaselineskip}\\
%        $\bar{H}$ & 2 & 1 & 3 & $1.075$\rule{0pt}{1.6\normalbaselineskip}\\
%        $\bar{S}$ & 1 & $0.894$ & 1 & $0.703$\rule{0pt}{1.6\normalbaselineskip} \\
%        $\bar{S}^{\dag}$ & 1 & $0.894$ & 3 & $1.075$\rule{0pt}{1.6\normalbaselineskip} \\
%        $\bar{I}\otimes\bar{I}$ & 4 & 1 & 4 & $0.931$\rule{0pt}{1.6\normalbaselineskip} \\
%        $\bar{H}\otimes\bar{H}$ & 4 & 1 & 6 & $1.075$\rule{0pt}{1.6\normalbaselineskip} \\
%        $(\bar{H}\otimes\bar{H})\bar{C}_{ZZ}$ & 2 & $0.894$ & 2 & $0.931$\rule{0pt}{1.6\normalbaselineskip} \\
%        $(\bar{H}\otimes\bar{H})\bar{C}_{YY}$ & 12 & 1 & 2 & $0.931$\rule{0pt}{1.6\normalbaselineskip}
%    \end{tabular}
    \includegraphics{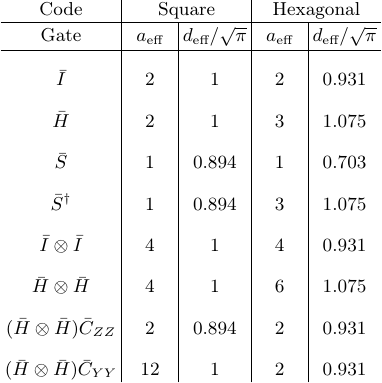}
    \label{tab:estimates2}
    \end{center}
\end{table}

\section{\label{sec:error_estimates}Analytical Error Estimates for Loss and Dephasing}

In the previous section, we developed an analytical formula \cref{eq:avg_fid_estimate} that estimates the average gate fidelity of a Clifford gate acting on approximate GKP codestates. In this section, we extend these methods to general noise channels such as loss and dephasing. The result of this is that we obtain a random displacement channel that approximates the effects of loss and dephasing on approximate GKP codestates. The obtained channel has two key advantages: first, random displacement channels are efficient to simulate even for multi-mode systems, and second, the resulting fidelity expressions agree quite well with the exact expressions, with less than a 1\% relative error for loss and less than 20\% for dephasing in the regime of interest. In the case of loss, this vastly outperforms previously-used random displacement channel approximations as used, for example, in GKP-qubit code concatenation studies such as Refs.~\cite{Noh20,Noh22},\ because it is able to capture the joint effect of loss acting upon \textit{approximate} GKP codestates. Finally, we also extend our methods to numerically analyze Kerr non-linearities that can arise from performing two-qubit gates in superconducting GKP qubits, although the formula we develop is not as convenient to analyze theoretically.

To obtain our results we use the GKP SSSD to formally write down the average gate fidelity of the \textit{logical} channel that represents the effect of the noise channel on the GKP code (Sec.~VII of Ref.~\cite{Shaw24-1}). Our resulting estimate is essentially the first-order approximation of that formal expression. Such an analytical error estimate offers both theoretical and practical insights. Our analytical error estimate is closely related to the \textit{twirling approximation} applied to the noise-plus-envelope channel~\cite{Menicucci14}, which has been previously used to model the envelope operator in simulations of GKP-qubit code concatenations~\cite{Vuillot19,Noh20,Noh22}. As such, our results both justify the twirling approximation and generalize it to arbitrary noise channels.
%To demonstrate this generality, we apply our estimate to the most common noise channels affecting bosonic modes: loss and dephasing; and we compare our estimates with numerically obtained error rates.
We can use our estimates to, for example, derive the optimal $\Delta_{\text{dB}}$ that minimizes the error given some amount of loss and/or dephasing. Importantly, our estimates are only valid when loss and dephasing are applied to \textit{approximate} GKP codes.
%These results may be of broader interest to those wishing to model realistic noise channels using random displacements, such as in GKP-qubit code concatenation studies.

We begin once again by defining the metric we use to quantify the error due to loss and dephasing. Similarly to in \cref{sec:gates}, we use the average gate fidelity~\cref{eq:avg_gate_fid_defn} to characterize the performance of the GKP code against noise as follows. Given a CV noise channel $\mathcal{N}$, we define a \textit{logical} noise channel
\begin{equation}\label{eq:logical_noise_channel}
    \mathcal{N}_{\mathcal{L}}=\mathcal{J}[R^{\dag}]\circ\mathcal{C}_{\mathcal{P}}\circ\mathcal{N}\circ\mathcal{J}[R],
\end{equation}
where $R=\big(\!\ket{\bar{0}}\!\bra{0}+\ket{\bar{1}}\!\bra{1}\!\big)^{\otimes n}$ is the ideal encoding map, and $\mathcal{C}_{\mathcal{P}}$ represents a round of ideal error correction over the patch $\mathcal{P}$. Comparing to the definition of the logical gate channel $\mathcal{E}_{U,\mathcal{P},\Delta}$ in \cref{eq:logical_channel} there are two key differences: the logical operator $\bar{U}$ in \cref{eq:logical_channel} is replaced with the noise channel $\mathcal{N}$ in \cref{eq:logical_noise_channel}, and the states are initially encoded in an ideal GKP codestate $\ket{\bar{\psi}}$ in \cref{eq:logical_noise_channel} instead of an approximate GKP codestate $\ket{\bar{\psi}_{\Delta,\text{o}}}$ in \cref{eq:logical_channel}. However, this is without loss of generality: to account for approximate codewords in \cref{eq:logical_noise_channel}, we simply include the envelope operator $e^{-\Delta^{2}a^\dag a}$ in $\mathcal{N}$.

We aim to derive an approximate expression for the average gate infidelity $\bar{F}$ of the logical noise map $\mathcal{N}_{\mathcal{L}}$. We present the full derivation in \cref{sec:gate_error_estimate} using the SSSD~\cite{Shaw24-1}, and focus on the main results here. All the expressions below depend on the \textit{characteristic function} $c(\vect{u},\vect{v})$ of the noise map $\mathcal{N}$, which we define such that
\begin{equation}\label{eq:char_func_defn}
    \mathcal{N}(\rho)=\iint\!d^{2}\vect{u}\,d^{2}\vect{v}\,c(\vect{u},\vect{v})\,T(\vect{u})\rho T(\vect{v})^{\dag}.
\end{equation}
Explicitly, one can obtain the characteristic function $c$ of a general CV channel from the characteristic functions $c_{i}(\vect{v})=\mathrm{tr}\big(E_{i}T(\vect{v})^{\dag}\big)/(2\pi)$ of any Kraus operator representation of $\mathcal{N}(\rho)=\sum_{i}E^{}_{i}\rho E_{i}^{\dag}$ through the equation $c(\vect{u},\vect{v})=\sum_{i}c_{i}(\vect{u})c_{i}(\vect{v})^{*}$. For Gaussian channels, one can also use the formulae presented in Appendix F of Ref.~\cite{Shaw24-1}{, although note that due to a different definition of the displacement operator the formulas differ in some places by a factor of $2\pi$}. The characteristic function also generalizes straightforwardly to multi-mode channels; however, in this section we focus on the single-mode case.

We make two assumptions in our derivation, that
\begin{enumerate}
    \item the noise map $\mathcal{N}$ is approximately proportional to a CPTP map, and
    \item the characteristic function $c$ decays sufficiently quickly as $|\vect{u}|,|\vect{v}|\rightarrow\infty$.
\end{enumerate}
The first assumption would be redundant if $\mathcal{N}$ were trace-preserving; however, since we incorporate the non-unitary envelope operator into $\mathcal{N}$, the assumption is valid only when $\Delta$ is small. The second assumption (which we discuss in more detail in \cref{sec:gate_error_estimate}) essentially requires that the noise channel can be written as a superposition of sufficiently small displacement operators; or, more intuitively, that $\mathcal{N}$ is ``close to the identity channel'', since the characteristic function of the identity channel is $\delta^{2}(\vect{u})\delta^{2}(\vect{v})$. Importantly, for this assumption to be valid for loss or dephasing, we must include the envelope operator (or any other regularizing operator) before the loss or dephasing channel.

Given the above assumptions, the average gate infidelity of the logical noise map \cref{eq:logical_noise_channel} is given by
\begin{equation}\label{eq:avg_gate_infidelity_ratio}
    1-\bar{F}\approx\frac{2}{3}\iint_{\mathrlap{\mathbb{R}^{2}\setminus\mathcal{P}}}\;d^{2}\vect{v}\,c(\vect{v},\vect{v})\bigg/\!\!\iint_{\mathbb{R}^{2}}\!\!\!d^{2}\vect{v}\,c(\vect{v},\vect{v})
\end{equation}
in the single-mode case. Note here that incorporating the envelope operator (or any other regularizing operator) into $\mathcal{N}$ is necessary since omitting the envelope operator would make $\mathcal{N}$ trace-preserving and the denominator would diverge.

Interestingly, since \cref{eq:avg_gate_infidelity_ratio} only depends on the diagonal, ``stochastic'' elements of the characteristic function $c(\vect{v},\vect{v})$, the result would be the same as if we had applied the \textit{twirling approximation} to the original noise channel. In particular, since the right-hand side of \cref{eq:avg_gate_infidelity_ratio} depends only on the diagonal elements of $c$, and these diagonal elements are real and non-negative, $\mathcal{N}$ could be replaced by a random displacement channel $\mathcal{E}_{p}$ over the probability distribution $p(\vect{v})\propto c(\vect{v},\vect{v})$ given by
\begin{equation}
    \mathcal{E}_{p}(\rho)=\iint d^{2}\vect{v}\,p(\vect{v})T(\vect{v})\rho T(\vect{v})^{\dag}.
\end{equation}
Such a channel may be obtained from $\mathcal{N}$ by uniformly twirling $\mathcal{N}$ over the group of displacement operators, i.e.
\begin{equation}\label{eq:CV_twirl_def}
    \mathcal{E}_{p}=\iint \frac{d^{2}\vect{v}}{2\pi}\mathcal{J}[T(\vect{v})^{\dag}]\circ\mathcal{N}\circ\mathcal{J}[T(\vect{v})],
\end{equation}
which we refer to the twirling approximation of $\mathcal{N}$. Note that the twirling approximation is a less-restrictive approximation than the two assumptions we made in arriving at \cref{eq:avg_gate_infidelity_ratio}; because, for example, in \cref{eq:avg_gate_infidelity_ratio} we have disregarded displacements that take you outside the patch $\mathcal{P}$ but do not cause a logical error, such as those corresponding to stabilizers.

The twirling approximation was first introduced in Ref.~\cite{Menicucci14} and is commonly used in studies on concatenated GKP-qubit codes \cite{Fukui18,Fukui23,Vuillot19,Noh20,Noh22} to approximate the envelope operator $e^{-\Delta^{2}a^\dag a}$ as a random Gaussian displacement noise channel with variance $\sigma^{2}=\Delta^{2}/2$. The twirling approximation can also be viewed as a CV generalization of the qubit \textit{Pauli twirling} operation that is used to approximate noise models in qubit QEC code simulations and to prevent the build-up of coherent errors in experiments~\cite{Cai19} -- although note that in the CV case the operation written in \cref{eq:CV_twirl_def} is unphysical as it requires the application of arbitrarily large displacements. The (CV) twirling approximation of the envelope operator has been justified theoretically~\cite{Noh20} and numerically~\cite{Hillmann22}. Therefore, this work justifies the twirling approximation using the SSSD and generalizes it to noise channels of interest that have not been analyzed using the twirling approximation before.

We can evaluate \cref{eq:avg_gate_infidelity_ratio} in the case when the diagonal elements of the characteristic function take the form
\begin{equation}\label{eq:diagonal_characteristic_function_ansatz}
    c(\vect{v},\vect{v})\propto e^{-|\vect{v}|^{2}/2\sigma^{2}},
\end{equation}
i.e.~the twirled channel $\mathcal{E}_{p}$ is a \textit{Gaussian} random displacement channel with variance $\sigma^{2}$. In this case, the approximations we have made are valid if $\sigma$ is much smaller than the length of the shortest logical operator. Then, we have
\begin{equation}\label{eq:avg_gate_infid_sigma}
    1-\bar{F}\approx\frac{2}{3}a\,\mathrm{erfc}\bigg(\frac{d}{2\sqrt{2}\sigma}\bigg),
\end{equation}
where $d$ is the distance and $a$ is the degeneracy of the patch $\mathcal{P}$, as described in \cref{sec:GKP}.

\cref{eq:avg_gate_infid_sigma} can be applied to several cases of interest. First, we consider a noise map consisting solely of the envelope operator itself, i.e.~$\mathcal{N}=\mathcal{J}[e^{-\Delta^{2}a^{\dag}a}]$. The diagonal elements of the characteristic function of the envelope operator take the form of \cref{eq:diagonal_characteristic_function_ansatz} with variance
\begin{equation}
    \sigma^{2}=\mathrm{tanh}(\Delta^{2}/2)\approx\Delta^{2}/2.
\end{equation}
As a sanity check, substituting $\sigma=\Delta/\sqrt{2}$ into \cref{eq:avg_gate_infid_sigma} indeed gives \cref{eq:avg_fid_estimate} with $n=1$. In words, this means that for sufficiently small values of $\Delta$, the logical action of the envelope operator on ideal codewords is approximately the same as the logical action of a Gaussian random displacement channel with variance $\sigma^{2}=\Delta^{2}/2$, as has already been utilized widely in GKP-qubit code concatenation studies.

However, we can also apply our estimate to other maps of interest. For example, consider the noise map given by the composition of the envelope operator with a loss of $\gamma$. We define loss as evolution under the master equation $\dot{\rho}=\kappa\big([a\rho,a^{\dag}]+[a,\rho a^{\dag}]\big)/2$ for some time $t$, where $\gamma=1-e^{-\kappa t}$. The characteristic function of $\mathcal{N}$ takes the form of \cref{eq:diagonal_characteristic_function_ansatz} with variance
\begin{subequations}\label{eq:loss_sigma_estimate}
\begin{align}
\sigma^{2}&=\sqrt{1-\gamma}\,\mathrm{tanh}\bigg(\!\frac{\Delta^{2}}{2}\!\bigg)+\frac{\gamma}{2}+\frac{(1-\sqrt{1-\gamma})^{2}}{2\,\mathrm{tanh}(\Delta^{2})}\label{eq:loss_exact_sigma_estimate}\\
&\approx \frac{\Delta^{2}}{2}+\frac{\gamma}{2}+\frac{\gamma^{2}}{8\Delta^{2}}.\label{eq:loss_approx_sigma_estimate}
\end{align}
\end{subequations}
Note that \cref{eq:loss_exact_sigma_estimate} is valid whenever $\sigma$ is much smaller than the length of the shortest logical operator (which is roughly $\sqrt{\pi}$ for most cases of interest), while the lowest-order approximation \cref{eq:loss_approx_sigma_estimate} is only valid in the slightly more specific regime where both $\gamma$ and $\Delta$ are also small. Indeed, the first of these approximations, $\sigma\ll\sqrt{\pi}$, is valid across a wide range of $\Delta$ and $\gamma$; for example, when $\Delta_{\text{dB}}=12$, $\sigma<\sqrt{\pi}/4$ whenever $\gamma<18\%$. Quantitatively, we show in \cref{fig:loss_error_comparison} that the estimate \cref{eq:loss_exact_sigma_estimate} agrees well with numerically-obtained values, with less than 1\% relative error in the regime of interest $10<\Delta_{\text{dB}}<18$ and $\gamma<10\%$.

Importantly, \cref{eq:loss_sigma_estimate} takes into account the combined effect of the envelope and loss operator acting together, represented by the $\gamma^{2}/8\Delta^{2}$ term which becomes \textit{larger} as $\Delta$ becomes small. This is because loss affects highly squeezed (small $\Delta$) GKP codestates more than less squeezed GKP codestates due to the large average photon number of highly squeezed GKP codestates. For fixed $\gamma$, we can (approximately) find the optimal level of GKP squeezing by minimizing \cref{eq:loss_sigma_estimate}, giving $\Delta_{\text{opt}}^{2}=-\ln{\sqrt{1-\gamma}}\approx \gamma/2$. Pleasingly, we can also write this in terms of the average photon number of the codestate $\bar{n}\approx 1/(2\Delta^{2})-1/2$, giving $\bar{n}_{\text{opt}}\approx 1/\gamma-1/2$. This tradeoff has already been noted in Ref.~\cite{Campagne19}. In \cref{sec:gate_error_estimate}, we also present a more detailed analysis that also considers other Gaussian noise channels such as gain and Gaussian random displacements.

\cref{eq:loss_sigma_estimate} can also be used to incorporate the effects of loss into studies into GKP-qubit code concatenations without the use of any additional computational resources. Explicitly, one can model the effect of loss on an approximate GKP codestate as a Gaussian random displacement channel with variance given by \cref{eq:loss_sigma_estimate}, which is a minor tweak to previous studies that have used a variance of $\Delta^{2}/2$.

Now we turn to analysing dephasing noise, defined by
\begin{equation}\label{eq:dephasing}
    \mathcal{D}_{\sigma_{\text{d}}}(\rho)=\frac{1}{\sqrt{2\pi\sigma_{\text{d}}^{2}}}\int_{\mathbb{R}}\!d\phi\,e^{-\phi^{2}/2\sigma_{\text{d}}^{2}}e^{i\phi a^{\dag}a}\rho e^{-i\phi a^{\dag}a}.
\end{equation}
In particular, we wish to consider the noise map $\mathcal{N}=\mathcal{D}_{\sigma_{\text{d}}}\circ\mathcal{J}[e^{-\Delta^{2}a^{\dag}a}]$. Since dephasing is a non-Gaussian channel, we cannot analyse $\mathcal{N}$ using the variance defined in \cref{eq:diagonal_characteristic_function_ansatz}, since the diagonal elements of the characteristic function take the form
\begin{multline}\label{eq:dephasing_diag_chi}
    c(\vect{v},\vect{v})=\frac{e^{\Delta^{2}}}{2(2\pi)^{5/2}\sigma_{\text{d}}}\int_{\mathbb{R}}\!d\phi\,\frac{e^{-\phi^{2}/2\sigma_{\text{d}}^{2}}}{\cosh(\Delta^{2})-\cos(\phi)}\\
    \times\exp\bigg(\!{-}\frac{1}{2}\frac{\sinh(\Delta^{2})}{\cosh(\Delta^{2})-\cos(\phi)}|\vect{v}|^{2}\!\bigg).
\end{multline}
For numerical simulations, it may be possible to sample random displacements approximately from the probability distribution defined by \cref{eq:dephasing_diag_chi} (which we leave for future research); however, for our analytic results we find it more convenient to work directly from \cref{eq:dephasing}. This allows us to consider unitary $e^{i\phi a^{\dag}a}$ rotation errors first, which are Gaussian, and then integrate the resulting average gate fidelity expressions over $\phi$. The details are shown in \cref{subsec:dephasing}, and we summarize the key results here.

To understand our results, first consider the situation where the envelope size $\Delta$ is fixed, and the amount of dephasing $\sigma_{\text{d}}$ is varied. Unlike loss, which has a linear contribution to infidelity for arbitrarily small $\gamma$ [as shown in \cref{eq:loss_sigma_estimate,eq:loss_Taylor_expand}], we find that dephasing does not have a first-order contribution to the infidelity for small $\sigma_{\text{d}}$. However, when $\sigma_{\text{d}}$ exceeds a critical amount of dephasing
\begin{equation}\label{eq:critical_dephasing}
    \sigma_{\text{d}}^{*}=\sqrt{2}\Delta^{3}/d,
\end{equation}
we find that dephasing has an exponential impact on the infidelity, comparable to the effect of loss. More precisely, in the \textit{subcritical} regime $\sigma^{}_{\text{d}}\ll\sigma_{\text{d}}^{*}$ we have
\begin{subequations}\label{eq:dephasing_infid}
\begin{equation}\label{eq:low_dephasing_infid}
    1-\bar{F}\approx\frac{2^{5/2}a\Delta^{4}}{3d^{2}\sqrt{\pi}\sqrt{\sigma_{\text{d}}^{*2}-\sigma^{2}_{\text{d}}}}\exp\bigg(\!{-}\frac{d^{2}}{4\Delta^{2}}\!\bigg),
\end{equation}
and in the \textit{supercritical} regime $\sigma^{}_{\text{d}}\gg\sigma_{\text{d}}^{*}$ we have
\begin{equation}\label{eq:high_dephasing_infid}
    1-\bar{F}\approx\frac{2^{7/4}a\sigma_{\text{d}}}{3\sqrt{\pi d\Delta(\sigma^{}_{\text{d}}-\sigma_{\text{d}}^{*})}}\exp\bigg(\!{-}\frac{\Delta d}{\sqrt{2}\sigma_{\text{d}}}\Big(1-\frac{\sigma_{\text{d}}^{*}}{2\sigma_{\text{d}}}\Big)\!\bigg).
\end{equation}
\end{subequations}
Importantly, in \cref{eq:low_dephasing_infid}, the only dependence of the infidelity on $\sigma_{\text{d}}$ is in the square root, while in \cref{eq:high_dephasing_infid}, $\sigma_{\text{d}}$ also appears in the exponent. In \cref{subsec:dephasing} we generalize \cref{eq:dephasing_infid} to include simultaneous loss and dephasing, and also derive an approximate expression for the infidelity in the critical case when $\sigma^{}_{\text{d}}=\sigma_{\text{d}}^{*}$. Our results agree reasonably well with numerics considering the crude nature of our approximations, with the relative error between numerical and analytical results being less than 20\% in most cases of interest, see \cref{fig:dephasing_error_comparison}.

Now, consider the slightly different situation where $\sigma_{\text{d}}$ is fixed and we wish to find the envelope size $\Delta$ that minimizes the infidelity. This can be done approximately by minimizing the exponent of \cref{eq:high_dephasing_infid}, giving $\Delta=\sqrt[3]{d\sigma_{\text{d}}}/\sqrt{2}$, which is equivalent to the condition $\sigma^{}_{\text{d}}=2\sigma_{\text{d}}^{*}$.

Therefore, to quickly assess the impact dephasing will have on a given GKP codestate, the single most important thing to consider is whether the dephasing is larger than or less than the critical dephasing $\sigma_{\text{d}}^{*}$. We show in \cref{sec:gate_error_estimate} that the critical dephasing when considering simultaneous loss and dephasing is given by
\begin{equation}\label{eq:critical_sigma_loss}
    \sigma_{\text{d}}^{*}=\frac{\sqrt{2}\Delta^{3}}{d}+\frac{\sqrt{2}\Delta\gamma}{d}+\frac{\gamma^{2}}{2\sqrt{2}d\Delta}.
\end{equation}
Evaluating $\sigma_{\text{d}}^{*}$ for $\gamma=1\%$, which is on the order of loss that is observed in current GKP experiments~\cite{Sivak23}, and $\Delta_{\text{dB}}=10$ and 12 (respectively) gives $\sigma_{\text{d}}^{*2}=0.077\%$ and $0.022\%$, which are an order of magnitude smaller than the current amounts of dephasing observed in the same experiment ($\sigma_{\text{d}}^{2}\approx 0.6\%$). On the other hand, it can be shown numerically that given $\sigma_{\text{d}}^{2}=0.6\%$ and $\gamma=1\%$, the optimal envelope size is $\Delta_{\text{dB}}\approx9.3$, which is remarkably close to the experimentally optimized envelope size $\Delta_{\text{dB}}\approx 9.4$ that was used in Ref.~\cite{Sivak23}.

Our methods can also be extended to analze the effects of a (unitary) Kerr nonlinearity $\mathrm{exp}\big(i\theta_{K}(a^{\dag}a)^{2}/2\big)$. In \cref{sec:gates} we explain how to implement GKP Clifford gates in superconducting circuits and explained why in practice Kerr nonlinearities can occur when implementing a gate, motivating our analysis which is presented in full in \cref{subsec:Kerr}. Our result is simple to implement numerically but is not particularly theoretically insightful because it involves a large summation, so here we simply summarize our numerical results.

Similarly to loss and dephasing, higher quality codestates are more highly affected by Kerr nonlinearity than lower quality codestates due to their higher average photon number, introducing a trade-off between the increased noise from the envelope operator and the Kerr nonlinearity. Qualitatively, when $\theta_{K}$ is small enough, the average gate infidelity remains almost identical to the Kerr-nonlinearity-free case, but when $\theta_{K}$ exceeds some critical amount (which depends on $\Delta$) the average gate infidelity increases significantly. For example, a codestate with $\Delta_{\mathrm{dB}}=10$ is barely affected by Kerr non-linearities of $\theta_{K}\leq\pi/4000$, but when $\theta_{K}=\pi/1000$ the average gate infidelity is more than 10 times higher than in the Kerr-nonlinearity-free case. For a higher quality codestate with $\Delta_{\mathrm{dB}}=12$, the Kerr nonlinearity must be well below $\theta_{K}=\pi/4000$ for the codestate to remain unaffected. However, we note that such requirements on $\theta_{K}$ may be too stringent: our analysis here compares the effect of Kerr nonlinearities acting on GKP codestates in the absence of loss and dephasing, which already can increase the average gate infidelity by orders of magnitude. To have a better idea of how much Kerr nonlinearity can be handled by a given experimental setup these noise sources would therefore need to be considered together, which is in principle possible using the methods presented here.

\section{\label{sec:measurements}Logical Pauli Measurements}

In order to implement Clifford circuits using only generalized controlled gates, it is necessary to also perform single-qubit logical measurements of all three Pauli operators $X,Y,Z$. To perform such read-out at the end of a computation, the simplest proposal is to use homodyne detection on the GKP mode. However, the efficiency of homodyne detection in the microwave regime is low, with state-of-the-art experiments achieving efficiencies on the order of 60\% to 75\%~\cite{Macklin15,Touzard19}.

Here, we analyze the effect of such inefficient homodyne detection on the logical readout failure rate of GKP codes. Moreover, we propose two {proof-of-principle} schemes that can improve the effective efficiency of logical readout, one scheme based on single-mode squeezing, and the second based on quadrature-quadrature coupling to a separate read-out mode (which can be implemented with the circuits shown in \cref{fig:gate_circuits}). Both of these schemes can be written as an effective homodyne read-out of the GKP mode with an effective efficiency $\eta_{\text{eff}}$ depending on the parameters used in the scheme. We show how to derive $\eta_{\text{eff}}$ in each of these schemes using the theory of quantum trajectories, and propose experimental parameters that would allow a high-efficiency fast measurement of the GKP mode.

Recall that we can write a logical Pauli operator as $\bar{\sigma}_{i}=e^{i\sqrt{\pi}s_{i}}$, where $s_{i}=-\bar{p},\bar{q}-\bar{p},\bar{q}$ for $i=1,2,3$ respectively [see \cref{eq:s_quadratures}], and we can write $s_{i}=r_{i}(q\cos\theta_{i}+p\sin\theta_{i})$ in polar coordinates. To measure $\bar{\sigma}_{i}$, we can measure the rotated quadrature $q'=s_{i}/r_{i}=q\cos\theta_{i}+p\sin\theta_{i}$, round the result to the nearest multiple of $b=\sqrt{\pi}/r_{i}$ (which we call the \textit{bin size} of the measurement), and interpret the result as $+1$ if it is an even multiple of $b$ and as $-1$ if it is an odd multiple.

As pointed out in Appendix B in Ref.~\cite{Shaw24-1}, binned measurement operators do not always ideally read out the logical Pauli operators of the GKP code. In the square GKP code, this occurs with the read-out of logical $\bar{Y}=T(\sqrt{\pi},\sqrt{\pi})$. To see this, note that a displacement error of $T(\sqrt{\pi}/4,-\sqrt{\pi}/4)$ would flip the outcome of a binned $\bar{Y}$ measurement even though such an error is correctable by a round of ideal error correction. To perform an \textit{ideal} Pauli $\bar{Y}$ measurement, we would need to simultaneously measure one of the stabilizers alongside $\bar{Y}$, and use minimum-weight decoding to infer the logical outcome. However, implementing such a scheme in practice requires the use of an additional GKP approximate codestate that increases the error rate enough that the scheme has at least as large an error rate as the binned measurement scheme (see \cref{sec:measurement_scheme_comparison}).

In the remainder of this section, we will (without loss of generality) only consider $\bar{Z}$ measurements of general GKP codes rotated such that $\beta_{1}=0$, as all other Pauli measurements are equivalent up to a re-scaling of the bin size and rotation of the measurement quadrature. This convention sets the bin size $b=\sqrt{\pi}/r_{3}=\alpha_{1}$ and measurement quadrature $q'=q$. Even though in this section we will analyze homodyne measurements in the context of end-of-the-line logical readout, the techniques used here can also be readily applied to improve the efficiency of the homodyne measurements used in the teleportation-based error-correction circuit~\cref{fig:tele}.

An interesting and subtle point is that applying the above binned $\bar{Z}$ measurement to approximate GKP codewords $\ket{\bar{\mu}_{\Delta}}$ is \textit{not} equivalent to maximum likelihood decoding of the measurement outcome, in which a measurement outcome $q=x$ is interpreted as a logical $+1$ outcome if and only if $\big|\prescript{}{q}{\braket{x|\bar{0}_{\Delta}}}\big|^{2}>\big|\prescript{}{q}{\braket{x|\bar{1}_{\Delta}}}\big|^{2}$. Such a maximum likelihood decoder cannot be exactly represented using a constant bin size as described above; however, it is well approximated by setting the bin size to $b=\cosh(\Delta^{2})\alpha_{1}$, where the correction $\cosh(\Delta^{2})$ {goes to }1 as $\Delta$ {goes to }0.
%This correction can be understood qualitatively by noting that the height of the central $q=0$ peak of $\prescript{}{q}{\braket{x|\bar{0}_{\Delta}}}$ is larger than the height of the innermost peaks of $\prescript{}{q}{\braket{x|\bar{1}_{\Delta}}}$, which occur at $q=\pm\,\mathrm{sech}(\Delta^{2})\alpha_{1}$. As a result, the two wavefunctions intersect at a value of $q$ greater than $\mathrm{sech}(\Delta^{2})\alpha_{1}/2$.
We note however that the differences between using a bin size of $b=\alpha_{1}$, $b=\cosh(\Delta^{2})\alpha_{1}$, and maximum likelihood decoding, are negligible for any values of $\Delta$ small enough to be useful in practice. We will nevertheless use a constant bin size $b=\cosh(\Delta^{2})\alpha_{1}$ in all subsequent results.

\begin{figure*}
\includegraphics{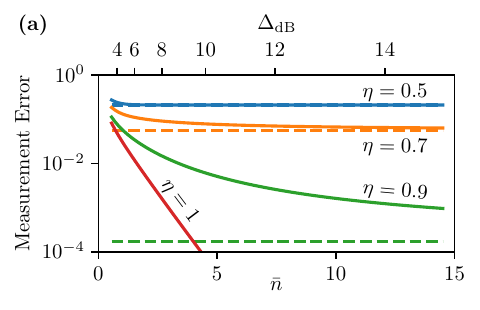}
\hspace*{\fill}
\includegraphics{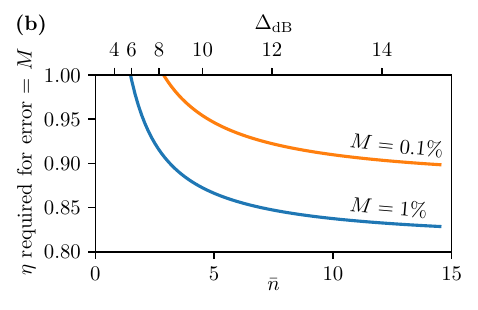}
\caption{The effect of inefficient homodyne measurements on the readout errors of approximate square GKP codestates. (a) Logical $Z$ measurement error $\big(P(0|1)+P(1|0)\big)/2$ of approximate GKP codestates (with average photon number $\bar{n}$ and squeezing $\Delta_{\mathrm{dB}}$) measured by a position measurement with efficiency $\eta$. For finite $\eta$, the limiting measurement error as the GKP codestate approaches the ideal limit ($\bar{n}\rightarrow\infty$) is plotted as a dashed line. (b) The efficiency required to reach a target logical $Z$ measurement error rate $M$, as a function of the GKP squeezing/photon number.}
\label{fig:measurement_error_plots}
\end{figure*}

Now, we consider inefficient homodyne detection of the position quadrature $q$, which can be described by the POVM elements
\begin{multline}\label{eq:inefficient_POVM}
    W_{\eta}(X)=\mathcal{N}\!\int dx~\ket{x}_{q}\!\bra{x}\exp\Big(-\frac{\eta}{1-\eta}(x-X)^{2}\Big),
\end{multline}
where $X\in\mathbb{R}$ is the recorded outcome of the measurement and $\mathcal{N}$ is a normalization constant. This POVM already accounts for the rescaling of the measurement outcome due to loss occurring from the inefficient measurement. We define the measurement error of the logical readout as $M_{\text{error}}=\big(P(1|0)+P(0|1)\big)/2$, where $P(1|0)$ is the probability of recording a $-1$ measurement outcome given the initial state was $\ket{\bar{0}_{\Delta}}$. This can be calculated by evaluating the integral
\begin{equation}
    P(1|0)=\sum_{t\in\mathbb{Z}}\int_{(2t+\frac{1}{2})b}^{(2t+\frac{3}{2})b}dX~\tr\big(W_{\eta}(X)\ket{\bar{0}_{\Delta}}\!\bra{\bar{0}_{\Delta}}\big).\label{eq:measurement_error}
\end{equation}

To evaluate the integral~\cref{eq:measurement_error}, we use the following expression for $\ket{\bar{0}_{\Delta}}$ in the position basis:
\begin{multline}\label{eq:0_bar_Delta}
    \prescript{}{q}{\braket{x|\bar{0}_{\Delta}}}\propto\sum_{s\in\mathbb{Z}}e^{2is^{2}\alpha_{1}\alpha_{2}}e^{-\frac{1}{2}(2s\alpha_{1})^{2}\mathrm{tanh}(\Delta^{2})}\\
    \times e^{-\frac{1}{2}\mathrm{coth}(\Delta^{2})[x-2s\alpha_{1}\mathrm{sech}(\Delta^{2})]^{2}}.
\end{multline}
\Cref{eq:0_bar_Delta} holds for general GKP codes, with the only restriction that $\vect{\alpha},\vect{\beta}$ are rotated such that $\beta_{1}=0$.

After substituting \cref{eq:0_bar_Delta} into \cref{eq:measurement_error} and evaluating the integrals over $X$ and $x$, this gives an exact infinite series for the error probability $P(1|0)$, which can be evaluated numerically by truncating the infinite series (see \cref{sec:measurement_error_estimate}). The values obtained using this semi-analytical method are indistinguishable from those obtained using a direct Fock space simulation, but the semi-analytic results can be used to probe smaller values of $\Delta$ than would be possible due to the truncation dimension of the numerical simulation. We plot the measurement error $M_{\text{error}}$ as a function of $\Delta$ and $\eta$ for the square GKP code in \cref{fig:measurement_error_plots}(a), and show the efficiency $\eta$ required to reach various target measurement error rates in \cref{fig:measurement_error_plots}(b).

Additionally, we derive in \cref{sec:measurement_error_estimate} the following approximation to the measurement error
\begin{equation}
    M_{\text{error}}\approx\mathrm{erfc}\bigg(\frac{1}{2}\alpha_{1}\Big(\Delta^{2}+\frac{1-\eta}{\eta}\Big)^{-1/2}\bigg),\label{eq:measurement_error_approx}
\end{equation}
which holds for small $\Delta$. These results demonstrate that $M_{\text{error}}$ is highly sensitive to the measurement efficiency $\eta$: even at $\eta=0.7$, which is close to the current state-of-the-art, the minimum achievable measurement error rate in the square GKP code is $M_{\text{error}}\approx 5.6\%$ as $\Delta_{\text{dB}}\rightarrow\infty$. This is consistent with results obtained in Ref.~\cite{Hillmann22} in their analysis of teleportation error-correction schemes and motivates the need to improve the effective efficiency of homodyne detection for use in GKP codes. Concretely, one needs a measurement efficiency of $\eta\approx 0.85$ to reach a measurement error of $1\%$, or $\eta \approx 0.92$ for $0.1\%$, using a square GKP code with $\Delta_{\text{dB}}\approx 12$.

\begin{figure}
\includegraphics{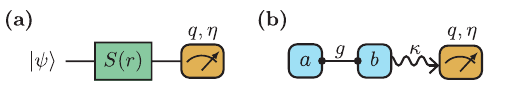}
\caption{Two position measurement schemes to improve the effective efficiency of the measurement. (a) The initial state is subjected to phase-sensitive amplification, equivalent to an application of the squeezing operator $S(r)=e^{\frac{r}{2}(a^{2}-a^{\dag 2})}$, followed by a position measurement with efficiency $\eta$. (b) The initial state (stored in cavity $a$) is coupled to a read-out mode initialized in the vacuum state (stored in cavity $b$) via the Hamiltonian $H/\hbar=-gq_{1}p_{2}$, which may be implemented using the circuit QED designs presented in \cref{sec:gates}. Simultaneously, the read-out mode's position quadrature is measured at a rate $\kappa$ via homodyne detection with efficiency $\eta$. In both schemes, we show that the whole scheme is equivalent to a single position measurement with an effective efficiency $\eta_{\text{eff}}$, which can be greater than the efficiency of the physical position measurement $\eta$ given the right parameters.}
\label{fig:measurement_schemes}
\end{figure}

To achieve such efficiencies, we consider two alternative schemes (illustrated in \cref{fig:measurement_schemes}) for performing homodyne detection to improve the effective efficiency of the measurement. In the first scheme, we simply consider applying a single-mode squeezing operator $S(r)=\exp\big(r(a^{\dag 2}-a^{2})/2\big)$ (where $r>0$ corresponds to squeezing the momentum quadrature and $r<0$ corresponds to squeezing the position quadrature), before performing homodyne detection with efficiency $\eta$. We quote the amount of squeezing in decibels as $S_{\text{dB}}=(20\log_{10}e)\,r$. Intuitively, amplifying the position quadrature should improve the effective efficiency of the measurement by improving the separation between different position eigenstates. We can write the effective POVM elements of the squeezing/measurement sequence as $V_{r,\eta}(X')= S^{\dag}(r)W_{\eta}(X')S(r)$, where $X'$ is the measurement outcome of the inefficient homodyne measurement \cref{eq:inefficient_POVM}. By defining $X=e^{-r}X'$, we obtain $V_{r,\eta}(X)= W_{\eta_{\text{eff}}}(X)$, where
\begin{equation}
    \eta_{\text{eff}}=\big(1+e^{-2r}(\eta^{-1}-1)\big)^{-1}.
\end{equation}
Using this, we can calculate the amount of amplification required to increase the effective efficiency of the measurement; for example, if the physical efficiency is $\eta=0.7$, we can achieve $\eta_{\text{eff}}=0.85$ (or a 1\% measurement error for $\Delta_{\text{dB}}=12$ dB) with a squeezing of $S_{\text{dB}}\approx 4$ dB, and $\eta_{\text{eff}}=0.92$ (0.1\% error) with $S_{\text{dB}}\approx7$ dB.

Although such levels of amplification are achievable, this scheme requires the GKP mode itself to be directly released into the measurement sequence, which either requires a change in the quality factor of the GKP mode itself, or requires coupling to a second readout mode. To combat this issue we consider a second scheme in \cref{fig:measurement_schemes}(b). We consider a high Q GKP mode with loss rate $\gamma$ coupled to a low Q read-out mode via a Hamiltonian $H=-gq_{1}p_{2}$ (where 1 refers to the GKP mode and 2 to the read-out mode). Such a Hamiltonian can be engineered using {the same circuits that implement generalized controlled gates as discussed in \cref{sec:gates}} and is identical to the Hamiltonian for a controlled-NOT between two square GKP qubits. Homodyne detection is performed on the position quadrature of the read-out mode with efficiency $\eta$, which we consider to be occurring at a rate $\kappa$ simultaneously with the coupling. To analyze this system we consider short timescales $t\ll 1/\gamma$ and neglect the effect of loss occurring on the GKP mode. In this regime, the quantity of interest is the time required to perform the measurement to a desired efficiency. Here, we use quantum trajectories to solve exactly for the POVM of the system using the method of Ref.~\cite{Warszawski20}. We relegate the details of the derivation of the POVM to \cref{sec:measurement_scheme_2_derivation} so that we can focus on the results of our analysis here.

The experimentally observed measurement outcome is given by the observed photocurrent $I(t)$ of the detector (see \cref{sec:measurement_scheme_2_derivation} for details). We find that the resulting POVM depends not on the entire measurement record from time $0$ to $t$, but only on one integral of the observed photocurrent
\begin{equation}\label{eq:X_integral}
    X=-\sqrt{\kappa/(8g^{2}\tau^{2}\eta)}\int_{0}^{t}dt'\,(1-e^{-\kappa t'/2})I(t'),
\end{equation}
where $\tau=t-(1-e^{-\kappa t/2})(3-e^{-\kappa t/2})/\kappa$. Given the initial state of the ancilla mode $\rho_{b}$, we can write a single-mode POVM element corresponding to a measurement outcome $X$ as
\begin{multline}\label{eq:measurement_scheme_2_general_POVM}
    T_{g,\kappa,\eta,t}(X)=\mathcal{N}\!\int\! dx\,\ket{x}_{q_{1}}\!\!\bra{x}\int\! d\tilde{x}\,{\vphantom{\ket{x}}}_{q_{2}}\!\!\braket{\tilde{x}|\rho_{b}|\tilde{x}}_{q_{2}}\\
    \times \exp\bigg\{-\frac{1}{C}\Big[X-x-\frac{1}{2g\tau}\big(1-e^{-\kappa t/2}\big)^{2}\tilde{x}\Big]^{2}\bigg\},
\end{multline}
where $\mathcal{N}$ is a normalization constant, and
\begin{equation}\label{eq:c_expression}
    C=\frac{\kappa\tau-\eta(1-e^{-\kappa t/2})^{4}}{4g^{2}\tau^{2}\eta}.
\end{equation}
We see now from \cref{eq:measurement_scheme_2_general_POVM} that $X$ indeed represents the estimate of the position of the state. Substituting $\rho_{b}=\ket{0}\!\bra{0}$ for the initial state of the ancilla results in a POVM that corresponds to an inefficient homodyne measurement [\cref{eq:inefficient_POVM}] with effective efficiency
\begin{equation}\label{eq:effective_efficiency}
    \eta_{\text{eff}}=\frac{4g^{2}\tau\eta}{4g^{2}\tau\eta+\kappa}.
\end{equation}

\begin{figure}
\includegraphics{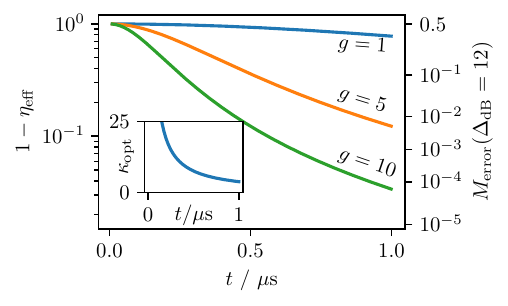}
\caption{(main) Effective measurement inefficiency $1-\eta_{\text{eff}}$ of the two-mode measurement scheme, as a function of the coupling strength $g/(2\pi$ MHz) and measurement time $t$, for a fixed physical efficiency $\eta=0.75$. The two-mode measurement scheme consists of coupling the GKP mode to an ancilla initiated in the vacuum state via the Hamiltonian $H=-gq_{1}p_{2}$, and performing homodyne detection for some finite time $t$ on the ancilla mode at a rate $\kappa$ and efficiency $\eta$. $\kappa$ is optimized to maximize the effective measurement efficiency. The right axis shows the corresponding logical measurement error of a GKP state with $\Delta_{\text{dB}}=12$ when measured with homodyne detection of efficiency $\eta_{\text{eff}}$. (inset) The optimal measurement rate $\kappa_{\text{opt}}/(2\pi$ MHz) as a function of the measurement time $t$.}
\label{fig:scheme_2_plots}
\end{figure}

We analyze \cref{eq:effective_efficiency} as follows. First, we optimize the measurement rate $\kappa$ to maximize the effective measurement efficiency for fixed $g,t,\eta$, noting that $\tau$ depends on $\kappa$. Interestingly, due to the QND nature of the measurement and the lack of noise on the GKP code, we find that the resulting $\kappa_{\text{opt}}$ depends only on the measurement time $t$ and is given approximately by $\kappa_{\text{opt}}/2\pi\approx 3.79/t$ (see inset to \cref{fig:scheme_2_plots}). The dependence of $\eta_{\text{eff}}$ on $\kappa$ is very weak near $\kappa_{\text{opt}}$, so a fine-tuning of $\kappa$ is not required to achieve competitive effective efficiencies. Next, we fix the physical efficiency of the measurement at $\eta=0.75$ and plot $\eta_{\text{eff}}$ as a function of $t$ and $g$ in \cref{fig:scheme_2_plots}. \Cref{fig:scheme_2_plots} demonstrates that for a coupling strength of $g=10\times2\pi$ MHz, one can achieve $\eta_{\text{eff}}=0.85$ (1\% read-out error for $\Delta_{\text{dB}}=12$ dB) in a measurement time $t\approx0.45\ \mu$s, and $\eta_{\text{eff}}=0.92$ (0.1\% error) after $t\approx0.63\ \mu$s.

We are interested in the required measurement time to achieve a given effective efficiency (and hence logical measurement error rate) for two reasons. First, if the measurement time is too long, loss on the GKP mode will become significant. Second, it is desirable for readout to be conducted on a time scale comparable to that of other superconducting architectures, such as those based on transmon qubits. In transmon qubits, logical readout typically takes on the order of hundreds of nanoseconds~\cite{Google23}, which is on the same order of magnitude as our estimates for the time for our measurement scheme. Stronger coupling strengths would help reduce the measurement time required in our scheme.

Using \cref{eq:measurement_scheme_2_general_POVM,eq:effective_efficiency}, one can determine the effect of changing other parameters such as the physical efficiency $\eta$ and the initial state of the ancilla mode $\rho_{b}$ on the effective efficiency $\eta_{\text{eff}}$. The effective efficiency has only a weak dependence on the physical efficiency; for example, increasing the physical efficiency of the measurement from $\eta=0.5$ to $\eta=1$ only results in a roughly 30\% decrease in the required measurement time to reach a given effective efficiency, independent of the choice of $g$ or target effective efficiency. For the ancilla initial state $\rho_{b}$, one would expect that starting with a vacuum state squeezed along the position quadrature would \textit{improve} the effective efficiency, as it reduces the additional noise added to the signal from the position quadrature of the GKP mode. In the infinite squeezing limit we substitute the $0$-position eigenstate $\rho_{b}=\ket{0}_{q_{2}}\!\bra{0}$ into \cref{eq:measurement_scheme_2_general_POVM}, giving an effective efficiency $\eta_{\text{eff}}=1/(C+1)$. However, this only results in a decrease of no more than 20\% in the required measurement time to reach a given $\eta_{\text{eff}}$, for $0.5<\eta<0.75$ and any choice of $g$ and $\eta_{\text{eff}}$. As such, the coupling strength $g$ is the most important factor (along with the GKP mode squeezing $\Delta$) in reducing the logical measurement error in GKP modes using this scheme.

As a final comment, we note that as $t\rightarrow\infty$, the state of the GKP mode is projected onto the $X$-position eigenstate, as shown in \cref{subsec:post_state}. It may then be easier to generate a new GKP codestate from this position eigenstate (or highly squeezed state in the finite $t$ regime) than if the final state of the oscillator was the vacuum state.

\section{Discussion and Conclusions}\label{sec:conc}

In this work, we have given concrete {methods to model realistic noise in superconducting GKP qubits, bridging the gap between experiment and theory.} We began in
%\cref{sec:gates} by presenting our scheme for performing Clifford circuits using generalized controlled gates implemented by four- or three-wave mixing circuits, where single-qubit Clifford gates are accounted for by updating the phase of a local oscillator in the implementation of each two-qubit gate. Next, in
\cref{sec:decoders} by presenting an average gate fidelity metric for quantifying the quality of Clifford gates and analyzing the performance of ideal logical gates when subjected to errors due to approximate GKP codestates. In \cref{sec:error_estimates} we presented a general method to analytically approximate the effect of loss and dephasing on GKP codes using the stabilizer subsystem decomposition~\cite{Shaw24-1}. Finally in \cref{sec:measurements}, we considered the effects of homodyne detection inefficiencies on the rate of logical measurement errors. We {analyzed} a scheme that can achieve feasible error rates even with a low measurement efficiency using quantum trajectories~\cite{Warszawski20}.

While our analysis of the performance of GKP Clifford gates {and of Kerr nonlinearities} gave significant insight into how to mitigate the spreading of errors due to the logical gates, {this does not necessarily exactly correspond to the noise that may occur due to non-ideal gate execution}. One could use the theoretical analysis conducted in Ref.~\cite{Zhang19} to determine the leading sources of error in the implementation of a given generalized controlled gate. One could then model these effects using the methods developed in \cref{sec:error_estimates} to produce an estimate of the average gate fidelity of each generalized controlled gate in the presence of realistic noise sources. We leave such an analysis to future work.

{It also remains an open question whether the insights gained from the modified decoding patches for GKP Clifford gates can be carried over to schemes using autonomous GKP error-correction~\cite{Royer20,DeNeeve22,Rojkov23} instead of teleportation-based error-correction. For example, perhaps our modified decoding patch could correspond to a modified autonomous error-correction scheme immediately following the gate, although we leave investigating such ideas to future work.}

To perform logical read-out of GKP codestates, we {analyzed} a scheme utilizing quadrature-quadrature coupling to an ancilla mode to implement {high-efficiency} logical read-out. However, this scheme requires the use of a low-Q read-out ancilla prepared in the vacuum state, increasing the overhead required for computation with GKP codes. {We found that the performance of} this scheme is highly sensitive to the strength $g$ of the quadrature-quadrature coupling between the GKP and ancilla modes, but the values of $g$ that are feasible in GKP experiments remain to be seen. Additionally, the high-Q GKP mode and low-Q readout ancilla must be coupled with a device with a large on-off ratio to prevent unwanted leakage from the GKP mode to the ancilla. These factors must be considered when assessing the viability of {this} scheme for specific experimental platforms.

\section{Acknowledgements}\label{sec:acknowledgements}

This work was supported by the Australian Research Council Centre of Excellence for Engineered Quantum Systems (EQUS, CE170100009).
During this project, MHS has been supported by an Australian Government Research Training Program (RTP) Scholarship and also by QuTech NWO funding 2020-2028 -- Part I ``Fundamental Research'', project number 601.QT.001-1, financed by the Dutch Research Council (NWO). MHS would also like to thank Barbara Terhal for providing feedback on a draft manuscript.
\bibliography{my_bib}
\appendix

\section{\label{sec:gates}\textcolor{black}{Phase-tracked single-qubit Clifford gates}}

In this {appendix}, we outline how to perform arbitrary Clifford circuits in GKP codes without physically implementing any single-qubit Clifford gates, reducing the spread of errors in the computation. In this scheme, single-qubit Clifford gates are tracked in software and absorbed into the two-qubit Clifford gates in the circuit. Instead, a larger set of two-qubit gates must be performed. We call these gates \textit{generalized controlled gates}, and this procedure is sometimes referred to as the \textquotedblleft Clifford frame\textquotedblright~\cite{Chamberland18}. Conveniently, all the generalized controlled gates can be implemented using a single piece of superconducting hardware -- either a transmon or a SNAIL -- with each gate differentiated by the phase of a local oscillator. This is advantageous since it reduces the number of physical gates that must be implemented, reducing the spread of errors in the circuit (as discussed in \cref{sec:decoders}). These results have already been presented in Ref.~\cite{Grimsmo21}, however we reiterate them here as they provide important context for the main text.

We now step through precisely how the single-qubit Clifford gates need to be tracked in order to implement a general quantum computation. We start with a universal quantum computing circuit comprising of state preparation in the $\ket{0}$ or $\ket{T}=(\ket{0}+e^{i\pi/4}\ket{1})/\sqrt{2}$ states, Pauli $Z$-measurements, and adaptive Hadamard, phase and controlled-$Z$ gates. One can rewrite such a circuit instead consisting of state preparation in the $\ket{0}$ or $\ket{T}$ states, Pauli $X$, $Y$ and $Z$-measurements, and generalized controlled gates (which must be performed adaptively). For Pauli matrices $\sigma_{i},\sigma_{j}$ ($i,j\in\{1,2,3\}$), we define the $\sigma_{i}$-controlled-$\sigma_{j}$ gate as
\begin{equation}
    C_{\sigma_{i}\sigma_{j}}=I\otimes I-\frac{1}{2}(I-\sigma_{i})\otimes(I-\sigma_{j}).\tag{\ref{eq:generalised_controlled_gate}}
\end{equation}
These gates can be interpreted as applying a $\sigma_{j}$ gate to the target (second) qubit if the control (first) qubit is a $-1$ eigenstate of $\sigma_{i}$, and doing nothing if it a $+1$ eigenstate. For example the $Z$-controlled-$Z$ gate is simply the controlled-$Z$ gate, and the $Z$-controlled-$X$ gate is the controlled-NOT gate.

To rewrite a Clifford circuit in terms of generalized controlled gates, we use the following fact: given a generalized controlled gate $C_{\sigma_{i}\sigma_{j}}$ and a single-qubit Clifford gate $A$, we have
\begin{equation}\label{eq:clifford_commute}
    C_{\sigma_{i}\sigma_{j}}(A\otimes I)=(A\otimes B)C_{\sigma_{i'}\sigma_{j}}\,,
\end{equation}
where $\sigma_{i'}$ is given by calculating $A^{\dag}\sigma_{i}A=\pm\sigma_{i'}$, and where $B=I$ if the sign is $+$ and $B=\sigma_{j}$ if the sign is $-$. \cref{eq:clifford_commute} can be used to commute the Hadamard and phase gates past the controlled-$Z$ gates in the original circuit. Subsequently, the remaining single-qubit Pauli and Clifford gates can be commuted past the $Z$-measurements, leaving $X$, $Y$ and $Z$ Pauli measurements (which are discussed in Sec.\ \ref{sec:measurements}) and removing the single-qubit Clifford gates entirely from the circuit.

To implement a generalized controlled gate $C_{\sigma_{i}\sigma_{j}}$ between two GKP modes $a$ and $b$ with a gate time $T$, we must engineer the Hamiltonian
\begin{subequations}\label{eq:gate_Hamiltonian}
\begin{align}
    H_{\sigma_{i}\sigma_{j}}&=-\frac{1}{T} s_{i}\otimes s_{j}\\
    &=-\frac{r_{i}r_{j}}{2T}(e^{-i(\theta_{i}+\theta_{j})}ab + e^{-i(\theta_{i}-\theta_{j})}ab^{\dagger} + \text{h.c.}),
\end{align}
\end{subequations}
where $s_{1}=-\bar{p},\ s_{2}=\bar{q}-\bar{p},\ s_{3}=\bar{q}$ are the logical quadrature operators introduced in \cref{eq:s_quadratures} and $r_{i},\theta_{i}$ are the polar coordinates $s_{i}=r_{i}(q\cos\theta_{i}+p\sin\theta_{i})$. We can interpret the Hamiltonian \cref{eq:gate_Hamiltonian} as quadrature-quadrature coupling; or, equivalently, as a phase-coherent superposition of beamsplitter and two-mode squeezing interactions. For the square GKP code, we have $r_{2}=\sqrt{2}>r_{1}=r_{3}=1$, so either the Hamiltonian strength or the gate time must be increased for generalized controlled gates involving $Y$. Alternatively, in systems where such a simultaneous interaction is not possible, the generalized controlled gate can be decomposed into a product of beamsplitter and single-mode squeezing interactions using the Euler decomposition of the symplectic matrix $\mathcal{S}(\bar{C}_{\sigma_{i}\sigma_{j}})$~\cite{Dutta95}.

\begin{figure}
    \centering
    \includegraphics{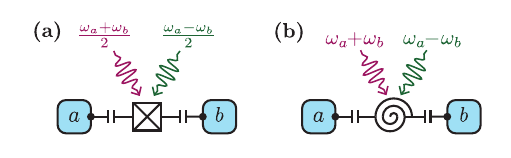}
    \caption{Schematic diagrams of (a) four-wave mixing, and (b) three-wave mixing between two GKP modes $a$ and $b$ with resonant frequencies $\omega_{a}$ and $\omega_{b}$. By phase coherently driving the non-linear coupling element with two microwave drives whose frequencies are given in \cref{eq:gate_Hamiltonian}, one can engineer the Hamiltonian required to perform each generalized controlled gate by altering the relative phase between the drives and the gate time. Supplemented with state preparation in the $\ket{0}$ or $\ket{T}$ states and measurement in any logical Pauli basis, this is sufficient to achieve universal quantum computing.}
    \label{fig:gate_circuits}
\end{figure}

To implement the Hamiltonian \cref{eq:gate_Hamiltonian} in superconducting circuits we can utilize either four-wave or three-wave mixing between GKP modes, as depicted in \cref{fig:gate_circuits}. Example elements that can be used to implement four- and three-wave mixing include an ancilla transmon or a SNAIL~\cite{Frattini17} (respectively). In both cases, the ancillary element is capacitively coupled to each of the microwave resonators housing the GKP modes. In both cases, two drive tones with frequencies $(\omega_{a}\pm\omega_{b})/2$ or $\omega_{a}\pm\omega_{b}$ are required  (for four- and three-wave mixing respectively), where $\omega_{a}$ and $\omega_{b}$ are the resonant frequencies of each mode. Finally, to implement each generalized controlled gate, one can simply change the relative phase between these two tones. As a consequence, single-qubit Cliffords can be implemented in software by tracking these relative phases.

In the following paragraphs, we give an intuitive and non-rigorous description of how to construct the Hamiltonian that results from such a four- or three-wave mixing circuit, and refer the reader to Ref.~\cite{Zhang19} for a more thorough derivation. We begin with a two-mode harmonic oscillator with Hamiltonian $H=\omega_{a}a^{\dag}a+\omega_{b}b^{\dag}b$. Four-wave mixing can be understood as adding to the system Hamiltonian any non-rotating terms consisting of the product of any four of the operators: $a, a^\dag, b, b^\dag$ and drive terms $V_{d}e^{\pm i(\omega_{d}t+\phi_{d})}$ (where $V_{d}$, $\omega_{d}$ and $\phi_{d}$ are the strength, frequency, and phase of any of the microwave drive tones). For example, we consider the term $a$ to be rotating with frequency $-\omega_{a}$ since $a\mapsto ae^{-i\omega_{a}t}$ in the rotating frame.

Using this heuristic, applying a drive with frequency $\omega_{1}=(\omega_{a}+\omega_{b})/2$ ensures that the term $V_{1}^{2}e^{2i(\omega_{1}t+\phi_{1})}ab$ (and its Hermitian conjugate) are non-rotating, thus providing a two-mode squeezing interaction with phase $2\phi_{1}$. This term acts as $V_{1}^{2}e^{2i\phi_{1}}ab+\text{h.c.}$ in the rotating frame, giving half of the interaction required in \cref{eq:gate_Hamiltonian}. Following similar logic, one can see that the beamsplitter terms in \cref{eq:gate_Hamiltonian} can be engineered with a second drive with frequency $\omega_{2}=(\omega_{a}-\omega_{b})/2$. The phase difference between each of the applied microwave tones and the oscillator determines the phase on the beamsplitter and two-mode squeezing terms, thereby specifying which generalized controlled gate is implemented.

However, four-wave mixing also introduces several unwanted terms into the Hamiltonian that reduce the gate fidelity. Kerr ($a^{\dag2} a^2$, $b^{\dag2} b^2$) and cross-Kerr ($a^\dag a b^\dag b$) terms will be added to the Hamiltonian even in the absence of microwave drives since these terms are always non-rotating. Moreover, the presence of the two drives also adds AC Stark shift terms such as $V_{1}^{2}a^{\dag}a$, which alter the resonant frequency of each cavity depending on the drive strength.

To avoid these unwanted Kerr, cross-Kerr, and AC Stark shift terms, one can instead use three-wave mixing, which can be implemented in circuit QED by replacing the transmon with a SNAIL~\cite{Frattini17}. Intuitively, three-wave mixing differs from four-wave mixing by only allowing the addition of non-rotating terms containing products of \textit{three} operators each. This avoids the introduction of Kerr, cross-Kerr and AC stark shift terms, while still allowing simultaneous beamsplitter and two-mode squeezing interactions when driven by tones with frequencies are $\omega_{1}=\omega_{a}+\omega_{b}$ and $\omega_{2}=\omega_{a}-\omega_{b}$. {In practice, it may be difficult to \textit{completely} eliminate Kerr and cross-Kerr terms, justifying our analysis of Kerr nonlinearity in \cref{sec:error_estimates,subsec:Kerr}.}

In summary: we have removed the need to explicitly perform single-qubit Clifford gates in superconducting GKP circuits. Instead, the single-qubit Clifford gates are tracked in software and implemented during two-qubit gates by altering the relative phase between drives in a three- or four-wave mixing circuit. Such a rewriting is favorable because single-qubit Clifford gates in general cause errors to spread during a circuit, as we explain in the following section. It is worth noting, however, that experimental implementations of such gates remain to be seen and may face practical difficulties to reach large coupling rates such as spurious non-linearities or additional dephasing.

Alongside the apparatus required to perform logical GKP gates, one also needs to prepare the GKP state~\cite{Terhal16,Royer20,Campagne19,Eickbusch22,DeNeeve22,Kolesnikow23}, which typically requires coupling to a qubit such as a transmon, and to implement non-Clifford gates either directly~\cite{Hastrup21,Royer22,nguyen2025fault} or via magic state preparation~\cite{Baragiola19}.

\section{Relating the Fidelity of Clifford Gates to that of the Identity Gate}\label{sec:Clifford_gate_estimate}

In this appendix, we derive the equation
\begin{equation}
    \bar{F}_{A,\mathcal{P},\Delta}=\bar{F}_{I,\mathcal{S}(\bar{A})^{-1}\mathcal{P},\Delta}\,,\tag{\ref{eq:Clifford_gate_estimate}}
\end{equation}
which equates the average gate fidelity of a GKP Clifford gate $\bar{A}$ decoded ideally over a patch $\mathcal{P}$ with the average gate fidelity of the identity gate decoded ideally over the modified patch $\mathcal{S}(\bar{A})^{-1}\mathcal{P}$.

To begin, recall the definition of the average gate fidelity
\begin{equation}
    \bar{F}(\mathcal{E},U)=\int d\psi\bra{\psi}U^\dag \mathcal{E}(\ket{\psi}\!\bra{\psi})U\ket{\psi},\tag{\ref{eq:avg_gate_fid_defn}}
\end{equation}
from which it can be trivially seen that $\bar{F}(\mathcal{E},U)=\bar{F}(\mathcal{J}[U^{\dag}]\circ\mathcal{E},I)$. Recall also our definition of the $n$-qubit to $n$-qubit channel
\begin{equation}
    \mathcal{E}_{U,\mathcal{P},\Delta}=\mathcal{J}[R^{\dag}]\circ\mathcal{C}_{\mathcal{P}}\circ\mathcal{J}[\bar{U}R_{\Delta}],\tag{\ref{eq:logical_channel}}
\end{equation}
where we defined $\bar{F}_{U,\mathcal{P},\Delta}=\bar{F}(\mathcal{E}_{U,\mathcal{P},\Delta},U)$ for shorthand. From \cref{eq:avg_gate_fid_defn} we can write this equivalently as $\bar{F}(\mathcal{J}[U^{\dag}]\circ\mathcal{E}_{U,\mathcal{P},\Delta},I)$.

Next, for a GKP Clifford gate $\bar{A}$, we claim that
\begin{equation}\label{eq:Clifford_through_decoder}
    \mathcal{J}[\bar{A}^{\dag}]\circ\mathcal{C}_{\mathcal{P}}\circ\mathcal{J}[\bar{A}]=\mathcal{C}_{\mathcal{S}(\bar{A})^{-1}\mathcal{P}}.
\end{equation}
On an intuitive level we can understand this as follows. From \cref{sec:GKP}, the decoder $\mathcal{C}_{\mathcal{P}}$ works by first measuring the stabilizer generators. This projects the state into a displaced ideal GKP codestate $T(\vect{v})\ket{\bar{\psi}}$, where $\vect{v}$ is revealed up to addition by a logical lattice vector $m\vect{\alpha}+n\vect{\beta}$ for some $m,n\in\mathbb{Z}$. Then, we choose the unique vector $\vect{k}$ that lies within the patch $\mathcal{P}$ and satisfies $\vect{k}=\vect{v}+m\vect{\alpha}+n\vect{\beta}$. Finally, we return the state to the codespace using $T(\vect{k})^{\dag}$.

Now consider $\mathcal{J}[\bar{A}^{\dag}]\circ\mathcal{C}_{\mathcal{P}}\circ\mathcal{J}[\bar{A}]$ as a single decoding step. Since the GKP Clifford gate $\bar{A}$ preserves the stabilizer group the information that is revealed about the state is the same, i.e.~we obtain $\vect{v}$ up to a logical lattice vector. However, the correction we apply is now $\bar{A}^{\dag}T(\vect{k})^{\dag}\bar{A}=T(S(\bar{A})^{-1}\vect{k})=T(\vect{k}')$, where $\vect{k}'$ is now the unique vector that lies within the patch $\mathcal{S}(\bar{A})^{-1}\mathcal{P}$ and satisfies $\vect{k}'=\vect{v}+m\vect{\alpha}+n\vect{\beta}$ (for $m,n\in\mathbb{Z}$). This operation we identify as simply being $\mathcal{C}_{\mathcal{S}(\bar{A})^{-1}\mathcal{P}}$, proving \cref{eq:Clifford_through_decoder}. Note that \cref{eq:Clifford_through_decoder} also follows from Eq.~(77) of Ref.~\cite{Shaw24-1}.

Finally, we have that
\begin{equation}
\bar{A}R=RA,\label{eq:Abar_R_commute}
\end{equation}
since $\bar{A}$ acts perfectly on the logical codespace. Putting this all together, we can write
\begin{subequations}
\begin{align}
    \bar{F}_{A,\mathcal{P},\Delta}&=\bar{F}(\mathcal{J}[A^{\dag}]\circ\mathcal{E}_{A,\mathcal{P},\Delta},I)\\
    &=\bar{F}(\mathcal{J}[A^{\dag}R^{\dag}]\circ\mathcal{C}_{\mathcal{P}}\circ\mathcal{J}[\bar{A}R_{\Delta}],I)\\
    &=\bar{F}(\mathcal{J}[R^{\dag}\bar{A}^{\dag}]\circ\mathcal{C}_{\mathcal{P}}\circ\mathcal{J}[\bar{A}R_{\Delta}],I)\\
    &=\bar{F}(\mathcal{J}[R^{\dag}]\circ\mathcal{C}_{\mathcal{S}(\bar{A})^{-1}\mathcal{P}}\circ\mathcal{J}[R_{\Delta}],I)\\
    &=\bar{F}_{I,\mathcal{S}(\bar{A})^{-1}\mathcal{P},\Delta},
\end{align}
\end{subequations}
proving \cref{eq:Clifford_gate_estimate}.

\section{Derivation of Average Gate Fidelity Estimates under Ideal Decoding}\label{sec:gate_error_estimate}

In this appendix, we present the derivation of the average gate fidelity estimates we used in \cref{sec:decoders} for GKP Clifford gates and in \cref{sec:error_estimates} for general noise maps under the assumption of ideal decoding. Our first task will be to derive a general formula [\cref{eq:avg_gate_infidelity_ratio}] for the average gate infidelity of a CV noise map $\mathcal{N}$ using the stabilizer subsystem decomposition~\cite{Shaw24-1}, assuming that $\mathcal{N}$ has a characteristic function that is sufficiently ``close to the identity'' (as we will explain in more detail later). Then, assuming that the noise map is Gaussian and symmetric in phase-space, we show that the average gate fidelity can be expressed using just three variables: the degeneracy $a$ and distance $d$ of the patch $\mathcal{P}$, and the variance $\sigma^{2}$ of the diagonal elements of the characteristic function of the noise map [\cref{eq:avg_gate_infid_sigma}]. We use this formula to analyze a wide range of relevant noise maps acting on approximate GKP codestates. When the only noise comes from approximate GKP codestates, we obtain \cref{eq:avg_fid_estimate}, which was used in \cref{sec:decoders}. However, we can also apply the formula to loss [\cref{eq:loss_sigma_estimate}], Gaussian random displacements, and gain. Finally, we extend our results to dephasing, which is a non-Gaussian channel, and derive the approximate expressions for the average gate fidelity in \cref{eq:dephasing_infid}.

Throughout this appendix, we will make frequent use of the entanglement fidelity $F_{e}$ of a (finite-dimensional) channel $\mathcal{E}$, defined as
\begin{equation}\label{eq:entanglement_fid}
    F_{e}=\braket{ME|(\mathcal{I}\otimes\mathcal{E})(\ket{ME}\!\bra{ME})|ME},
\end{equation}
where $\ket{ME}$ is a maximally entangled state across two copies of the finite-dimensional Hilbert space, and $\mathcal{I}$ is the identity channel. $F_{e}$ is related to the average gate fidelity \cref{eq:avg_gate_fid_defn} by the relationship~\cite{Nielsen02Simple}
\begin{equation}\label{eq:Nielsen_simple}
    \bar{F}=\frac{dF_{e}+1}{d+1},
\end{equation}
where $d$ is the dimension of the Hilbert space. For clarity, we will also define both the GKP \textit{stabilizer} lattice
\begin{equation}\label{eq:GKP_stab_lattice}
    \mathcal{L}_{\text{stab}}=\{2m\vect{\alpha}+2n\vect{\beta}\>|\>m,n\in\mathbb{Z}\},
\end{equation}
and the GKP \textit{logical} lattice
\begin{equation}
    \mathcal{L}_{\text{log}}=\{m\vect{\alpha}+n\vect{\beta}\>|\>m,n\in\mathbb{Z}\},\tag{\ref{eq:GKP_logical_lattice}}
\end{equation}
the latter of which coincides with the definition of $\mathcal{L}$ in \cref{sec:GKP}. Note that $\mathcal{L}_{\text{stab}}\subset\mathcal{L}_{\text{log}}$, since we can think of each stabilizer as implementing a logical identity gate.

\subsection{Stabilizer Subsystem Decomposition}\label{subsec:SSD}

We begin by summarizing the properties of the GKP stabilizer subsystem decomposition (SSSD) that we will use in the rest of this appendix. We refer the reader to Ref.~\cite{Shaw24-1}, particularly sections II and VII, for a more thorough introduction.

The SSSD decomposes the CV Hilbert space $\mathcal{H}$ into a tensor product of two Hilbert spaces $\mathcal{L}$ (the logical subsystem) and $\mathcal{S}$ (the stabilizer subsystem) as follows. First, consider an ideal GKP codestate $\ket{\bar{\psi}}$ that has been displaced by a vector $\vect{v}$, and label it
\begin{equation}\label{eq:stabilizer_states}
    \ket{\psi,\vect{v}}=T(\vect{v})\ket{\bar{\psi}}.
\end{equation}
One can show that for any patch $\mathcal{P}$ that is a primitive cell of the logical lattice $\mathcal{L}_{\text{log}}$, the set
\begin{equation}
    \big\{\!\ket{\mu,\vect{v}}\big|\;\mu\in\{0,1\},\vect{v}\in\mathcal{P}\big\}
\end{equation}
forms a basis of $\mathcal{H}$. Therefore, we can replace the comma in \cref{eq:stabilizer_states} with a tensor product, which defines the decomposition
\begin{align}
    \mathcal{H}&=\mathcal{L}\otimes\mathcal{S},&\ket{\psi,\vect{v}}=\ket{\psi}\otimes\ket{\vect{v}},
\end{align}
for $\vect{v}\in\mathcal{P}$.

The stabilizer subsystem $\mathcal{S}$ encodes the possible stabilizer measurement outcomes of a state, and is isomorphic to the \textit{full} Hilbert space $\mathcal{H}$, analogous to how if you split the real number line in two, you are left with two real number lines. $\mathcal{S}$ has a quasi-periodic structure, such that when an ideal codestate is displaced across a boundary of $\mathcal{P}$, a logical Pauli operator is applied to the logical subsystem. The logical subsystem $\mathcal{L}$ is isomorphic to $\mathbb{C}^{2}$ and represents the logical GKP information stored in the CV state. In particular, the partial trace of a CV state over the stabilizer subsystem $\mathcal{S}$ is equivalent to performing a round of ideal decoding over the patch $\mathcal{P}$ and ``forgetting'' the stabilizer measurement outcomes.

Recall that given a CV noise channel $\mathcal{N}$, we define the logical noise channel (which maps qubit density matrices to qubit density matrices) as
\begin{equation}
    \mathcal{N}_{\text{log}}=\mathcal{J}[R^{\dag}]\circ\mathcal{C}_{\mathcal{P}}\circ\mathcal{N}\circ\mathcal{J}[R],\tag{\ref{eq:logical_noise_channel}}
\end{equation}
where $R=\ket{\bar{0}}\!\bra{0}+\ket{\bar{1}}\!\bra{1}$. The logical noise channel can be written in terms of the SSSD as
\begin{equation}\label{eq:logical_noise_channel_defn_SSD}
    \mathcal{N}_{\text{log}}(\rho)=\mathrm{tr}_{\mathcal{S}}\big(\mathcal{N}(\rho\otimes\ket{\vect{0}}\!\bra{\vect{0}})\big).
\end{equation}
In Section VII.A of Ref.~\cite{Shaw24-1} we show that, in the single-mode case,
\begin{multline}\label{eq:single_mode_logical_noise_channel_SSD}
    \mathcal{N}_{\text{log}}(\rho)=\;\;\sum_{\mathclap{m_{1},m_{2},n_{1},n_{2}\in\mathbb{Z}}}\;\;(i^{m_{1}m_{2}}X^{m_{1}}Z^{m_{2}})\,\rho\, (i^{n_{1}n_{2}}X^{n_{1}}Z^{n_{2}})\\
    {\times}\bigg(\!\int_{\mathcal{P}}\!\!d^{2}\vect{v}\,e^{i\vect{v}^{T}\Omega\left(\vphantom{2^2}(m_{1}-n_{1})\vect{\alpha}+(m_{2}-n_{2})\vect{\beta}\right)/2}\\
    \times c\big(\vect{v}+m_{1}\vect{\alpha}+m_{2}\vect{\beta},\vect{v}+n_{1}\vect{\alpha}+n_{2}\vect{\beta}\big)\!\bigg),
\end{multline}
where
\begin{equation}
    \Omega=\begin{bmatrix}0&1\\-1&0\end{bmatrix}
\end{equation}
and $c(\vect{u},\vect{v})$ is the characteristic function of $\mathcal{N}$ as defined in \cref{eq:char_func_defn}. {Note that due to a different definition of the displacement operator the formulae differ in some places by a factor of $2\pi$.} Intuitively, one can obtain \cref{eq:single_mode_logical_noise_channel_SSD} from \cref{eq:logical_noise_channel_defn_SSD} by first writing $\mathcal{N}(\rho)$ in terms of displacements applied to the ideal codestate, then applying the quasi-periodic boundary conditions of the subsystem decomposition, and then taking the partial trace over $\mathcal{S}$.

The results obtained above generalize straightforwardly to multi-mode GKP codes as we now show (for more details about multi-mode GKP codes we refer the reader to Refs.~\cite{Gottesman01,Harrington04,Conrad22,Royer22}, and Sections III and IV of Ref.~\cite{Shaw24-1} for the multi-mode SSSD). In the present paper, the details of the multi-mode construction are not important, and we simply include the multi-mode case to demonstrate the generality of our results. To set this up, we define $n$-mode displacement operators by
\begin{equation}
    T(\vect{v})=\bigotimes_{j=1}^{n}T(v_{j},v_{j+n})
\end{equation}
for $\vect{v}\in\mathbb{R}^{2n}$. As such, the characteristic function $c(\vect{u},\vect{v})$ of a CV noise channel $\mathcal{N}$ takes vectors with length $2n$ as input. Then, suppose we have $k$ qubits in our multi-mode GKP code with logical lattice $\mathcal{L}_{\text{log}}$. Let the generators of this lattice $\{\vect{\ell}_{i}\}_{i=1}^{2n}$ be structured such that $T(\vect{\ell}_{1}),\dots,T(\vect{\ell}_{k})$ represent the $k$ logical $X$ operators and $T(\vect{\ell}_{n+1}),\dots,T(\vect{\ell}_{n+k})$ represent the $k$ logical $Z$ operators of the code. Then, we can write the logical noise channel $\mathcal{N}_{\text{log}}$ corresponding to a CV noise channel $\mathcal{N}$ as
\begin{multline}\label{eq:logical_noise_channel_SSD}
    \mathcal{N}_{\text{log}}(\rho)=\;\sum_{\mathclap{\vect{m},\vect{n}\in\mathbb{Z}^{2n}}}\;P(\vect{m})\rho P(\vect{n})\\
    {\times}\bigg(\!\int_{\mathcal{P}}\!\!d^{2n}\vect{v}\,c\big(\vect{v}{+}\vect{\ell}_{\vect{m}},\vect{v}{+}\vect{\ell}_{\vect{n}}\big)e^{i\vect{v}^{T}\Omega(\vect{\ell}_{\vect{m}}-\vect{\ell}_{\vect{n}})/2}\!\bigg),
\end{multline}
where we defined the $k$-qubit Pauli operators 
\begin{equation}
    P(\vect{m})=\bigotimes_{j=1}^{k}i^{m_{i}m_{i+n}}X^{m_{i}}Z^{m_{i+n}}
\end{equation}
and the matrix
\begin{equation}
    \Omega=\begin{bmatrix}0_{n}&I_{n}\\-I_{n}&0_{n}\end{bmatrix}
\end{equation}
in block form. \cref{eq:logical_noise_channel_SSD} is a straightforward generalization of \cref{eq:single_mode_logical_noise_channel_SSD}. \cref{eq:logical_noise_channel_SSD} will be the starting point of the rest of our derivations.

\subsection{Approximating the Entanglement Infidelity}\label{subsec:derivation}

Now we are ready to derive \cref{eq:avg_gate_infidelity_ratio}. Given a (trace-preserving) quantum channel $\mathcal{E}$, we define the $\chi$ matrix of $\mathcal{E}$ as
\begin{equation}\label{eq:chi_mat_defn}
    \mathcal{E}(\rho)=\sum_{P,P'}\chi_{P,P'}P\rho P',
\end{equation}
where the sum is over all $n$-qubit Pauli operators. Then, the entanglement fidelity of the channel is given by
\begin{equation}\label{eq:fid_chi}
    F_{e}=\chi_{I,I}.
\end{equation}

Now suppose that we only have access to a non-normalized map $\tilde{\mathcal{E}}=N\mathcal{E}$ with $\chi$ matrix components $\tilde{\chi}_{P,P'}$, where $N$ is some positive real constant and $\mathcal{E}$ is trace-preserving. We can still obtain the entanglement fidelity of $\mathcal{E}$ from $\tilde{\chi}$ using the equation
\begin{align}\label{eq:non-normalized_entanglement_fidelity}
    F_{e}&=\tilde{\chi}_{I,I}\bigg/\sum_{P}\tilde{\chi}_{P,P}.
\end{align}

In the context of GKP codes, we are interested in finding the entanglement fidelity (and, hence, the average gate fidelity) of the logical noise channel associated with the envelope operator $e^{-\Delta^{2}a^{\dag}a}$ followed by a (trace-preserving) error channel $\mathcal{E}$. We call the composition of the two maps the noise map $\mathcal{N}(\rho)=\mathcal{E}(e^{-\Delta^{2}a^\dag a}\rho e^{-\Delta^{2}a^{\dag}a})$, which is not trace-preserving since the envelope operator is non-unitary. Importantly, the entanglement fidelity is not well defined for non-trace-preserving maps, so we instead consider an ``orthonormalized'' channel $\mathcal{N}_{\text{o}}$, see Appendix G of Ref.~\cite{Shaw24-1} for one way of defining $\mathcal{N}_{\text{o}}$ from $\mathcal{N}$. Fortunately, for small $\Delta$, the norm of the approximate encoded state $\tr(e^{-2\Delta^{2}a^\dag a}\bar{\rho})$ is approximately independent of the qubit state $\rho$. So, we approximate $\mathcal{N}$ to be proportional to $\mathcal{N}_{\text{o}}$ with constant of proportionality $N$. The logical channel $\mathcal{N}_{\text{log},\text{o}}$ associated with $\mathcal{N}_{\text{o}}$ via \cref{eq:logical_noise_channel_SSD} is trace-preserving and has a well-defined entanglement fidelity $F_{e}$. We can then approximately obtain $F_{e}$ from the non-trace-preserving logical map $\mathcal{N}_{\text{log}}$ associated with $\mathcal{N}$ using \cref{eq:non-normalized_entanglement_fidelity}.

Next, consider any non-normalized CV map $\mathcal{N}$ that is ``sufficiently close to the identity map''. We take this to mean that the characteristic function $c(\vect{u},\vect{v})$ decays sufficiently quickly to 0 as $|\vect{u}|$ and $|\vect{v}|$ increase; note that the characteristic function of the identity map is $c(\vect{u},\vect{v})=\delta^{2n}(\vect{u})\delta^{2n}(\vect{v})$. In particular, we will assume that whenever there is an integral of $c(\vect{u},\vect{v})$, we can ignore any terms where \textit{either} $\vect{u}$ or $\vect{v}$ lie outside the Voronoi cell $\mathcal{V}_{\text{stab}}$ of the \textit{stabilizer} lattice $\mathcal{L}_{\text{stab}}$.

From \cref{eq:logical_noise_channel_SSD}, we can write the entanglement infidelity exactly as
\begin{subequations}\label{eq:precise_starting_point}
\begin{align}
    1{-}F_{e}&=\sum_{P\neq I}\tilde{\chi}_{P,P}\bigg/\sum_{P}\tilde{\chi}_{P,P},\label{eq:precise_starting_point_chis}\\
    \tilde{\chi}_{P,P}&=\!\!\sum_{\vect{s},\vect{t}\in\mathcal{L}_{\text{stab}}}\!\!\!\!\bigg(\!\int_{\mathcal{P}}\!d^{2n}\vect{v}\,e^{i(\vect{v}+\vect{\ell}_{P})^{T}\Omega(\vect{s}{-}\vect{t})}\nonumber\\
    &\hspace{2.25 cm}\times c(\vect{v}{+}\vect{\ell}_{P}{+}\vect{s},\vect{v}{+}\vect{\ell}_{P}{+}\vect{t})\!\bigg),\label{eq:chi_PP_exact}
\end{align}
\end{subequations}
where we denote $\vect{\ell}_{P}$ as any representative of $\mathcal{L}_{\text{log}}$ such that $T(\vect{\ell}_{P})$ applies a logical $P$ to the GKP code; in the single-mode case we simply have $\vect{\ell}_{P}=\vect{0},\vect{\alpha},\vect{\alpha}+\vect{\beta},\vect{\beta}$ respectively for $P=I,X,Y,Z$.

One could of course numerically evaluate $1-F_{e}$ by using the exact expression \cref{eq:precise_starting_point} and truncating the sums over $\vect{s},\vect{t}$ to some finite number; however here we wish to arrive at a simpler, more intuitive expression for the infidelity. To do this, note that $c(\vect{v},\vect{v}+\vect{s})$ can be ignored for any non-zero $\vect{s}$ since $\vect{v}$ and $\vect{v}+\vect{s}$ can't be both in the Voronoi cell $\mathcal{V}_{\text{stab}}$ of the stabilizer lattice $\mathcal{L}_{\text{stab}}$. So, we can approximate \cref{eq:chi_PP_exact} as
\begin{equation}\label{eq:chi_PP_approx}
    \tilde{\chi}_{P,P}\approx\!\!\sum_{\vect{s}\in\mathcal{L}_{\text{stab}}}\!\int_{\mathcal{P}}\!d^{2n}\vect{v}\,c(\vect{v}{+}\vect{\ell}_{P}{+}\vect{s},\vect{v}{+}\vect{\ell}_{P}{+}\vect{s}),
\end{equation}
and the denominator of \cref{eq:precise_starting_point_chis} as
\begin{equation}
    \sum_{P}\tilde{\chi}_{P,P}\approx\int_{\mathbb{R}^{2n}}\! d^{2n}\vect{v}\,c(\vect{v},\vect{v}).
\end{equation}
We can also use \cref{eq:chi_PP_approx} to write the numerator of \cref{eq:precise_starting_point_chis} as
\begin{subequations}
    \begin{align}
        \sum_{P\neq I}\tilde{\chi}_{P,P}&\approx\sum_{P\neq I}\,\sum_{\vect{s}\in\mathcal{L}_{\text{stab}}}\!\int_{\mathcal{P}}\!d^{2n}\vect{v}\,c(\vect{v}{+}\vect{\ell}_{P}{+}\vect{s},\vect{v}{+}\vect{\ell}_{P}{+}\vect{s}),\label{eq:numerator_approx_1}\\
        &\approx\int_{\mathrlap{\mathbb{R}^{2n}\setminus\mathcal{P}}}\;\;d^{2n}\vect{v}\,c(\vect{v},\vect{v}),\label{eq:numerator_approx_2}
    \end{align}
\end{subequations}
where to go from \cref{eq:numerator_approx_1} to \cref{eq:numerator_approx_2}, we have added the sum 
\begin{equation}
\sum_{\substack{\vect{s}\in\mathcal{L}_{\text{stab}}\\\vect{s}\neq\vect{0}}}\!\!\!c(\vect{v}+\vect{s},\vect{v}+\vect{s})
\end{equation}
which is small since $\vect{v}+\vect{s}$ lies outside $\mathcal{V}_{\text{stab}}$ for any $\vect{v}\in\mathcal{P}$~\footnote{We note here that strictly speaking this statement relies on $\mathcal{P}$ (which is an arbitrary primitive cell of $\mathcal{L}_{\text{log}}$) being contained entirely within $\mathcal{V}_{\text{stab}}$ (the Voronoi cell of $\mathcal{L}_{\text{stab}}$), which is true for most cases of interest.}.

Therefore, to calculate the entanglement fidelity we now simply need to determine the ratio
\begin{equation}\label{eq:entanglement_infidelity_ratio}
    1-F_{e}\approx\int_{\mathrlap{\mathbb{R}^{2n}\setminus\mathcal{P}}}\;d^{2n}\vect{v}\,c(\vect{v},\vect{v})\bigg/\!\!\int_{\mathbb{R}^{2n}}\!\!\!d^{2n}\vect{v}\,c(\vect{v},\vect{v}),
\end{equation}
where $\setminus$ denotes the ``set-minus'' symbol. Alternatively, we can trivially convert this to the average gate infidelity using \cref{eq:Nielsen_simple}, giving
\begin{equation}
    1-\bar{F}\approx\frac{2^{n}}{2^{n}+1}\int_{\mathrlap{\mathbb{R}^{2n}\setminus\mathcal{P}}}\;d^{2n}\vect{v}\,c(\vect{v},\vect{v})\bigg/\!\!\int_{\mathbb{R}^{2n}}\!\!\!d^{2n}\vect{v}\,c(\vect{v},\vect{v}),\tag{\ref{eq:avg_gate_infidelity_ratio}$'$}
\end{equation}
where the prefactor is $2/3$ when $n=1$.

\subsection{Gaussian Characteristic Functions}\label{subsec:gaussian_char}

To continue our analysis, we assume that the diagonal elements of the characteristic function take the form
\begin{equation}
    c(\vect{v},\vect{v})\propto e^{-|\vect{v}|^{2}/2\sigma^{2}},\tag{\ref{eq:diagonal_characteristic_function_ansatz}}
\end{equation}
where we call $\sigma^{2}$ the ``variance'' of the noise map. Then, for small $\sigma$, we can apply \cref{eq:entanglement_infidelity_ratio} to obtain
\begin{equation}\label{eq:entanglement_fidelity_int_over_P}
    1-F_{e}\approx\sigma^{-2n}\!\int_{\mathbb{R}^{2n}\setminus\mathcal{P}}\!\frac{d^{2n}\vect{v}}{(2\pi)^{n}}e^{-|\vect{v}|^{2}/2\sigma^{2}}.
\end{equation}

\begin{figure}
    \centering
    \includegraphics{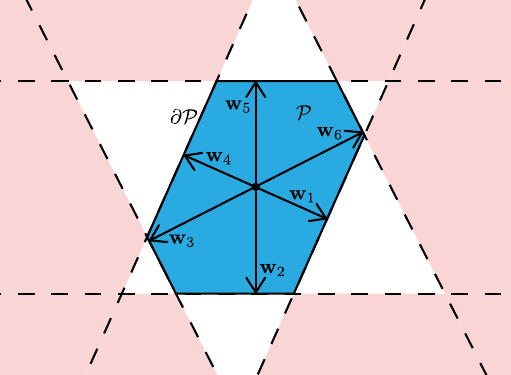}
    \caption{Diagram representing the regions of integration in \cref{eq:patch_integral_upper_bound}. There are six local shortest vectors $\vect{w}_{1},\dots,\vect{w}_{6}$ on the boundary $\partial\mathcal{P}$ which each have length $d_{1}/2,\dots,d_{6}/2$. Due to the $\pi$-rotational symmetry of $\mathcal{P}$, the vectors $\vect{w}_{i}$ come in pairs of vectors of equal length. The approximation of the integral over $\mathbb{R}^{2n}\setminus\mathcal{P}$ as a sum of integrals in \cref{eq:patch_integral_upper_bound} overcounts the red shaded regions in the figure.}
    \label{fig:patch_approximation}
\end{figure}

To approximate the integral over $\mathbb{R}^{2n}\setminus\mathcal{P}$ in \cref{eq:entanglement_fidelity_int_over_P}, we must now consider the geometry of the patch $\mathcal{P}$ and its boundary $\partial\mathcal{P}$, as depicted in \cref{fig:patch_approximation}. Consider the length of each vector $\vect{v}\in\partial\mathcal{P}$ as one travels around the boundary $\partial\mathcal{P}$. Label each local minimum in the length of $\vect{v}$ as $d_{i}/2$, and the corresponding vector $\vect{w}_{i}$. We will call $\{d_{i}\}$ the set of local distances of the patch. Then, the integral \cref{eq:entanglement_fidelity_int_over_P} is upper bounded by
\begin{align}
    &\frac{1}{(2\pi\sigma^{2})^{n}}\!\!\int_{\mathrlap{\mathbb{R}^{2n}\setminus\mathcal{P}}}\;\;d^{2n}\vect{v}\,e^{-|\vect{v}|^{2}/2\sigma^{2}}\nonumber\\
    &<\frac{1}{(2\pi\sigma^{2})^{n}}\sum_{i}\int_{d_{i}/2}^{\infty}\!\!dv_{1}e^{-v_{1}^{2}/2\sigma^{2}}\!\!\int_{\mathrlap{\mathbb{R}^{2n-1}}}\;\;d^{2n-1}\vect{v}\,e^{-|\vect{v}|^{2}/2\sigma^{2}}\label{eq:patch_integral_upper_bound}\\
    &=\frac{1}{2}\sum_{i}\mathrm{erfc}\big(d_{i}/(2\sqrt{2}\sigma)\big),\label{eq:patch_integral_sum}
\end{align}
where $\mathrm{erfc}(x)$ is the complementary error function.

One can understand \cref{eq:patch_integral_upper_bound} as follows. The region of integration $\mathbb{R}^{2n}\setminus\mathcal{P}$ can be written as the union of regions $\mathcal{R}_{i} = \{\vect{v}\in\mathbb{R}^{2n}\mid\vect{v}^{T}\vect{w}_{i}>(d_{i}/2)^{2}\}$. Each integral in the right-hand side of \cref{eq:patch_integral_upper_bound} is simply an integral of $e^{-|\vect{v}|^{2}/2\sigma^{2}}$ over the region $\mathcal{R}_{i}$ where we have performed an orthogonal change of variables such that the component $v_{1}$ points in the direction of $\vect{w}_{i}$. Therefore, the sum of the integrals over each region $\mathcal{R}_{i}$ is greater than the integral over $\mathbb{R}^{2n}\setminus\mathcal{P}$ as it overcounts any region of $\mathbb{R}^{2n}$ that is in the intersection of at least two regions $\mathcal{R}_{i}\cap\mathcal{R}_{j}$ (see \cref{fig:patch_approximation}). However, the intersection of these regions is further away from the origin than $\partial\mathcal{P}$ and their contribution to the integral is exponentially suppressed. As such, the upper bound in \cref{eq:patch_integral_upper_bound} becomes approximately equal to the desired integral as $\sigma\rightarrow0$.

As $\sigma\rightarrow0$ we can make additional approximations to the sum in \cref{eq:patch_integral_sum}. In particular, we can neglect all the terms in the sum except for when $d_{i}$ is minimal. We define the \textit{distance} of the patch
\begin{equation}
    d=\min_{i}(d_{i})=2\min_{\vect{v}\in\partial\mathcal{P}}|\vect{v}|,
\end{equation}
where the factor of 2 ensures that when $\mathcal{P}$ is the Voronoi cell, the distance of the patch coincides with the length of the shortest (non-zero) displacement in the logical Pauli group. We define the \textit{degeneracy} $a$ of the patch as half the number of vectors $\vect{v}\in\partial\mathcal{P}$ with length $d$. Assuming the patch is centered at the origin, $a$ is an integer. With these definitions, we finally arrive at our entanglement infidelity estimate
\begin{equation}\label{eq:entanglement_infid_sigma}
    1-F_{e}\approx a\,\mathrm{erfc}\big(d/(2\sqrt{2}\sigma)\big),
\end{equation}
or, equivalently,
\begin{equation}
    1-\bar{F}\approx \frac{2^{n}a}{2^{n}+1}\mathrm{erfc}\big(d/(2\sqrt{2}\sigma)\big).\tag{\ref{eq:avg_gate_infid_sigma}$'$}
\end{equation}

From \cref{eq:entanglement_infid_sigma} we see that three terms contribute to the infidelity of the logical noise map. First is the variance $\sigma$ of the noise map, which we calculate for various noise maps of interest below. Since $\mathrm{erfc}$ is a monotonically decreasing function, reducing $\sigma$ always decreases the infidelity of the noise map. The other two terms are the distance $d$ and degeneracy $a$ of the patch $\mathcal{P}$. The distance of the patch is maximized (and hence the infidelity minimized) when $\mathcal{P}$ is the Voronoi cell of the GKP lattice, in which case it is equal to the length of the shortest non-trivial logical Pauli operator of the code. Voronoi cells typically exhibit more symmetry than arbitrary patches and consequently can have degeneracies greater than 1.

Finally, a brief note of caution. In some cases, the approximation of \cref{eq:patch_integral_sum} as \cref{eq:entanglement_infid_sigma} is not very good because the distance $d$ is not much smaller than the next smallest local distance $d_{i}$. While \cref{eq:entanglement_infid_sigma} will still hold asymptotically, it may only provide a good approximation for \textit{very} small $\sigma$. In these cases, it is best to simply use \cref{eq:patch_integral_sum} to estimate the infidelity. Moreover, \cref{eq:patch_integral_sum} has the advantage that it is \textit{always} an upper-bound on the integral \cref{eq:entanglement_fidelity_int_over_P}. If the patch $\mathcal{P}$ is the Voronoi cell of some lattice $\mathcal{L}$, then finding the set of local distances $\{d_{i}\}$ amounts to finding the lengths of all the \textit{Voronoi-relevant} vectors of $\mathcal{L}$, for instance using the algorithms presented in Ref.~\cite{Agrell02}.

\subsection{Envelope Operator}\label{subsec:envelope}

We spend the rest of this appendix using \cref{eq:entanglement_infid_sigma} to calculate the entanglement infidelities of various noise maps of interest. To start, we consider the errors arising only from the approximate codewords themselves; i.e.~we choose the noise map $\mathcal{N}(\rho)=e^{-\Delta^{2}a^{\dag}a}\rho e^{-\Delta^{2}a^{\dag}a}$. The characteristic function is given by
\begin{equation}\label{eq:envelope_characteristic_function}
    c(\vect{u},\vect{v})\propto \mathrm{exp}\bigg(\!{-}\frac{1}{4}\coth(\Delta^{2}/2)\big(|\vect{u}|^{2}+|\vect{v}|^{2}\big)\!\bigg),
\end{equation}
where the constant of proportionality is irrelevant since it is canceled in the ratio \cref{eq:entanglement_infidelity_ratio}. The diagonal elements $c(\vect{v},\vect{v})$ therefore take the form of \cref{eq:diagonal_characteristic_function_ansatz} with variance $\sigma^{2}=\mathrm{tanh}(\Delta^{2}/2)\approx\Delta^{2}/2$. As such we can immediately apply this to \cref{eq:entanglement_infid_sigma} to obtain the entanglement infidelity of the trace-preserving logical map $\mathcal{N}_{\text{log},\text{o}}$ as
\begin{equation}\label{eq:infid_Delta}
    1-F_{e}\approx a\,\mathrm{erfc}(d/2\Delta),
\end{equation}
which directly gives \cref{eq:avg_fid_estimate}.

\begin{figure*}
\includegraphics{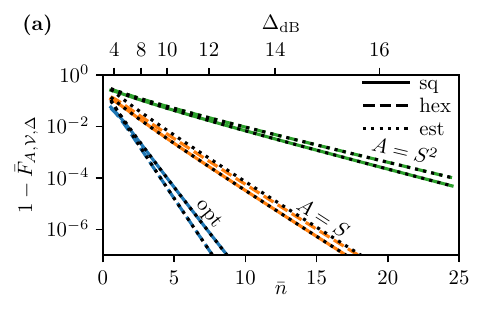}
\hspace*{\fill}
\includegraphics{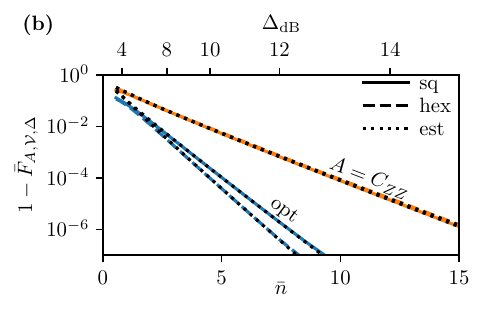}
\caption{Comparison between the theoretical estimated average gate fidelity from \cref{eq:avg_fid_estimate} and the numerically calculated average gate fidelity from \cref{fig:gate_error_plots}, under the assumption of ideal error-correction. Estimated and numerical average gate infidelity $1-\bar{F}_{A,\mathcal{V}_{\mathcal{L}},\Delta}$ of (a) single-qubit, and (b) two-qubit Clifford gates $A=I,S,S^{2},C_{Z}$ applied to the square and hexagonal GKP codes, decoded over the Voronoi cell $\mathcal{V}_{\mathcal{L}}$ of their respective lattices. Note that for the hexagonal $C_{ZZ}$ plots in (b) and (d), a limited number of data points are included due to the time taken to run the simulations.}
\label{fig:gate_error_estimate_comparison}
\end{figure*}

\cref{eq:infid_Delta} is a useful equation to test the validity of our approximations over various patch geometries. In \cref{fig:gate_error_estimate_comparison} we compare the estimated average gate infidelities from \cref{eq:infid_Delta} with the numerical average gate infidelities as plotted in \cref{fig:gate_error_plots} for the envelope operator decoded over a Clifford-gate-deformed patch $\mathcal{P}$. The estimated values are obtained by using the values for the distance and degeneracy of each patch in \cref{tab:estimates} and substituting this into \cref{eq:infid_Delta}. The estimated values typically show good agreement with the numerical values, with relative errors below 1\% in the parameter regime of interest. One exception is that of the hexagonal $S$ gate, which has a larger relative error of between 10\% and 25\% in the regime of interest, even though the error tends to 0 as $\Delta_{\text{dB}}$ increases. We attribute this increased error to the unusual geometry of the deformed patch $\mathcal{S}(\bar{S}_{\text{hex}})^{-1}\mathcal{V}_{\text{hex}}$. The degeneracy of the patch is 2, but the two sides of the patch that contribute to the equal distance are nearly parallel. This means that the approximation of the integral \cref{eq:patch_integral_upper_bound} is less accurate because it overcounts a region of $\mathbb{R}^{2}$ that contains vectors with length only slightly larger than $d/2$. Intuitively, one can imagine the red shaded regions in \cref{fig:patch_approximation} if two adjacent sides of the hexagon are nearly parallel. The red region would then contain relatively short vectors that correspond to terms $e^{-|\vect{v}|^{2}/(2\sigma^{2})}$ in the integral. These $e^{-|\vect{v}|^{2}/(2\sigma^{2})}$ terms are, in turn, only significantly smaller than the leading order terms $e^{-(d/2)^{2}/(2\sigma^{2})}$ for \textit{very} small $\sigma$. Nevertheless, our estimate still provides the correct asymptotic scaling for the infidelity, and the relative error is less than 25\% in the regime of interest.

\subsection{Phase-covariant Gaussian Channels}

One advantage of our analysis is that we can also easily apply \cref{eq:avg_gate_infid_sigma} to more general noise maps. For example, consider the noise map given by the composition
\begin{equation}\label{eq:Gaussian_noise_map_composed}
    \mathcal{N}=\mathcal{E}_{\tau,\nu}\circ\mathcal{J}[e^{-\Delta^{2}a^{\dag}a}],
\end{equation}
where $\mathcal{E}_{\tau,\nu}$ is a single-mode \textit{phase-covariant} Gaussian channel defined as follows. First, we write the expectation values and covariance matrix of an arbitrary single-mode density matrix as
\begin{align}
\vect{\mu}&=\begin{bmatrix}\langle q\rangle\\\langle p\rangle\end{bmatrix},&
V&=\begin{bmatrix}\langle q^{2}\rangle&\frac{1}{2}\langle qp+pq\rangle\\\frac{1}{2}\langle qp+pq\rangle&\langle p^{2}\rangle\end{bmatrix}-\vect{\mu}\vect{\mu}^{T},
\end{align}
as is standard in the study of Gaussian quantum states~\cite{Eisert05}. Then, $\mathcal{E}_{\tau,\nu}$ is the channel that transforms these matrices via
\begin{align}
\vect{\mu}&\mapsto \tau\vect{\mu},&V\mapsto\tau^{2}V+\nu,
\end{align}
where $\tau,\nu$ are non-negative real numbers that satisfy $\nu\geq|\tau^{2}-1|/2$ to ensure $\mathcal{E}_{\tau,\nu}$ is completely positive. Roughly speaking, $\tau$ represents how much phase-insensitive squeezing ($\tau<1$) or amplification ($\tau>1$) is applied to the quadratures, and $\nu$ represents how much Gaussian noise is added to the quadratures. Three commonly-studied channels take this form:
\begin{itemize}
\item loss: $\tau=\sqrt{1-\gamma}$, $\nu=\gamma/2$, $\gamma\in[0,1]$;
\item Gaussian random displacements (GRD): $\tau=1$, $\nu=\sigma_{G}^{2}$, $\sigma_{G}\geq0$; and
\item gain: $\tau=\sqrt{g}$, $\nu=(g-1)/2$, $g\geq1$.
\end{itemize}
Note that setting $\tau=1$, $\nu=0$ results in the identity channel. These channels are called phase covariant since they commute with the rotation operators $e^{i\theta a^{\dag}a}$.

We have previously written down the characteristic function of phase-covariant Gaussian channels in Appendix F of Ref.~\cite{Shaw24-1}, although again note some differences in factors of $2\pi$ due to the different definitions of displacement functions. To compose this with the envelope operator, we use the composition formula for characteristic functions
\begin{multline}
    c(\vect{u},\vect{v})=\iint \!d^{2n}\vect{\tilde{u}}d^{2n}\vect{\tilde{v}}\,c_{1}(\vect{u}-\vect{\tilde{u}},\vect{v}-\vect{\tilde{v}})c_{2}(\vect{\tilde{u}},\vect{\tilde{v}})\\
    \times e^{-i\vect{\tilde{u}}^{T}\Omega\vect{u}/2}e^{i\vect{\tilde{v}}^{T}\Omega\vect{v}/2},
\end{multline}
where $c$ is the characteristic function of the channel $\mathcal{N}=\mathcal{N}_{1}\circ\mathcal{N}_{2}$, and $c_{1}$ and $c_{2}$ are the characteristic channels of $\mathcal{N}_{1}$ and $\mathcal{N}_{2}$, respectively. With this,

we can write the diagonal elements of the characteristic function of $\mathcal{N}$ [\cref{eq:Gaussian_noise_map_composed}] in the form of \cref{eq:diagonal_characteristic_function_ansatz} with variance
\begin{subequations}\label{eq:variance_tau_nu}
\begin{align}
    \sigma^{2}&=\tau\mathrm{tanh}(\Delta^{2}/2)+\nu+\frac{(1-\tau)^{2}}{2}\coth(\Delta^{2})\label{eq:exact_variance_tau_nu}\\
    &\approx \frac{\Delta^{2}}{2}+\nu+\frac{(1-\tau)^{2}}{2\Delta^{2}}.\label{eq:approx_variance_tau_nu}
\end{align}
\end{subequations}
We can interpret the variance as being the sum of three terms: one directly from the envelope operator ($\Delta^{2}/2$), one from the noise added by the channel ($\nu$), and one final term that has a quadratic dependence on the phase-insensitive squeezing/amplification $|1-\tau|$ and an inverse dependence on $\Delta^{2}$. Intuitively, this final term arises because amplifying or squeezing phase space affects GKP codes more severely the wider the GKP state is spread in phase space (i.e.~for small $\Delta$).

We pause briefly here to precisely analyze the assumptions under which the infidelity estimate given by \cref{eq:entanglement_infid_sigma,eq:exact_variance_tau_nu} is valid. As stated previously in \cref{subsec:derivation}, we require that the modulus of the characteristic function $|c(\vect{u},\vect{v})|$ is negligibly small whenever either $\vect{u}$ or $\vect{v}$ is longer than the length of the shortest logical Pauli operator $\vect{\ell}_{\text{min}}$ (which is roughly $\sqrt{\pi}$ for most cases of interest), since $\vect{\ell}_{\text{min}}$ is the shortest vector on the boundary of $\mathcal{V}_{\text{stab}}$. In particular, since $\mathcal{N}$ is a Gaussian channel, we can always write
\begin{equation}
    c(\vect{u},\vect{v})\propto\exp\bigg(-\frac{1}{2}\begin{bmatrix}\vect{u}^{T},\vect{v}^{T}\end{bmatrix}M\begin{bmatrix}\vect{u}\\\vect{v}\end{bmatrix}\bigg),
\end{equation}
where $M$ is a $4n\times4n$ complex matrix. Then, the condition on $c$ is satisfied if the smallest eigenvalue $\lambda_{\text{min}}$ of the real part of $M$ satisfies $\lambda_{\text{min}}\gg |\vect{\ell}_{\text{min}}|^{-2}$. To see this, first note that for $|c(\vect{u},\vect{v})|$ to be negligibly small, we need the real part of $\begin{bmatrix}\vect{u}^{T},\vect{v}^{T}\end{bmatrix}M\begin{bmatrix}\vect{u}\\\vect{v}\end{bmatrix}$ to be large. Then, we have
\begin{equation}
    \begin{bmatrix}\vect{u}^{T},\vect{v}^{T}\end{bmatrix}M\begin{bmatrix}\vect{u}\\\vect{v}\end{bmatrix}\geq\lambda_{\text{min}}\max\big(|\vect{u}|^{2},|\vect{v}|^{2}\big)\gg1,
\end{equation}
whenever $\max\big(|\vect{u}|^{2},|\vect{v}|^{2}\big)\geq|\vect{\ell}_{\text{min}}|^{2}$, from which the condition follows. For the noise map given in \cref{eq:Gaussian_noise_map_composed}, the smallest eigenvalue of $M$ is $1/(2\sigma^{2})$, so our approximations are valid whenever $\sigma$ is much smaller than $|\vect{\ell}_{\text{min}}|/\sqrt{2}\sim \sqrt{\pi/2}$. This is a fairly loose requirement and, in practice, is valid for a wide range of parameters. The further approximation \cref{eq:approx_variance_tau_nu} is only valid in the slightly more specific regime where $|1-\tau|,\nu,\Delta$ each are also small.

Assuming $\tau$ and $\nu$ are fixed, one can easily show that $\sigma^{2}$ is minimized when $\Delta^{2}=|\ln\tau|\approx|1-\tau|$, which represents the optimal $\Delta$ that minimizes the infidelity of the noise map. This optimal $\Delta$ depends only on the amount of quadrature ``squeezing'' or ``amplification'' applied by the Gaussian channel, and not on the additional noise $\nu$ added by the channel.

If we set $\mathcal{E}_{\tau,\nu}$ in \cref{eq:Gaussian_noise_map_composed} to a GRD channel with $\nu=\sigma^{2}_{G}$, we obtain $\sigma^{2}=\sigma_{G}^{2}+\tanh(\Delta^{2}/2)$; in other words, the variance from the Gaussian random displacements adds linearly to the variance from the envelope map. Moreover, since $\tau=1$, the optimal envelope size is $\Delta=0$, i.e.~an ideal GKP codeword.

\begin{figure}
\includegraphics{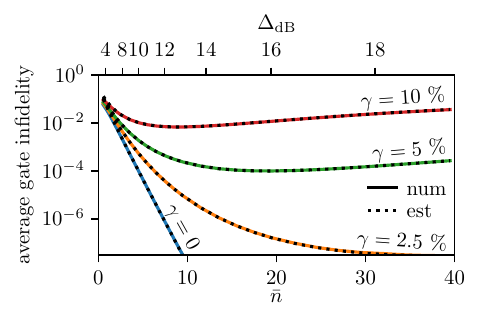}
\caption{Comparison between the theoretical estimated average gate fidelity from \cref{eq:diagonal_characteristic_function_ansatz,eq:loss_exact_sigma_estimate} and the numerically calculated average gate fidelity from Fig.~5(a) of Ref.~\cite{Shaw24-1}. Estimated and numerical average gate infidelity $1-\bar{F}$ of the logical noise map corresponding to the envelope operator followed by loss for the single-mode square GKP code.}
\label{fig:loss_error_comparison}
\end{figure}

In contrast, setting $\mathcal{E}_{\tau,\nu}$ to a loss channel gives
\begin{subequations}
\begin{align}
\sigma^{2}&=\sqrt{1-\gamma}\,\mathrm{tanh}\bigg(\!\frac{\Delta^{2}}{2}\!\bigg)+\frac{\gamma}{2}+\frac{(1-\sqrt{1-\gamma})^{2}}{2\,\mathrm{tanh}(\Delta^{2})}\tag{\ref{eq:loss_exact_sigma_estimate}}\\
&\approx \frac{\Delta^{2}}{2}+\frac{\gamma}{2}+\frac{\gamma^{2}}{8\Delta^{2}}.\tag{\ref{eq:loss_approx_sigma_estimate}}
\end{align}
\end{subequations}
for small $\Delta,\gamma$, as discussed in \cref{sec:error_estimates}. The optimal envelope size here is given by $\Delta^{2}=-\mathrm{ln}\sqrt{1-\gamma}\approx\gamma/2$.

In \cref{fig:loss_error_comparison}, we assess the validity of the average gate fidelity estimate from \cref{eq:entanglement_infid_sigma,eq:loss_exact_sigma_estimate} with the numerically obtained fidelities in Fig.~5(a) of Ref.~\cite{Shaw24-1}. Similarly to in \cref{fig:gate_error_estimate_comparison}, we see that our estimates perform well, with the relative error between the numerical and estimated values being less than 1\% within the region of interest.

For comparison, note that the entanglement infidelity of loss acting on the trivial bosonic encoding into the Fock states $\ket{0}$ and $\ket{1}$ is $1-F_{e}=1-\big((1+\sqrt{1-\gamma})/2\big)^{2}\approx \gamma/2$. In contrast, approximate GKP codestates have average entanglement fidelity given by \cref{eq:entanglement_infid_sigma,eq:loss_approx_sigma_estimate}, which can be expanded to first order in $\gamma$ to give
\begin{equation}\label{eq:loss_Taylor_expand}
1-F_{e}\approx a\,\mathrm{erfc}\big(d/(2\Delta)\big)+\gamma\frac{ad}{2\sqrt{\pi}\Delta^{3}}e^{-d^{2}/(2\Delta)^{2}},
\end{equation}
although note that for typical values of $\Delta$, the approximation \cref{eq:loss_Taylor_expand} is an underestimate of the infidelity and is valid only for even smaller values of $\gamma$ than for \cref{eq:loss_approx_sigma_estimate}. We can now see that the error-corrected infidelity is \textit{greater} than the trivial encoding infidelity for \textit{very} small $\gamma$ where the constant term from $\Delta$ dominates, but is less than the trivial encoding for slightly larger $\gamma$ due to the exponential-in-$\Delta$ suppression of the linear term in $\gamma$. For much larger $\gamma$, the approximation \cref{eq:loss_Taylor_expand} becomes invalid, eventually leading to a second cross-over where the error-corrected infidelity is larger than the trivial infidelity.

It is straightforward to show that the composition of two channels $\mathcal{E}_{\tau_{2},\nu_{2}}\circ\mathcal{E}_{\tau_{1},\nu_{1}}=\mathcal{E}_{\tau,\nu}$ is given by $\tau=\tau_{1}\tau_{2}$, $\nu=\tau_{2}^{2}\nu_{1}+\nu_{2}$. For example, post-composing loss with a gain channel with $g=1/(1-\gamma)$ results in a Gaussian random displacement channel with $\nu=\sigma_{G}^{2}=\gamma/(1-\gamma)$~\cite{Noh18}. To first order in $\gamma$, we have $\nu\approx\gamma$ and therefore the noise associated with this channel is given by $\sigma^{2}\approx\Delta^{2}/2+\gamma$. Comparing this to \cref{eq:loss_approx_sigma_estimate}, we see that the linear contribution of $\gamma$ to the variance is doubled, but there is no longer an inverse dependence on $\Delta^{2}$. Converting loss into a Gaussian random displacement noise channel via this method is therefore advantageous whenever $\Delta^{2}<\gamma/4$ (approximately).

We can extend this analysis to find the optimal amount of gain $g$ given $\Delta,\gamma$ that minimizes the noise. The composition of loss and gain has parameters $\tau=\sqrt{g(1-\gamma)}$ and $\nu=\big(\gamma+(g-1)(\gamma+1)\big)/2$. In the low noise regime where $g-1$ and $\gamma$ are small, we have
\begin{equation}
\sigma^{2}\approx \frac{\Delta^{2}}{2}+\frac{g-1}{2}+\frac{\gamma}{2}+\frac{(g-1-\gamma)^{2}}{8\Delta^{2}}.
\end{equation}
We can find the optimal gain by minimizing $\sigma^{2}$, which leads exactly to
\begin{equation}\label{eq:optimal_gain}
g_{\text{opt}}=\frac{1-\gamma}{\big(e^{\Delta^{2}}-\gamma e^{-\Delta^{2}}\big)^{2}}\approx 1+\gamma-2\Delta^{2}.
\end{equation}
As $\Delta\rightarrow0$, we have $g_{\text{opt}}\rightarrow1/(1-\gamma)$, which is the amount of gain required to recover the Gaussian random-displacement channel. However, in the large $\Delta$ regime, we are restricted by the fact that $g$ must be greater than or equal to 1. Therefore, \cref{eq:optimal_gain} is only valid when $g_{\text{opt}}\geq1$, or, equivalently, whenever $\Delta^{2}\leq\gamma/2$ (approximately). If $\Delta^{2}>\gamma/2$, the optimal amount of gain to apply after a loss channel is 0; in words, the reduction in error rate obtained by (partially) canceling the phase-insensitive squeezing from the loss is \textit{not} enough to overcome the extra noise from gain.

\subsection{Dephasing}\label{subsec:dephasing}

So far, we have only considered Gaussian noise models, such as the envelope operator, loss, Gaussian random displacements, and gain. However, we also wish to model (single-mode) dephasing, which is a non-Gaussian noise channel defined by
\begin{equation}
    \mathcal{D}_{\sigma_{\text{d}}}(\rho)=\frac{1}{\sqrt{2\pi\sigma_{\text{d}}^{2}}}\int_{\mathbb{R}}\!d\phi\,e^{-\phi^{2}/2\sigma_{\text{d}}^{2}}e^{i\phi a^{\dag}a}\rho e^{-i\phi a^{\dag}a}.\tag{\ref{eq:dephasing}}
\end{equation}
Since dephasing is non-Gaussian, we cannot apply \cref{eq:entanglement_infid_sigma} directly. However, unitary rotations $e^{i\phi a^{\dag}a}$ \textit{are} Gaussian, so we will instead begin by analyzing the logical performance of unitary rotations before discussing the effects of dephasing.

For unitary rotations we consider the noise map $\mathcal{N}=\mathcal{E}_{\tau,\nu}\circ\mathcal{J}[e^{i\phi a^{\dag}a}e^{-\Delta^{2}a^{\dag}a}]$, where we note that $e^{i\phi a^{\dag}a}$ commutes with $\mathcal{E}_{\tau,\nu}$. Since the noise map is Gaussian, the diagonal elements of the characteristic function of $\mathcal{N}$ are again a mean-zero Gaussian function [\cref{eq:diagonal_characteristic_function_ansatz}], with variance given by
\begin{subequations}\label{eq:variance_rot}
\begin{align}
    \sigma^{2}&=\tau\tanh(\Delta^{2}/2)+\nu+\frac{(1-\tau)^{2}}{2\tanh(\Delta^{2})}+\frac{2\tau\sin^{2}(\phi/2)}{\sinh(\Delta^{2})}\label{eq:exact_variance_rot}\\
    &\approx\frac{\Delta^{2}}{2}+\nu+\frac{(1-\tau)^{2}+\phi^{2}}{2\Delta^{2}}.\label{eq:approx_variance_rot}
\end{align}
\end{subequations}
Compared to \cref{eq:variance_tau_nu}, there is only new term, which is inversely proportional to $\Delta^{2}$. Intuitively, this is because rotations affect more highly-squeezed (lower $\Delta$) GKP codestates since the states are more widely-spread in phase space, increasing the impact of rotations on the state. If we substitute $\mathcal{E}_{\tau,\nu}$ for loss, we obtain
\begin{equation}\label{eq:loss_and_rotation}
    \sigma^{2}\approx\frac{\Delta^{2}}{2}+\frac{\gamma}{2}+\frac{\gamma^{2}}{8\Delta^{2}}+\frac{\phi^{2}}{2\Delta^{2}}.
\end{equation}
Note that, to the lowest order, there are no terms in \cref{eq:loss_and_rotation} that are linear in $\phi$.

Unfortunately, we cannot analyze dephasing errors in terms of a single variance since the channel is non-Gaussian and therefore the diagonal elements of the characteristic function~\cref{eq:dephasing_diag_chi} are not Gaussian-distributed. However, we can use \cref{eq:dephasing} and the linearity of $F_{e}$ with respect to $\mathcal{E}$ to write the entanglement infidelity of the dephasing map as an integral of the entanglement infidelities of the rotation maps by an angle $\phi$. More precisely, we consider the noise map
\begin{subequations}
\begin{align}
    \mathcal{N}&=\mathcal{E}_{\tau,\nu}\circ\mathcal{D}_{\sigma_{\text{d}}}\circ\mathcal{J}[e^{-\Delta^{2}a^{\dag}a}]\\
    &=\frac{1}{\sqrt{2\pi\sigma_{\text{d}}^{2}}}\int_{\mathbb{R}}\!d\phi\,e^{-\phi^{2}/2\sigma_{\text{d}}^{2}}\mathcal{E}_{\tau,\nu}\circ\mathcal{J}[e^{i\phi a^{\dag}a}e^{-\Delta^{2}a^{\dag}a}],\label{eq:dephasing_decomposed}
\end{align}
\end{subequations}
where we note that $\mathcal{D}_{\sigma_{\text{d}}}$ commutes with $\mathcal{E}_{\tau,\nu}$. To recover the noise map discussed in the main text we simply set $\tau=1,\nu=0$ such that $\mathcal{E}_{\tau,\nu}$ is the identity map. From \cref{eq:dephasing_decomposed,eq:entanglement_infid_sigma} we can write the entanglement infidelity estimate as
\begin{equation}\label{eq:dephasing_infid_integral}
    1-F_{e}\approx\frac{a}{\sqrt{2\pi\sigma_{\text{d}}^{2}}}\int_{\mathbb{R}}\!d\phi\,e^{-\phi^{2}/2\sigma_{\text{d}}^{2}}\,\mathrm{erfc}\bigg(\frac{d}{2\sqrt{2}\sigma(\phi)}\bigg),
\end{equation}
where we've written $\sigma(\phi)$ in place of \cref{eq:exact_variance_rot}. Such an expression is valid since the entanglement fidelity $F_{e}$ is linear in the quantum channel it takes as input.

Next, we wish to obtain an approximate expression for \cref{eq:dephasing_infid_integral} in the regime where both $\sigma_{\text{d}}$ and $\sigma(\phi)$ are small. To do this, we use Laplace's method, which states that
\begin{equation}\label{eq:Laplaces_method}
    \int_{a}^{b}\!dx\,h(x)e^{Mg(x)}\approx\sqrt{\frac{2\pi}{M|g''(x_{0})|}}h(x_{0})e^{Mg(x_{0})}
\end{equation}
as $M\rightarrow\infty$, where $x_{0}\in(a,b)$ is the unique input that maximizes $g(x)$ over the interval $[a,b]$. Intuitively, \cref{eq:Laplaces_method} comes from approximating $h(x)e^{Mg(x)}$ as a Gaussian function $h(x_{0})e^{Mg(x_{0})-M|g''(x_{0})|(x-x_{0})^{2}\!/2}$ and integrating this over $\mathbb{R}$. To apply Laplace's method to \cref{eq:dephasing_infid_integral}, we first perform an asymptotic expansion of $\mathrm{erfc}(x)\approx e^{-x^{2}}\!/(x\sqrt{\pi})$, giving
\begin{equation}
    1-F_{e}\approx\frac{2a}{\pi d\sigma_{\text{d}}}\int_{\mathbb{R}}\!d\phi\,\sigma(\phi)e^{-\phi^{2}/2\sigma^{2}_{\text{d}}-d^{2}/8\sigma(\phi)^{2}}.
\end{equation}
To make contact with \cref{eq:Laplaces_method}, we write
\begin{align}
    g(\phi)&=-\frac{\phi^{2}}{2}-\frac{d^{2}\sigma_{\text{d}}^{2}}{8\sigma(\phi)^{2}},&
    h(\phi)&=\frac{2a}{\pi d\sigma_{\text{d}}}\sigma(\phi),
\end{align}
and $M=1/\sigma_{\text{d}}^{2}$. Next, to find the maximum of $g(\phi)$ we use the approximate expression for $\sigma(\phi)$ in \cref{eq:approx_variance_rot}. For convenience, we work in the regime where the low order expansion of $\sigma(\phi)$ is valid, and introduce the notation
\begin{align}
    \sigma(\phi)^{2}&=\sigma_{\text{g}}^{2}+\frac{\phi^{2}}{2\Delta^{2}},&\sigma_{\text{g}}^{2}=\frac{\Delta^{2}}{2}+\nu+\frac{(1-\tau)^{2}}{2\Delta^{2}},
\end{align}
where $\sigma_{\text{g}}^{2}$ is (approximately) the variance of the noise introduced by $\mathcal{E}_{\tau,\nu}\circ\mathcal{J}[e^{-\Delta^{2}a^{\dag}a}]$ as in \cref{eq:approx_variance_tau_nu}. Note that in the main text, we only consider the case where $\tau=1,\nu=0$ (i.e.~noise is only coming from the envelope operator and dephasing), in which case we have $\sigma_{\text{g}}=\Delta/\sqrt{2}$.

With this notation, the turning points of $g(\phi)$ occur at
\begin{align}\label{eq:g_turning_points}
    \phi&=0,&\phi&=\pm\frac{\sqrt{d\Delta}}{\sqrt[4]{2}}\sqrt{\sigma^{}_{\text{d}}-\sigma_{\text{d}}^{*}},&\sigma_{\text{d}}^{*}&=\frac{2\sqrt{2}\Delta\sigma_{\text{g}}^{2}}{d},
\end{align}
where the latter two solutions in \cref{eq:g_turning_points} only appear if the argument of the square root is positive, i.e.~$\sigma^{}_{\text{d}}>\sigma_{\text{d}}^{*}$. We call $\sigma_{\text{d}}^{*}$ the ``critical'' dephasing; when $\tau=1,\nu=0$, we have $\sigma_{\text{d}}^{*}=\sqrt{2}\Delta^{3}/d$, as discussed in the main text in \cref{eq:critical_dephasing}. 

At this point it is important to separate our analysis into three cases:
\begin{enumerate}
    \item the ``subcritical'' case when $\sigma^{}_{\text{d}}<\sigma_{\text{d}}^{*}$ and $g(\phi)$ has a single turning point $\phi=0$,
    \item the ``supercritical'' case when $\sigma^{}_{\text{d}}>\sigma_{\text{d}}^{*}$ and $g(\phi)$ has three turning points, and
    \item the ``critical'' case when $\sigma^{}_{\text{d}}=\sigma_{\text{d}}^{*}$ and $g(\phi)$ has a vanishing second derivative at $\phi=0$.
\end{enumerate}
From the second derivative of $g(\phi)$ it is easy to show that in the subcritical case $\phi=0$ is a global maximum, in the supercritical case $\phi=0$ is a local minimum and the remaining two stationary points both achieve the global maximum (since $g$ is even), and in the critical case $\phi=0$ is a global maximum with $g''(0)=0$.
%Intuitively, in the subcritical and critical cases, the largest contributor to the infidelity comes from small angles around $\phi=0$, so Laplace's approximation treats the error function as a Gaussian function of $\phi$ centered around $\phi=0$. In the supercritical case, the noise from dephasing becomes significant enough that the largest contributor to the error comes at the non-zero angles given in \cref{eq:g_turning_points}.

We begin by applying Laplace's method to the subcritical case. Here, the global maximum is $\phi_{0}=0$, so we can directly apply \cref{eq:Laplaces_method} to obtain
\begin{equation}\label{eq:subcritical_infid}
    1-F_{e}\approx \frac{8a\sigma_{\text{g}}^{3}\Delta}{d^{2}\sqrt{\pi}\sqrt{\sigma_{\text{d}}^{*2}-\sigma_{\text{d}}^{2}}}\exp\bigg({-}\frac{d^{2}}{8\sigma_{\text{g}}^{2}}\bigg).
\end{equation}
Note that the denominator of \cref{eq:subcritical_infid} diverges at $\sigma^{}_{\text{d}}=\sigma_{\text{d}}^{*}$, so this approximation is only valid when $\sigma_{\text{d}}$ is sufficiently lower than $\sigma_{\text{d}}^{*}$.

In the supercritical case, since we do not have a unique maximum of $g(\phi)$ over $\phi\in\mathbb{R}$, we first use the fact that $g$ is even to write
\begin{equation}
    1-F_{e}\approx 2\int_{0}^{\infty}\!dx\,h(x)e^{Mg(x)}.
\end{equation}
Now, $g(\phi)$ has a unique maximum in $[0,\infty)$, so we can apply Laplace's method, giving
\begin{equation}\label{eq:supercritical_infid}
    1-F_{e}\approx\frac{2^{3/4}a\sigma_{\text{d}}}{\sqrt{\pi d\Delta}\sqrt{\sigma^{}_{\text{d}}-\sigma_{\text{d}}^{*}}}\exp\bigg(\!{-}\frac{\Delta d}{\sqrt{2}\sigma_{\text{d}}}\Big(1-\frac{\sigma_{\text{d}}^{*}}{2\sigma_{\text{d}}}\Big)\!\bigg).
\end{equation}
Again, note that the denominator diverges at $\sigma^{}_{\text{d}}=\sigma_{\text{d}}^{*}$, so this approximation is valid only when $\sigma_{\text{d}}$ is sufficiently larger than $\sigma_{\text{d}}^{*}$.

In the critical case, we cannot apply Laplace's method since $g''(0)=0$. Instead, we make a fourth-order approximation of $g(\phi)$ and evaluate the integral using
\begin{equation}
    \int_{\mathbb{R}}\!dx\,e^{-ax^{4}}=\frac{2\Gamma(5/4)}{\sqrt[4]{a}}.
\end{equation}
This gives an expression for the infidelity
\begin{equation}\label{eq:critical_infid}
    1-F_{e}\approx\frac{2^{7/4}\Gamma(5/4)a\sqrt{\sigma_{\text{g}}}}{\pi\sqrt{d}}\mathrm{exp}\bigg(\!{-}\frac{d^{2}}{8\sigma_{\text{g}}^{2}}\!\bigg),
\end{equation}
which has no dependence on $\sigma_{\text{d}}$ since we have set it equal to $\sigma_{\text{d}}^{*}$. We note that unlike the sub- and super-critical cases, \cref{eq:critical_infid} is an underestimate of the infidelity, which arises from the fourth-order expansion of $g(\phi)$. We have already discussed the main qualitative features of \cref{eq:subcritical_infid,eq:supercritical_infid,eq:critical_infid} in the main text.

\begin{figure*}
\includegraphics{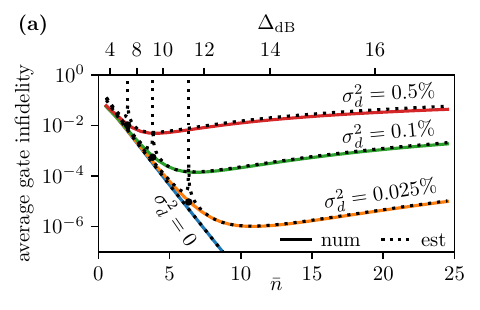}
\hspace*{\fill}
\includegraphics{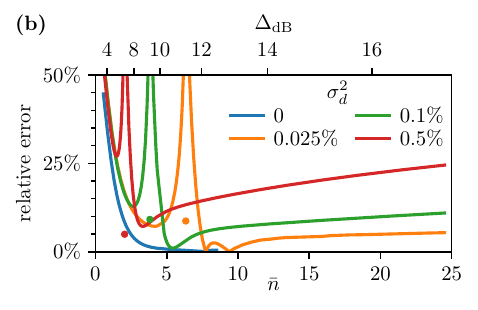}
\caption{Comparison between the theoretical estimated average gate fidelity from \cref{eq:subcritical_infid,eq:supercritical_infid,eq:critical_infid} and the numerically calculated average gate fidelity from Fig.~6(a) of Ref.~\cite{Shaw24-1}. (a) Estimated and numerical average gate infidelity $1-\bar{F}$ of the logical noise map corresponding to the envelope operator followed by dephasing for the single-mode square GKP code, with no additional Gaussian noise (i.e.~$\sigma_{\text{g}}=\Delta/\sqrt{2}$). The critical infidelity estimates are plotted as black circles. (b) Relative error $|1-\text{num}/\text{est}|$ between the estimated average gate infidelity (est) and the numerical average gate infidelity (num), for the same logical noise map. Values are only plotted when the numerical average gate infidelity is above $10^{-7}$ to avoid effects due to numerical inaccuracy. The critical infidelity estimates are plotted as independent data points.}
\label{fig:dephasing_error_comparison}
\end{figure*}

We compare the three approximate expressions in \cref{eq:subcritical_infid,eq:supercritical_infid,eq:critical_infid} to the numerically evaluated dephasing error rates given in Fig.~6(a) of Ref.~\cite{Shaw24-1} in \cref{fig:dephasing_error_comparison}, with $\sigma_{\text{g}}$ set to $\Delta/\sqrt{2}$, i.e.~the noise map consists only of the envelope operator and dephasing. The biggest difference in the performance of the dephasing noise estimates compared to the Clifford gate and loss estimates in \cref{fig:gate_error_estimate_comparison,fig:loss_error_comparison} is from the asymptotic behavior of the sub- and supercritical estimates. Indeed, the approximate expressions are useless in a region around the critical dephasing rate. This is not a serious issue since one could in principle interpolate between the three estimates \cref{eq:subcritical_infid,eq:supercritical_infid,eq:critical_infid} in this region to avoid the asymptote. Notwithstanding this issue, our estimates perform reasonably well but notably worse than our other estimates: in most regimes of interest the relative error is below 25\% compared with the numerical values.

Finally, we calculate the optimal $\Delta$ in the case where $\sigma_{\text{d}}=\Delta/\sqrt{2}$. We can see from \cref{fig:dephasing_error_comparison}(a) that the optimal infidelity is achieved in the supercritical regime~\cref{eq:supercritical_infid}. We can approximately find the optimal $\Delta$ by minimizing the exponent, which leads to $\Delta_{\text{opt}}\approx\sqrt[3]{d\sigma_{\text{d}}}/\sqrt{2}$; or, equivalently, $\sigma_{\text{d}}^{}\approx2\sigma_{\text{d}}^{*}$. Calculating an analogous expression in the presence of a phase-covariant Gaussian channel $\mathcal{E}_{\tau,\nu}$ is less insightful since the minimum of the infidelity can occur in either the sub- or supercritical regimes depending on the values of $\tau$ and $\nu$; therefore it is easier to simply use a simple numerical optimization to find the minimum as stated in the main text.

\subsection{Kerr Nonlinearity}\label{subsec:Kerr}

Finally, we consider Kerr nonlinearity, which is a source of error in superconducting devices that may arise for example from the the coupling elements that we proposed using in the two-qubit gates in \cref{sec:gates}. We define a Kerr nonlinearity as the unitary operator
\begin{equation}\label{eq:Kerr_nonlinearity}
    e^{i\theta_{K}(a^{\dag}a)^{2}/2}.
\end{equation}
To analyze the effect of Kerr nonlinearity on approximate GKP codestates using the methods from this section, we make use of the identity
\begin{equation}
    e^{i\pi n^{2}/(2M)}=\frac{e^{i\pi/4}}{\sqrt{2M}}\sum_{j=0}^{2M-1}e^{-i\pi j^{2}/(2M)}e^{i\pi nj/M},
\end{equation}
which holds for any non-negative integers $M$ and $n$. With this, we can rewrite any Kerr nonlinearity with $\theta_{K}=\pi/M$ for some integer $M$ as a coherent sum of rotation operators:
\begin{subequations}\label{eq:Kerr_as_rotations}
\begin{align}
    e^{i\pi(a^{\dag}a)^{2}/(2M)}&=\sum_{n=0}^{\infty}e^{i\pi n^{2}/(2M)}\ket{n}\!\bra{n}\\
    &=\frac{e^{\frac{i\pi}{4}}}{\sqrt{2M}}\sum_{j=0}^{2M-1}e^{-i\pi j^{2}/(2M)}e^{i\pi nj/M}\ket{n}\!\bra{n}\\
    &=\frac{e^{\frac{i\pi}{4}}}{\sqrt{2M}}\sum_{j=0}^{2M-1}e^{-i\pi j^{2}/(2M)}e^{i\pi j (a^\dag a)/M}.
\end{align}
\end{subequations}
The sum \cref{eq:Kerr_as_rotations} is not particularly enlightening to theoretically analyze in the low $\theta_{K}$ limit, because the number of terms in the sum increases even though the unitary operator approaches the identity. Therefore, we briefly outline how one can use \cref{eq:Kerr_as_rotations} to numerically estimate the effect of Kerr nonlinearity on GKP codestates and present our results.

\begin{figure}
\centering
\includegraphics{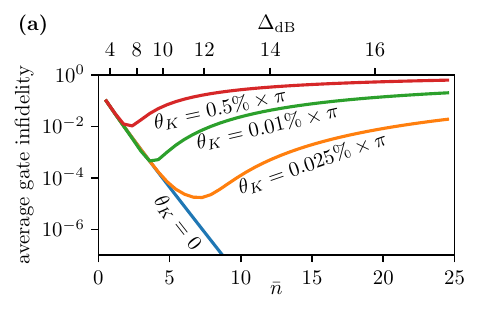}
\caption{Estimated theoretical average gate infidelity $1-\bar{F}$ of the logical noise map corresponding to the envelope operator followed by Kerr nonlinearity~\cref{eq:Kerr_nonlinearity} for the single-mode square GKP code.}
\label{fig:Kerr_nonlinearity}
\end{figure}

First, we write the characteristic function of the Kerr nonlinearity and the envelope operator $e^{i\pi(a^{\dag}a)^{2}/(2M)-\Delta^{2}a^{\dag}a}$ as
\begin{subequations}
\begin{align}
    c(\vect{v})&=\frac{e^{\frac{i\pi}{4}}}{2\pi\sqrt{2M}}\sum_{j=0}^{2M-1}f_{j}(\vect{v}),\\
    f_{j}(\mathbf{v})&=\frac{e^{-i\frac{\pi}{2M}j^{2}}}{1-e^{i\pi j/M-\Delta^{2}}}e^{\frac{1}{4}\mathrm{coth}\big(\frac{i\pi j}{2M}-\frac{\Delta^{2}}{2}\big)|\mathbf{v}|^{2}},
\end{align}
\end{subequations}
which comes from applying \cref{eq:Kerr_as_rotations} and using the characteristic function of the envelope operator \cref{eq:envelope_characteristic_function} extended to complex $\Delta$. Then, to calculate the average gate fidelity using \cref{eq:avg_gate_infidelity_ratio}, we need the diagonal elements of the characteristic function of the channel $\mathcal{J}[e^{i\pi(a^{\dag}a)^{2}/(2M)-\Delta^{2}a^{\dag}a}]$, given by
\begin{equation}
    c(\vect{v},\vect{v})=|c(\vect{v})|^{2}=\frac{1}{8\pi^{2} M}\sum_{j,k=0}^{2M-1}f_{j}(\vect{v})f_{k}(\vect{v})^{*}.
\end{equation}
Now, each term $f_{j}(\vect{v})f_{k}(\vect{v})^{*}$ is a Gaussian function of $\vect{v}$, which means that we can leverage the methods from \cref{subsec:gaussian_char} to approximate the contribution to the average gate fidelity from each term in the sum. This directly leads to our results which are shown in \cref{fig:Kerr_nonlinearity} and discussed in \cref{sec:error_estimates}.

\section{Non-ideal error correction after Clifford Gates}\label{sec:approx_Clifford_gates}

In this appendix, we extend the approximate analytical Clifford gate fidelity~\cref{eq:avg_fid_estimate} to incorporate the effects of approximate codestates in the QEC circuit. More precisely, in the $\Delta\rightarrow0$ limit, we wish to derive the expression for the average gate infidelity
\begin{equation}
    1-\bar{F}_{A,\mathcal{P},\Delta}\approx \frac{2^{n}a_{\text{eff}}}{2^{n}+1}\mathrm{erfc}\big(d_{\text{eff}}/(2\sqrt{2}\Delta)\big),\tag{\ref{eq:avg_fid_estimate_approx}}
\end{equation}
where $a_{\text{eff}}$ and $d_{\text{eff}}$ are the effective degeneracy and distance of a modified patch $\mathcal{P}'$. To do this, we will first use the twirling approximation to derive our noise model for quantum error correction with approximate codewords~\cref{eq:logical_channel_approx}. Then, we will derive \cref{eq:avg_fid_estimate_approx}, including a description of how to obtain the effective distance and degeneracy assuming the patch $\mathcal{P}'$ is chosen to be optimal. We conclude with a few interesting observations: first, to explain why in the hexagonal code one should perform a $\bar{H}$ gate before each round of QEC; second, a description of how the square $\bar{C}_{YY}$ gate is related to the $D_{4}$ lattice; and third, a brief commentary on how performing single-qubit gates after a square $\bar{C}_{ZZ}$ gate improves the effective distance.

\begin{figure}
    \includegraphics[]{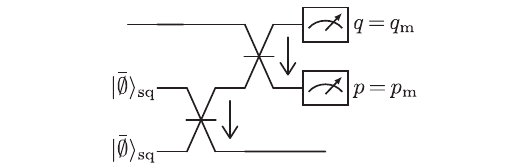}
    \caption{Square GKP passive teleportation error-correction circuit. The GKP qunaught state and the 50-50 beamsplitter gates used in the circuit are defined in \cref{eq:sq_qunaught,eq:bs} respectively. The arrows adjacent to the beamsplitter symbols are there simply to indicate which of the modes picks up the minus sign in \cref{eq:bs}.}\label{fig:tele_sq_circuit}
\end{figure}

We begin by describing our model for QEC with approximate codestates. We are motivated by the passive teleportation QEC circuit~\cite{Walshe20}, shown for the square GKP code in \cref{fig:tele_sq_circuit}. This circuit uses two ancillary modes each of which is initialized in the GKP qunaught state
\begin{equation}\label{eq:sq_qunaught}
    \ket{\bar{\emptyset}}_{\text{sq}}\propto\sum_{s\in\mathbb{Z}}\ket{\sqrt{2\pi}s}_{q}.
\end{equation}
Note that the qunaught state is the unique simultaneous $+1$-eigenstate of the operators $T(\sqrt{2\pi},0)$ and $T(0,\sqrt{2\pi})$. Then, two 50-50 beamsplitter gates are applied, which we define as
\begin{align}\label{eq:bs}
    U_{\text{bs}}&=e^{i\pi(p_{1}q_{2}-q_{1}p_{2})/4},&\mathcal{S}_{\text{bs}}&=\frac{1}{\sqrt{2}}\begin{bmatrix}1&-1&0&0\\1&1&0&0\\0&0&1&-1\\0&0&1&1\end{bmatrix}.
\end{align}
Finally, the first and second modes are read-out using a homodyne measurement of the first and second modes.

The overall effect of the circuit in \cref{fig:tele_sq_circuit} can be described by the measurement operators
\begin{equation}
    M(q_{\text{m}},p_{\text{m}})=\Pi_{\text{GKP,sq}}T\big(\sqrt{2}q_{\text{m}},\sqrt{2}p_{\text{m}}\big)^{\dag},
\end{equation}
where $\Pi_{\text{GKP,sq}}$ is the projector onto the square GKP codespace. Importantly, $\sqrt{2}q_{\text{m}}$ (respectively, $\sqrt{2}p_{\text{m}}$) reveals the position (momentum) eigenvalues of the input state modulo $\sqrt{\pi}$. Conveniently, the state is projected into an ideal GKP codestate automatically, but a logical operator may have been applied due to the displacement operator. Therefore, the vector $\sqrt{2}[q_{\text{m}},p_{\text{m}}]^{T}$ must be decoded using a patch $\mathcal{P}$ to determine the most likely Pauli error that was applied. Subsequently, the logical correction can either be applied physically or in software by updating the Pauli frame of the state.

\begin{figure*}
    \includegraphics[]{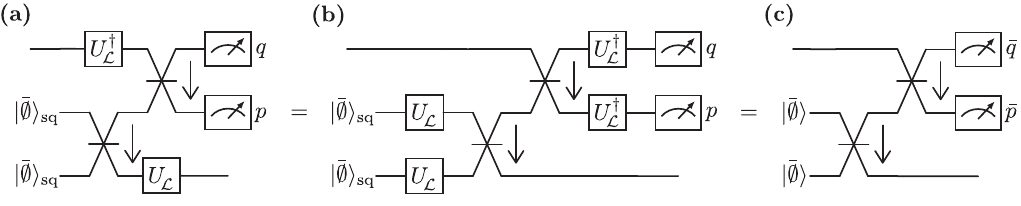}
    \caption{Teleportation QEC Circuit for a general GKP code with logical lattice $\mathcal{L}$. (a) A general GKP code can be decoded by unitarily transforming to a square GKP code via $U_{\mathcal{L}}^{\dag}$, then performing the square teleportation QEC circuit \cref{fig:tele_sq_circuit}, and then transforming back to the general code with $U_{\mathcal{L}}$. (b) and (c), rearrangements of the circuit using the identities described in the main text.}\label{fig:tele_general_circuit}
\end{figure*}

It is straightforward to generalize the square error correction circuit to a general circuit by introducing the Gaussian unitary operator $U_{\mathcal{L}}$, which transforms a square GKP codestate into a general one with lattice $\mathcal{L}$. In particular, $U_{\mathcal{L}}$ is the unitary operator corresponding to the symplectic matrix $M_{\mathcal{L}}=[\vect{\alpha},\vect{\beta}]/\sqrt{\pi}$. A general GKP code can then be decoded by first transforming it to a square GKP code with $U_{\mathcal{L}}^{\dag}$, then performing the circuit in \cref{fig:tele_sq_circuit}, and then transforming back to the general GKP code with $U_{\mathcal{L}}$.

In \cref{fig:tele_general_circuit}, we show how the general GKP error-correction circuit can be rewritten solely in terms of passive linear elements (i.e.~beamsplitters and single-mode rotations). In particular, we use the fact that for any Gaussian unitary $U$, we have
\begin{equation}
    (U\otimes U)U_{\text{bs}}=U_{\text{bs}}(U\otimes U),
\end{equation}
which can be shown easily using the symplectic matrix $\mathcal{S}_{\text{bs}}$ in \cref{eq:bs}. The result is the circuit in \cref{fig:tele_general_circuit}(c), which uses general GKP qunaught states, 50-50 beamsplitters, and logical quadrature measurements. In particular, the general GKP qunaught state is defined as $\ket{\bar{\emptyset}}=U_{\mathcal{L}}\ket{\bar{\emptyset}}_{\text{sq}}$ and is the unique simultaneous $+1$-eigenstate of the operators $T(\sqrt{2}\vect{\alpha})$ and $T(\sqrt{2}\vect{\beta})$.

\begin{figure*}
    \includegraphics[]{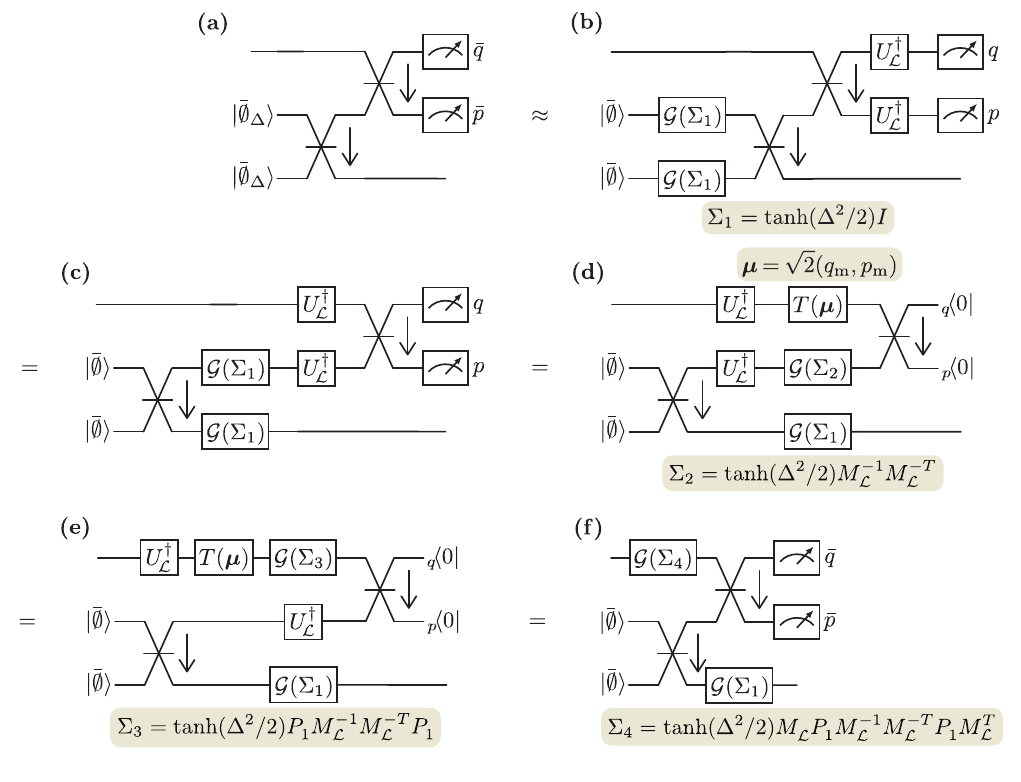}
    \caption{Teleportation QEC circuit with approximate codewords. (a) The ideal teleportation QEC circuit \cref{fig:tele_general_circuit}(c) with the ideal codestates replaced with approximate ones. As described in the main text, we can apply the twirling approximation and use circuit identities to rewrite circuit (a) as circuit (f), which consists of an ideal teleportation QEC circuit preceded and followed by Gaussian random displacement channels.}\label{fig:tele_approx_circuit}
\end{figure*}

Now that we have an ideal circuit that describes the error correction cycle in \cref{fig:tele_general_circuit}(c), we can derive the model of approximate error-correction~\cref{eq:logical_channel_approx} by replacing the ideal qunaught states with approximate ones. In \cref{fig:tele_approx_circuit} we show how this can be approximated as an ideal circuit preceded and followed by a Gaussian random displacement (GRD) channel, defined as
\begin{multline}
    \mathcal{G}(\Sigma)(\rho)\\
    =\frac{1}{\sqrt{\det\Sigma}}\!\iint\! \frac{d^{2}\vect{v}}{(2\pi)^{n}}e^{-\vect{v}^{T}\Sigma^{-1}\vect{v}/2}\,T(\vect{v})\rho T(\vect{v})^{\dag}\tag{\ref{eq:GRD}}
\end{multline}
for a covariance matrix $\Sigma$ that we now derive.

We derive the following similar methods to Ref.~\cite{Rozpedek23}. First going from \cref{fig:tele_approx_circuit}(a) to (b), we apply the twirling approximation to the envelope operator $e^{-\Delta^{2}a^{\dag}a}$ by taking only the diagonal elements of its characteristic function, as done in~\cref{subsec:envelope}. More precisely, we replace the envelope operator on each mode with the GRD channel $\mathcal{G}(\tanh(\Delta^{2}/2)I)$. One particularly useful property of Gaussian random displacement channels is that, for a Gaussian unitary $U$ with symplectic representation $\mathcal{S}(U)$, we have
\begin{equation}\label{eq:GRD_Gaussian_CR}
    \mathcal{J}[U]\circ\mathcal{G}(\Sigma)\circ\mathcal{J}[U^{\dag}]=\mathcal{G}\big(\mathcal{S}(U)\Sigma\mathcal{S}(U)^{T}\big).
\end{equation}
We can use \cref{eq:GRD_Gaussian_CR} to proceed from \cref{fig:tele_approx_circuit}(b) to (d), first applying it to the beamsplitter \cref{eq:bs} and then to the general GKP encoding Gaussian unitary $U_{\mathcal{L}}$. Note that in \cref{fig:tele_approx_circuit} we have also rewritten the $q$ and $p$ measurements as a displacement $T(\vect{\mu})$ (that depends on the measurement outcome) followed by a projection onto the $0$-position and momentum eigenstates. While such a rearrangement does not make physical sense, mathematically it is convenient since the resulting two-mode projection is a projection onto the maximally entangled state
\begin{equation}
    \ket{ME}=\sum_{n\in\mathbb{Z}}\ket{n}\ket{n},
\end{equation}
written in the Fock basis. This maximally entangled state has the property that for any bosonic operator $O$, $O\otimes I\ket{ME}=I\otimes O^{T}\ket{ME}$, where the transpose is with respect to the Fock space representation of $O$. In particular, we have $q^{T}=q$, $p^{T}=-p$, and $T(v_{1},v_{2})^{T}=T(-v_{1},v_{2})$. As such, it is straightforward to show from \cref{eq:GRD} that
\begin{equation}\label{eq:GRD_ME}
    \big(\mathcal{I}\otimes \mathcal{G}(\Sigma)\big)(\ket{ME}\!\bra{ME})=\big(\mathcal{G}(P_{1}\Sigma P_{1})\otimes\mathcal{I}\big)(\ket{ME}\!\bra{ME}),
\end{equation}
where $P_{1}=\mathrm{diag}(-1,1)$ is a reflection in the $y$-axis, and $\mathcal{I}=\mathcal{J}[I]$ is the identity superoperator. Applying \cref{eq:GRD_ME} leads us to \cref{fig:tele_approx_circuit}(e), from which panel (f) follows by commuting the Gaussian random displacement channel through the displacement and remaining Gaussian unitary $U_{\mathcal{L}}^{\dag}$ via \cref{eq:GRD_Gaussian_CR}.

The result is that the approximate error-correction circuit \cref{fig:tele_approx_circuit}(a) can be written approximately as an ideal error-correction circuit preceded by a Gaussian random displacement channel with covariance matrix
\begin{equation}
    \Sigma_{\text{QEC}}=\tanh(\Delta^{2}/2)M^{}_{\mathcal{L}}P^{}_{1}M^{-1}_{\mathcal{L}}M^{-T}_{\mathcal{L}}\!P^{}_{1}M^{T}_{\mathcal{L}},\tag{\ref{eq:Sigma_QEC}}
\end{equation}
and followed by a Gaussian random displacement channel with covariance $\mathrm{tanh}(\Delta^{2}/2)I$. Note that the form of $\Sigma_{\text{QEC}}$ is specific to the teleportation-based model of QEC that we consider, and would be different for Steane error-correction circuits.

In rectangular GKP codes, $\Sigma_{\text{QEC}}=\tanh(\Delta^{2}/2)I$ since the matrix $M_{\mathcal{L}}$ is diagonal and commutes with $P_{1}$. However, in general these matrices do not commute, leading to an asymmetric distribution of errors due to error correction. This is a problem for the hexagonal GKP code, which has three equal-length stabilizer generators, only two of which are measured. As a result, $\Sigma_{\text{QEC}}$ is asymmetric and therefore does not reflect the hexagonal symmetry of the code. We will show how to solve this problem using logical gates soon.

Since our model of approximate QEC is now an ideal QEC round preceded and followed by a Gaussian random displacement channel, we can incorporate approximate QEC into our existing logical channel definitions with the addition of the channel $\mathcal{G}(\Sigma_{\text{QEC}})$. In particular, we have
\begin{equation}
    \mathcal{E}_{U,\mathcal{P},\Delta}=\mathcal{J}[R^{\dag}]\circ\mathcal{C}_{\mathcal{P}}\circ\mathcal{G}(\Sigma_{\text{QEC}})\circ\mathcal{J}[\bar{U}R_{\Delta}],\tag{\ref{eq:logical_channel_approx}}
\end{equation}
as described in the main text. Note that since we are not interested in the state of the system after the round of error correction, we discard the second Gaussian random displacement channel in \cref{fig:tele_approx_circuit}(f).

Now, we wish to approximate the average gate fidelity $\bar{F}_{A,\mathcal{P},\Delta}$ for a Clifford gate $A$. Ultimately, we wish to derive \cref{eq:avg_fid_estimate_approx}, and explain how to obtain the effective $d_{\text{eff}}$. Our strategy will be to rewrite this fidelity in the language of \cref{sec:error_estimates,sec:gate_error_estimate} by considering the average gate fidelity $\bar{F}(\mathcal{N}_{\mathcal{L}},I)$ of a logical noise channel
\begin{equation}
    \mathcal{N}_{\mathcal{L}}=\mathcal{J}[R^{\dag}]\circ\mathcal{C}_{\mathcal{P}}\circ\mathcal{N}\circ\mathcal{J}[R].\tag{\ref{eq:logical_noise_channel}}
\end{equation}
To do this, first recall the definition of the average gate fidelity
\begin{equation}
    \bar{F}(\mathcal{E},U)=\int d\psi\bra{\psi}U^\dag \mathcal{E}(\ket{\psi}\!\bra{\psi})U\ket{\psi}.\tag{\ref{eq:avg_gate_fid_defn}}
\end{equation}
By performing a change of variables $\ket{\phi}=U\ket{\psi}$ we see that $\bar{F}(\mathcal{E},U)=\bar{F}(\mathcal{E}\circ\mathcal{J}[U^{\dag}],I)$. Applying this to \cref{eq:logical_channel_approx} gives
\begin{subequations}
    \begin{align}
        \bar{F}_{A,\mathcal{P},\Delta}&=\bar{F}\big(\mathcal{J}[R^{\dag}]\circ\mathcal{C}_{\mathcal{P}}\circ\mathcal{G}(\Sigma_{\text{QEC}})\circ\mathcal{J}[\bar{A}R_{\Delta}A^{\dag}],I\big)\label{eq:approx_Clifford_step1}\\
        &\propto\bar{F}\big(\mathcal{J}[R^{\dag}]\circ\mathcal{C}_{\mathcal{P}}\circ\mathcal{G}(\Sigma_{\text{QEC}})\nonumber\\
        &\qquad\qquad\qquad\quad\circ\mathcal{J}[\bar{A}e^{-\Delta^{2}a^{\dag}a}\bar{A}^{\dag}R],I\big),\label{eq:approx_Clifford_step2}
    \end{align}
\end{subequations}
where to go from \cref{eq:approx_Clifford_step1} to \cref{eq:approx_Clifford_step2} we have used $R_{\Delta}\propto e^{-\Delta^{2}a^{\dag}a}R$,\footnote{Here we briefly note that in the main text and for our numerics, this equation is not strictly true. This is because we orthogonalize the codestates after applying the envelope operator. However, in the context of approximate analytical formulae, these issues are negligible since we are in the regime of small $\Delta$.} along with \cref{eq:Abar_R_commute}. Therefore, we have
\begin{equation}
    \bar{F}_{A,\mathcal{P},\Delta}=\bar{F}(\mathcal{N}_{\mathcal{L}},I)
\end{equation}
with the logical noise channel $\mathcal{N}_{\mathcal{L}}$ given by \cref{eq:logical_noise_channel} and
\begin{equation}\label{eq:approx_effective_N}
    \mathcal{N}=\mathcal{G}(\Sigma_{\text{QEC}})\circ\mathcal{J}[\bar{A}e^{-\Delta^{2}a^{\dag}a}\bar{A}^{\dag}].
\end{equation}

Under the assumption that $\Delta$ is small, we can apply the twirling approximation to the envelope operator in \cref{eq:approx_effective_N}, allowing us to write $\mathcal{N}$ approximately as a Gaussian random displacement channel:
\begin{subequations}
    \begin{align}
        \mathcal{N}&=\mathcal{G}(\Sigma_{\text{QEC}})\circ\mathcal{J}[\bar{A}e^{-\Delta^{2}a^{\dag}a}\bar{A}^{\dag}]\\
        &\approx\mathcal{G}(\Sigma_{\text{QEC}})\circ\mathcal{J}[\bar{A}]\circ\mathcal{G}\big(\!\tanh(\Delta^{2}/2)\big)\circ\mathcal{J}[\bar{A}^{\dag}]\\
        &=\mathcal{G}(\Sigma_{\text{QEC}})\circ\mathcal{G}\big(\!\tanh(\Delta^{2}/2)\mathcal{S}(\bar{A})\mathcal{S}(\bar{A})^{-1}\big)\\
        &=\mathcal{G}(\Sigma),
    \end{align}
\end{subequations}
with
\begin{equation}\label{eq:Sigma_approx_QEC}
    \Sigma=\tanh(\Delta^{2}/2)\big(M^{\vphantom{-1}}_{\mathcal{L}}P^{\vphantom{-1}}_{1}M_{\mathcal{L}}^{-1}M_{\mathcal{L}}^{-T}P^{\vphantom{-1}}_{1}M_{\mathcal{L}}^{T}+\mathcal{S}(\bar{A})\mathcal{S}(\bar{A})^{T}\big).
\end{equation}
Intuitively, \cref{eq:Sigma_approx_QEC} contains contributions both from errors that occur before the logical gate (second term), and from errors that occur after the logical gate due to approximate error correction (first term).

Now, we can apply the methods of \cref{sec:gate_error_estimate} to approximate the average gate fidelity $\bar{F}(\mathcal{N}_{\mathcal{L}},I)$. Specifically, since $\mathcal{N}$ is a Gaussian random displacement channel, its characteristic function is simply
\begin{equation}
    c(\vect{u},\vect{v})=\frac{e^{-\vect{v}^{T}\Sigma^{-1}\vect{v}/2}}{(2\pi)^{n}\sqrt{\mathrm{det}\Sigma}}\delta^{2n}(\vect{u}-\vect{v}).
\end{equation}
Therefore, applying \cref{eq:entanglement_infidelity_ratio} gives the entanglement infidelity as
\begin{equation}\label{eq:ent_infid_Sigma_int}
    1-F_{e}\approx\frac{1}{\sqrt{\mathrm{det}\Sigma}}\int_{\mathbb{R}^{2n}\setminus\mathcal{P}}\!\frac{d^{2n}\vect{v}}{(2\pi)^{n}}e^{-\vect{v}^{T}\Sigma^{-1}\vect{v}/2}.
\end{equation}
At this point, we define $\Sigma=2\tanh(\Delta^{2}/2)\Sigma_{0}$ and $\sigma^{2}=2\tanh(\Delta^{2}/2)\approx\Delta^{2}$. This definition is chosen such that $\Sigma_{0}=I$ for the identity gate in the square GKP code. Now, $\Sigma_{0}$ is a positive definite matrix, so we can always write $\Sigma_{0}=BB^{T}$ for some matrix $B$. Performing the change of variables $\vect{v}\mapsto B\vect{v}$ gives
\begin{subequations}
\begin{align}
    1-F_{e}&\approx\sigma^{-2n}\!\int_{\mathbb{R}^{2n}\setminus B^{-1}\mathcal{P}}\!\frac{d^{2n}\vect{v}}{(2\pi)^{n}}e^{-|\vect{v}|^{2}/2\sigma^{2}}\\
    &<\frac{1}{2}\sum_{i}\mathrm{erfc}\big(d_{i}/(2\sqrt{2}\sigma)\big)\label{eq:ent_fid_estimate_relevant}\\
    &\approx a_{\text{eff}}\,\mathrm{erfc}\big(d_{\text{eff}}/(2\sqrt{2}\sigma)\big),\label{eq:ent_fid_estimate_approx}
\end{align}
\end{subequations}
where $\{d_{i}\}$ is the set of local distances as described in \cref{subsec:gaussian_char}, while $d_{\text{eff}}$ and $a_{\text{eff}}$ are the distance and degeneracy of the deformed patch $B^{-1}\mathcal{P}$ (respectively). From now on we will refer to these as the ``effective'' distance and degeneracy, which reflect how the distance and degeneracy of the patch are altered due to approximate error-correction. Note that the deformed patch $B^{-1}\mathcal{P}$ is a primitive cell of the deformed lattice $B^{-1}\mathcal{L}$, which is in general not equal to $\mathcal{L}$. To write \cref{eq:ent_fid_estimate_approx} in the form of \cref{eq:avg_fid_estimate_approx} we simply apply \cref{eq:Nielsen_simple} to write the entanglement fidelity as an average gate fidelity, and use $\sigma=\sqrt{2\tanh(\Delta^{2}/2)}\approx\Delta$.

In the case where $\mathcal{P}$ is optimally chosen, it is straightforward to calculate the effective distance. In particular, the optimal choice of $\mathcal{P}$ is such that $B^{-1}\mathcal{P}$ is the Voronoi cell $\mathcal{V}_{B^{-1}\mathcal{L}}$ of the deformed lattice $B^{-1}\mathcal{L}$. In this case, the effective distance is simply the length of the shortest vector in $B^{-1}\mathcal{L}$, while the effective degeneracy is half the number of such shortest vectors in the lattice. This can also be done by calculating the shortest vector in the original lattice $\mathcal{L}$ under the metric $B^{-T}\!B^{-1}=\Sigma_{0}^{-1}$, which avoids needing to explicitly calculate the matrix $B$.

We summarize the distance and degeneracies of various patches in the ideal QEC case in \cref{tab:estimates_full} and in the approximate QEC case in \cref{tab:estimates2_full}, which are more detailed versions of \cref{tab:estimates,tab:estimates2} given in the main text. In particular, note that in \cref{tab:estimates_full}, the distances and degeneracies represent the performance of the na\"ive patch, and the distance and degeneracy can be improved to be that of the identity gate by modifying the patch according to \cref{eq:modified_patch_ideal}. In contrast, in \cref{tab:estimates2_full}, the patch has already been modified and therefore the effective distances and degeneracies represent the optimal performance of the code against the gate. For more discussion, see the main text.

It is worth briefly reiterating the discussion at the end of \cref{subsec:gaussian_char}, in particular, that sometimes \cref{eq:ent_fid_estimate_approx} is not a good approximation of \cref{eq:ent_fid_estimate_relevant} for too large $\sigma$. This occurs when the distance $d_{\text{eff}}$ is not much shorter than the next shortest local distance in the set $\{d_{i}\}$. Therefore when giving our rough estimates here and in the main text, we use \cref{eq:ent_fid_estimate_approx}, where the set $\{d_{i}\}$ can be found by calculating the set of \textit{Voronoi-relevant} vectors of the lattice with the algorithms given in Ref.~\cite{Agrell02}.

\begin{figure}
    \includegraphics{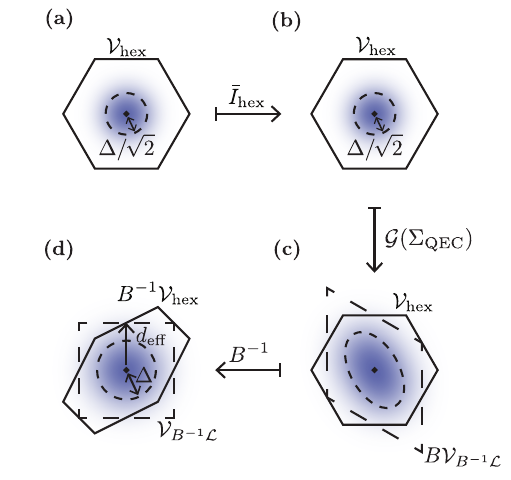}
    \caption{Visual representation of the optimal patch and effective distance of the $\bar{I}_{\text{hex}}$ gate when approximate error-correction is considered. (a) An approximate GKP state, represented (using the twirling approximation) as an ideal state that has incurred Gaussian random displacement error with standard deviation ${\sim} \Delta/\sqrt{2}$. (b) is the same as (a) since $\bar{I}_{\text{hex}}$ does nothing. (c) $\mathcal{G}(\Sigma_{\text{QEC}})$ is applied to incorporate the effects of approximate error correction. The optimal correction patch (dashed) can be found by performing a (non-orthogonal) transformation $B^{-1}$ that reshapes the Gaussian noise to be symmetrical with standard deviation ${\sim} \Delta$. (d) The optimal patch is the Voronoi cell of the transformed lattice, $\mathcal{V}_{B^{-1}\mathcal{L}}$.}\label{fig:hex_approx_transformation}
\end{figure}

\begin{figure}
    \includegraphics{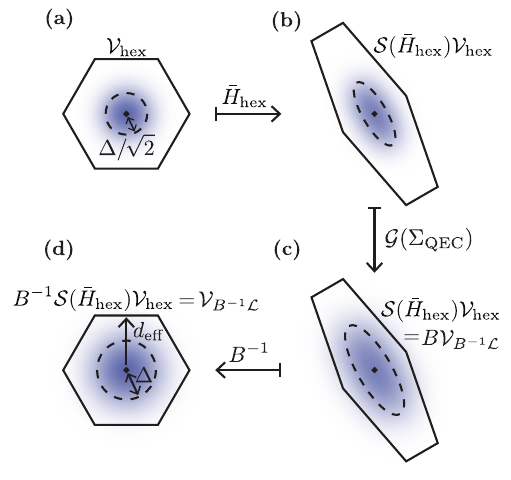}
    \caption{Visual representation of the optimal patch and effective distance of the $\bar{H}_{\text{hex}}$ gate when approximate error-correction is considered. The panels (a) to (d) represent the same information as in \cref{fig:phase_approx_transformation,fig:hex_approx_transformation}. In this case, the noise added by $\mathcal{G}(\Sigma_{\text{QEC}})$ has the same shape as the noise after it is spread by $\bar{H}_{\text{hex}}$, and that therefore, $B$ can be chosen to be $\mathcal{S}(\bar{H}_{\text{hex}})$. As a result the transformed lattice $B^{-1}\mathcal{L}$ in (d) is equal to the original hexagonal lattice $\mathcal{L}$.}\label{fig:hex_had_approx_transformation}
\end{figure}

Next, we explain why performing a logical Hadamard gate $\bar{H}$ outperforms the identity gate $\bar{I}$ in the hexagonal GKP code. Intuitively, this is because the asymmetric spreading of the noise due to error correction in \cref{eq:Sigma_QEC} is canceled out by the action of $\bar{H}$. In fact, this $\bar{H}$ cancellation occurs in \textit{any} GKP code for which $|\vect{\alpha}|=|\vect{\beta}|$ as we now show. When $\bar{A}=\bar{H}$, we have
\begin{subequations}
\begin{align}
    \Sigma_{0}&=\big(M^{\vphantom{-1}}_{\mathcal{L}}P^{\vphantom{-1}}_{1}M_{\mathcal{L}}^{-1}M_{\mathcal{L}}^{-T}P^{\vphantom{-1}}_{1}M_{\mathcal{L}}^{T}+\mathcal{S}(\bar{H})\mathcal{S}(\bar{H})^{T}\big)/2\\
    &=M^{\vphantom{-1}}_{\mathcal{L}}\big(P^{\vphantom{-1}}_{1}M_{\mathcal{L}}^{-1}M_{\mathcal{L}}^{-T}P^{\vphantom{-1}}_{1}\nonumber\\
    &\qquad+\mathcal{S}(\bar{H}_{\text{sq}})M_{\mathcal{L}}^{-1}M_{\mathcal{L}}^{-T}\mathcal{S}(\bar{H}_{\text{sq}})^{T}\big)M_{\mathcal{L}}^{T}/2
\end{align}
\end{subequations}
with $\mathcal{S}(\bar{H}_{\text{sq}})$ given in \cref{eq:symplectic_sq_Hadamard}. Using $M_{\mathcal{L}}=[\vect{\alpha},\vect{\beta}]/\sqrt{\pi}$ we can directly calculate
\begin{subequations}
    \begin{align}
        P^{\vphantom{-1}}_{1}M_{\mathcal{L}}^{-1}M_{\mathcal{L}}^{-T}P^{\vphantom{-1}}_{1}&=\frac{1}{\pi}\begin{bmatrix}|\vect{\beta}|^{2}&\vect{\alpha}\cdot\vect{\beta}\\\vect{\alpha}\cdot\vect{\beta}&|\vect{\alpha}|^{2}\end{bmatrix},\label{eq:P1_matrices}\\
        \mathcal{S}(\bar{H}_{\text{sq}})M_{\mathcal{L}}^{-1}M_{\mathcal{L}}^{-T}\mathcal{S}(\bar{H}_{\text{sq}})^{T}&=\frac{1}{\pi}\begin{bmatrix}|\vect{\alpha}|^{2}&\vect{\alpha}\cdot\vect{\beta}\\\vect{\alpha}\cdot\vect{\beta}&|\vect{\beta}|^{2}\end{bmatrix},\label{eq:had_matrices}
    \end{align}
\end{subequations}
and therefore \cref{eq:P1_matrices,eq:had_matrices} are equal if $|\vect{\alpha}|=|\vect{\beta}|$. In this case, we therefore have
\begin{align}
    \Sigma_{0}&=\mathcal{S}(\bar{H})\mathcal{S}(\bar{H})^{T},&B&=\mathcal{S}(\bar{H}).
\end{align}
Importantly, the transformation matrix $B$ corresponds to an \textit{ideal} Clifford gate $\mathcal{S}(\bar{H})$, so the transformation leaves the lattice invariant, $B^{-1}\mathcal{L}=\mathcal{L}$. After the transformation, the noise is symmetrically distributed with variance $\sigma=\sqrt{2\tanh(\Delta^{2}/2)}$, and the effective distance $d_{\text{eff}}$ and degeneracy $a_{\text{eff}}$ are the same as those of the original lattice $\mathcal{L}$. In the hexagonal code the $\bar{H}$ gate therefore achieves the optimal performance, since the hexagonal lattice is the densest sphere-packing in 2D (see \cref{fig:hex_approx_transformation,fig:hex_had_approx_transformation}).

Intriguingly, the rotational symmetry of the hexagonal code means that the logical Hadamard $\bar{H}$ gate is not the only gate with ideal performance. To be precise, the rotations $\bar{H}^{\vphantom{\dag}}_{\text{hex}}\bar{S}_{\text{hex}}^{\dag}=e^{i\pi a^{\dag}a/3}$ and $\bar{S}^{\vphantom{\dag}}_{\text{hex}}\bar{H}^{\dag}_{\text{hex}}=e^{-i\pi a^{\dag}a/3}$, when acting on codestates with a symmetric noise distribution, cannot spread any errors. As such, the gates $\bar{H}^{\vphantom{\dag}}_{\text{hex}}e^{i\pi a^{\dag}a/3}=\bar{S}_{\text{hex}}^{\dag}$ and $\bar{H}^{\vphantom{\dag}}_{\text{hex}}e^{-i\pi a^{\dag}a/3}=\bar{H}^{\vphantom{\dag}}_{\text{hex}}\bar{S}^{\vphantom{\dag}}_{\text{hex}}\bar{H}_{\text{hex}}^{\dag}$ also achieve the optimal average gate fidelity, alongside the logical $\bar{H}_{\text{hex}}$ gate. This is \textit{not} true for, for example, the $\bar{S}_{\text{hex}}$ gate, as shown in \cref{tab:estimates2_full}. Nevertheless, the gates $S^{\dag}$ and $H$ generate the single-qubit Clifford group, and it is straightforward to show that any single-qubit Clifford gate can be implemented up to Pauli gates in no more than two rounds of QEC, without compromising the performance of the hexagonal GKP code.

Next, we comment on the surprising observation that the effective distance of the $\bar{C}_{YY}$ gate is the same as that of the identity gate $\bar{I}$ for the square GKP code. First, we calculate the matrix $\Sigma_{0}$, which in this case is simply
\begin{equation}
    \Sigma_{0}=\big(I+\mathcal{S}(\bar{C}_{YY})\mathcal{S}(\bar{C}_{YY})^{T}\big)/2=\begin{bmatrix}2&1&1&0\\1&2&0&1\\1&0&2&-1\\0&1&-1&2\end{bmatrix}.
\end{equation}
We are interested in the properties of the lattice $B^{-1}\mathcal{L}_{\text{log}}=\sqrt{\pi}B^{-1}\mathbb{Z}^{4}$, where $B^{-1}$ is any matrix that satisfies $BB^{T}=\Sigma_{0}$. One such matrix is given by
\begin{equation}\label{eq:D4_B}
    B=\frac{1}{\sqrt{2}}\begin{bmatrix}1&-1&-1&-1\\1&1&-1&-1\\-1&-1&-1&-1\\1&1&1&-1\end{bmatrix}.
\end{equation}

Now consider the $D_{4}$ lattice, which is generated by the vectors $(1,1,0,0),(1,-1,0,0),(0,1,-1,0),(0,0,1,-1)$. Scaled in this standard way, the shortest vector in $D_{4}$ has length $\sqrt{2}$, and there are 24 vectors in $D_{4}$ with length $\sqrt{2}$. Rescaling by this $\sqrt{2}$ and performing a change of basis, we can write the $D_{4}$ lattice as
\begin{equation}
    D_{4}/\sqrt{2}=B^{-1}\mathbb{Z}^{4},
\end{equation}
where $B$ is precisely the same matrix as in \cref{eq:D4_B}. Therefore we can conclude that $B^{-1}\mathcal{L}_{\text{log}}$ is proportional to the $D_{4}$ lattice, which is known to be the densest lattice-packing in 4D.

Finally, we comment that in the square GKP code, the effective distance of the $\bar{C}_{ZZ}$ and $\bar{C}_{ZY}$ gate can be improved by applying a single-qubit gate $\bar{S}\bar{H}$ to each mode immediately following the generalized controlled gate. In both cases, the effective distance of the combined gates is 0.966$\sqrt{\pi}$, which is greater than the distances of 0.894$\sqrt{\pi}$ and 0.816$\sqrt{\pi}$ for the $C_{ZZ}$ and $C_{ZY}$ gates respectively. However, the advantages are only felt for somewhat small values of $\Delta$ because the increase in $d_{\text{eff}}$ comes with an increase in $a_{\text{eff}}$. Indeed, the cross-over point only occurs at 12.2 dB for the $C_{ZZ}$ gate and 9.77 dB for the $C_{ZY}$ gate. We leave the consequences of these observations to future work.

\begin{table}[]
    \caption{Extension of \cref{tab:estimates}: Ideal Clifford gate patches without taking into account the modification of the patch. With such a modification the effective degeneracy and distances are all equal to those of the identity gate.}
    \begin{center}
%    \begin{tabular}{c|>{\centering\arraybackslash}m{0.8cm}|c|>{\centering\arraybackslash}m{0.8cm}|c}
%        Code & \multicolumn{2}{c|}{Square} & \multicolumn{2}{c}{Hexagonal} \\
%        \hline
%        Gate&$a$& $d/\sqrt{\pi}$ & $a$ & $d/\sqrt{\pi}$\rule{0pt}{1em}\rule[-4pt]{0pt}{1em} \\
%        \hline
%        $\bar{I}$ & 2 & 1 & 3 & $\sqrt{2/\sqrt{3}}\approx1.075$\rule{0pt}{1.6\normalbaselineskip}\\
%        $\bar{H}$ & 2 & 1 & 2 & $\sqrt{2\sqrt{3}/7}\approx0.703$\rule{0pt}{1.6\normalbaselineskip}\\
%        $\bar{S}$ & 1 & $1/\sqrt{2}\approx0.707$ & 2 & $\sqrt{2\sqrt{3}/7}\approx0.703$\rule{0pt}{1.6\normalbaselineskip} \\
%        $\bar{S}^{2}$ & 1 & $1/\sqrt{5}\approx0.447$ & 1 & $\sqrt{2\sqrt{3}/19}\approx0.427$\rule{0pt}{1.6\normalbaselineskip} \\
%        $\bar{S}^{4}$ & 1 & $1/\sqrt{17}\approx0.243$ & 1 & $\sqrt{2\sqrt{3}/67}\approx0.227$\rule{0pt}{1.6\normalbaselineskip} \\
%        $\bar{I}\otimes\bar{I}$ & 4 & 1 & 6 & $\sqrt{2/\sqrt{3}}\approx1.075$\rule{0pt}{1.6\normalbaselineskip} \\
%        $\bar{H}\otimes\bar{H}$ & 4 & 1 & 4 & $\sqrt{2\sqrt{3}/7}\approx0.703$\rule{0pt}{1.6\normalbaselineskip} \\
%        $\bar{C}_{ZZ}$ & 2 & $1/\sqrt{2}\approx0.707$ & 2 & \rule[-6pt]{0pt}{4pt}$\sqrt{2\sqrt{3}/7}\approx0.703$\rule{0pt}{1.6\normalbaselineskip} \\
%        $\bar{C}_{ZY}$ & 1 & $1/\sqrt{3}\approx0.577$ & 2 & \rule[-6pt]{0pt}{4pt}$\sqrt{2\sqrt{3}/7}\approx0.703$\rule{0pt}{1.6\normalbaselineskip} \\
%        $\bar{C}_{YY}$ & 4 & $1/\sqrt{3}\approx0.577$ & 2 & \rule[-6pt]{0pt}{4pt}$\sqrt{2\sqrt{3}/7}\approx0.703$\rule{0pt}{1.6\normalbaselineskip}
%    \end{tabular}
    \includegraphics{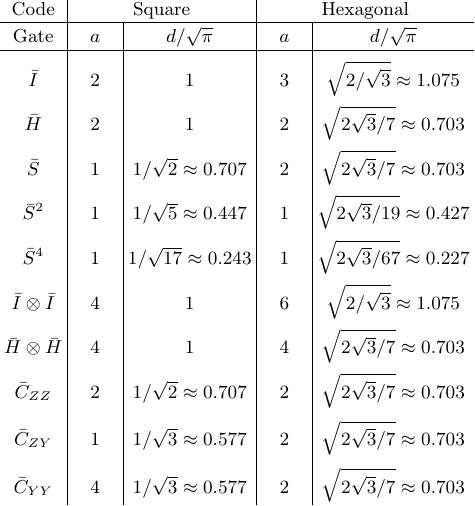}
    \label{tab:estimates_full}
    \end{center}
\end{table}

\begin{table}[]
    \begin{center}
    \caption{Extension of \cref{tab:estimates2}: Clifford gate patches taking into account approximate error correction. For notational brevity, we have written $R=SH$.}
    \includegraphics{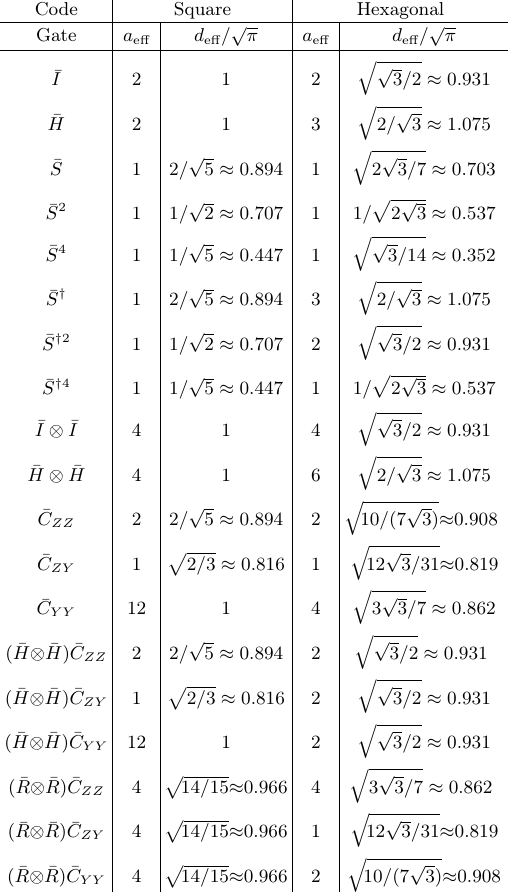}
    \label{tab:estimates2_full}
    \end{center}
\end{table}

\section{Characteristic Function of Ideal Pauli Operators}\label{sec:EC_chi}

In this appendix, we derive the characteristic function of the ideal logical Pauli operators of a single-mode GKP qubit code. The resulting expression provides a useful way of performing the SSSD partial trace in Fock basis simulations that may be of independent interest.

First, recall that the displacement operators $T(\vect{v})$ form an orthogonal operator basis of $\mathcal{H}$ and therefore arbitrary operators can be written
\begin{equation}\label{eq:O_chi}
O=\iint\frac{dv_{1}\,dv_{2}}{2\pi}\chi_{O}(v_{1},v_{2})T(v_{1},v_{2})^\dag,
\end{equation}
where $\chi_{O}(v_{1},v_{2})=\mathrm{tr}\big(OT(v_{1},v_{2})\big)$ is the \textit{characteristic function} of the operator $O$. We define the stabilizer lattice $\mathcal{L}_{\text{stab}}$ and the logical lattice $\mathcal{L}_{\text{log}}$ as
\begin{subequations}
\begin{align}
\mathcal{L}_{\text{stab}}&=\{2(n_{1}\vect{\alpha}+n_{2}\vect{\beta})\mid n_{1},n_{2}\in\mathbb{Z}\},\\
\mathcal{L}_{\text{log}}&=\{n_{1}\vect{\alpha}+n_{2}\vect{\beta}\mid n_{1},n_{2}\in\mathbb{Z}\}=\mathcal{L}_{\text{stab}}/2.
\end{align}
\end{subequations}
These lattices are defined such that $T(\vect{v})$ is a stabilizer whenever $\vect{v}\in\mathcal{L}_{\text{stab}}$, and a (possibly trivial) logical Pauli operator whenever $\vect{v}\in\mathcal{L}_{\text{log}}$. In the language of lattice theory, the stabilizer lattice (divided by a factor of $\sqrt{2\pi}$) is a symplectically integral lattice, and the logical lattice (divided by $\sqrt{2\pi}$) is the symplectic dual of the stabilizer lattice (divided by $\sqrt{2\pi}$).

As discussed in \cref{subsec:SSD}, $\vect{\alpha},\vect{\beta}$ and a decoding patch $\mathcal{P}$ together uniquely specify a GKP SSSD of $\mathcal{H}\cong\mathcal{L}\otimes\mathcal{S}$ into a logical ($\mathcal{L}$) and stabilizer ($\mathcal{S}$) subsystem, for more details see Ref.~\cite{Shaw24-1}. One motivation for considering the SSSD is because the partial trace over the stabilizer subsystem $\mathrm{tr}_{\mathcal{S}}$ is equivalent to performing a round of ideal error correction and decoding over the patch $\mathcal{P}$, and ``forgetting'' the stabilizer measurement outcomes obtained during the error correction.

It is worth noting that the logical Pauli operators $\bar{X}=T(\vect{\alpha}),\bar{Y}=T(\vect{\alpha}+\vect{\beta}),\bar{Z}=T(\vect{\beta})$ are not Hermitian operators and therefore have complex eigenvalues. However, when performing a logical Pauli measurement, the result ought to be $\pm1$ in keeping with the eigenvalues of a qubit Pauli operator. To resolve this discrepancy we introduce the logical Pauli measurement operators, which we define in terms of the subsystem decomposition as
\begin{align}
X_{\text{m}}&=X\otimes I,&Y_{\text{m}}&=Y\otimes I,&Z_{\text{m}}&=Z\otimes I.
\end{align}
These operators are both Hermitian and unitary; moreover, they correspond to performing a round of ideal error correction followed by a logical Pauli measurement using $\bar{X},\bar{Y}$ or $\bar{Z}$ (which have eigenvalues $\pm1$ on the ideal GKP codespace). One reason for studying these logical Pauli measurement operators is that they can be used to obtain the partial trace over the stabilizer subsystem via
\begin{equation}\label{eq:partial_trace_from_Pm}
\mathrm{tr}_{\mathcal{S}}(\rho)=\frac{1}{2}\Big(I\mathrm{tr}(\rho)+X\mathrm{tr}(\rho X_{\text{m}})+Y\mathrm{tr}(\rho Y_{\text{m}})+Z\mathrm{tr}(\rho Z_{\text{m}})\Big).
\end{equation}

An equivalent characterization of the logical Pauli measurement operator $P_{\text{m}}$ is by its action on displaced ideal GKP codestates $T(\vect{k})\ket{\bar{\pm}_{P}}$, where the qubit state $\ket{\pm_{P}}$ is a $\pm1$ eigenvalue of the Pauli operator $P$. If $\vect{k}\in\mathcal{P}$, then the logical state of $T(\vect{k})\ket{\bar{\pm}_{P}}$ in the subsystem decomposition is $\ket{\pm_{P}}$, and therefore we have $P_{\text{m}}T(\vect{k})\ket{\bar{\pm}_{P}}=\pm T(\vect{k})\ket{\bar{\pm}_{P}}=T(\vect{k})\bar{P}\ket{\bar{\pm}_{P}}$, where $\bar{P}$ is a logical Pauli operator (i.e.~a displacement). Indeed, since the states $T(\vect{k})\ket{\bar{\pm}_{P}}$ form an orthogonal basis for $\vect{k}\in\mathcal{P}$, $\pm\in\{+,-\}$ and fixed $P$, we can write
\begin{equation}\label{eq:Pm_starting_point}
P_{\text{m}}=N\sum_{\pm}\int_{\mathcal{P}}\!d^{2}\vect{k}\,T(\vect{k})\bar{P}\ket{\bar{\pm}_{P}}\!\bra{\bar{\pm}_{P}}T(\vect{k})^{\dag},
\end{equation}
where the constant $N$ is included to account for the fact that the displaced codestates are not normalized.

Next, note that $\sum_{\pm}\ket{\bar{\pm}_{P}}\!\bra{\bar{\pm}_{P}}$ is the projector onto the ideal GKP codespace, $\Pi_{\text{GKP}}$, and can be written in terms of displacement operators as
\begin{equation}\label{eq:GKP_projector}
\Pi_{\text{GKP}}=\sum_{\vect{s}\in\mathcal{L}_{\text{stab}}}\!T(\vect{s}),
\end{equation}
since every state in the codespace is a $+1$ eigenstate of every stabilizer. Substituting \cref{eq:GKP_projector} into \cref{eq:Pm_starting_point} gives
\begin{equation}
P_{\text{m}}=N\!\!\!\sum_{\vect{s}\in\mathcal{L}_{\text{stab}}}\!\!\!\bigg(\int_{\mathcal{P}}\!d^{2}\vect{k}\,e^{-i\omega(\vect{k},\vect{s}+\vect{\ell}_{P})}\!\bigg)T(\vect{\ell}_{P})T(\vect{s}),
\end{equation}
where we've written $\vect{\ell}_{X,Y,Z}=\vect{\alpha},\vect{\alpha}+\vect{\beta},\vect{\beta}$, respectively.

We can evaluate the constant $N$ using the unitarity of $P_{\text{m}}$. Indeed from \cref{eq:Pm_starting_point} we have
\begin{multline}\label{eq:N_starting_point}
P_{\text{m}}^{}P_{\text{m}}^{\dag}=N^{2}\;\sum_{\mathclap{\vect{s},\vect{s}'\in\mathcal{L}_{\text{stab}}}}\;\, T(\vect{s}')\\
\times\iint_{\mathcal{P}}\!d^{2}\vect{k}\,d^{2}\vect{k}'\,e^{i\omega(\vect{s}+\vect{\ell}_{P},\vect{k}-\vect{k}')} e^{i\omega(\vect{s}',\vect{k}'+\vect{\ell}_{P})},
\end{multline}
where we have also used $T(\vect{s})T(\vect{s}')=T(\vect{s}+\vect{s}')$ for $\vect{s},\vect{s}'\in\mathcal{L}_{\text{stab}}$. Next, we can evaluate the sum over $\vect{s}$ for $\vect{k},\vect{k}'\in\mathcal{P}$ using the Poisson summation formula:
\begin{subequations}\label{eq:psf}
\begin{align}
\hspace{0.25 cm}&\hspace{-0.25 cm}\sum_{\vect{s}\in\mathcal{L}_{\text{stab}}}e^{i\omega(\vect{s},\vect{k}-\vect{k}')}\nonumber\\
&=\;\;\sum_{\mathclap{n_{1},n_{2}\in\mathbb{Z}}}e^{2i\omega(\vect{\alpha},\vect{k}-\vect{k}')n_{1}+2i\omega(\vect{\beta},\vect{k}-\vect{k}')n_{2}}\label{eq:psf1}\\
&=\;\;\sum_{\mathclap{m_{1},m_{2}\in\mathbb{Z}}}\;\;\delta\big(\omega(\vect{\alpha},\vect{k}{-}\vect{k}')/\pi {+} m_{1}\big)\,\delta\big(\omega(\vect{\beta},\vect{k}{-}\vect{k}')/\pi {+} m_{2}\big)\label{eq:psf2}\\
&=\;\;\sum_{\mathclap{m_{1},m_{2}\in\mathbb{Z}}}\;\;\delta^{2}\!\bigg(\frac{1}{\pi}\!\begin{bmatrix}\vect{\alpha}^{T}\\\vect{\beta}^{T}\end{bmatrix}\!\begin{bmatrix}0&1\\-1&0\end{bmatrix}\!(\vect{k}-\vect{k}'+m_{1}\vect{\beta}-m_{2}\vect{\alpha})\!\bigg)\label{eq:psf3}\\
&=\pi\!\sum_{\vect{\ell}\in\mathcal{L}_{\text{log}}}\!\delta^{2}(\vect{k}-\vect{k}'+\vect{\ell})\label{eq:psf4}\\
&=\pi\delta^{2}(\vect{k}-\vect{k}').\label{eq:psf5}
\end{align}
\end{subequations}
From \cref{eq:psf1} to \cref{eq:psf2} we have used the Poisson summation formula, and \cref{eq:psf3} is simply a rearrangement of \cref{eq:psf2} using
\begin{equation}
\omega(\vect{v},\vect{w})=\vect{v}^{T}\begin{bmatrix}0&1\\-1&0\end{bmatrix}\vect{w},
\end{equation}
and $\omega(\vect{\alpha},\vect{\beta})=\pi$. To arrive at \cref{eq:psf4} we use the property of Dirac delta functions that
\begin{equation}
\delta^{n}(M\vect{v})=\frac{1}{|\det M|}\delta^{n}(\vect{v}).
\end{equation}
In our case we have
\begin{equation}
\det M=\det\bigg(\frac{1}{\pi}\!\begin{bmatrix}\vect{\alpha}^{T}\\\vect{\beta}^{T}\end{bmatrix}\!\begin{bmatrix}0&1\\-1&0\end{bmatrix}\!\bigg)=\frac{1}{\pi},
\end{equation}
giving \cref{eq:psf4}. Finally, \cref{eq:psf5} is obtained by using the fact that $\vect{k},\vect{k}'\in\mathcal{P}$ and $\mathcal{P}$ is a primitive cell of the logical lattice $\mathcal{L}_{\text{log}}$. Applying \cref{eq:psf} to \cref{eq:N_starting_point} gives
\begin{equation}\label{eq:N_mid_point}
P^{}_{\text{m}}P_{\text{m}}^{\dag}=\pi N^{2}\,\sum_{\mathclap{\vect{s}'\in\mathcal{L}_{\text{stab}}}}\;\bigg(\!\int_{\mathcal{P}}\!d^{2}\vect{k}\,e^{i\omega(\vect{s}',\vect{k})}\!\bigg)e^{i\omega(\vect{s}',\vect{\ell}_{P})}T(\vect{s}').
\end{equation}
The integral over $\vect{k}\in\mathcal{P}$ can be evaluated by first noting that since $\omega(\vect{\ell},\vect{s})$ is a multiple of $2\pi$ for any vector $\vect{\ell}\in\mathcal{L}_{\text{log}}$, adding a logical vector $\vect{\ell}$ to $\vect{k}$ does not change the integrand. Using this logic, this means that the integrand does not depend on the primitive cell $\mathcal{P}$ of the lattice $\mathcal{L}_{\text{log}}$ that defines the region of integration\footnote{One can also see this by considering the cell transformations described in Sec V.A. of Ref.~\cite{Shaw24-1}.} -- and we can choose $\mathcal{P}$ to be the most convenient for our present calculation. Writing $\vect{k}=k_{\alpha}\vect{\alpha}+k_{\beta}\vect{\beta}$ and $\vect{s}'=2s'_{\alpha}\vect{\alpha}+2s'_{\beta}\vect{\beta}$ for $s'_{\alpha},s'_{\beta}\in\mathbb{Z}$, we therefore have
\begin{subequations}
\begin{align}
\int_{\mathcal{P}}\!d^{2}\vect{k}\,e^{i\omega(\vect{s}',\vect{k})}&=\pi\int_{\mathrlap{-1/2}}^{\mathrlap{1/2}}dk_{\alpha}\int_{\mathrlap{-1/2}}^{\mathrlap{1/2}}dk_{\beta}\;e^{i\omega(2s'_{\alpha}\vect{\alpha}+2s'_{\beta}\vect{\beta},k_{\alpha}\vect{\alpha}+k_{\beta}\vect{\beta})}\\
&=\pi\int_{\mathrlap{-1/2}}^{\mathrlap{1/2}}dk_{\alpha}\,e^{-2i\pi k_{\alpha}s_{\beta}'}\int_{\mathrlap{-1/2}}^{\mathrlap{1/2}}dk_{\beta}\,e^{2i\pi k_{\beta}s_{\alpha}'}\\
&=\pi\delta_{s_{\alpha}',0}\delta_{s_{\beta}',0}=\pi\delta_{\vect{s}',\vect{0}}.
\end{align}
\end{subequations}
Substituting into \cref{eq:N_mid_point} finally gives
\begin{equation}
P_{\text{m}}^{}P_{\text{m}}^{\dag}=\pi^{2}N^{2}T(\vect{0})=I\quad\Rightarrow\quad N=1/\pi.
\end{equation}

We can finally write our result in full:
\begin{center}
\setlength{\fboxsep}{1 em}
\fbox{\parbox{0.9\linewidth}{\vskip -2 ex
\begin{equation}\label{eq:Pm_chi}
P_{\text{m}}=\sum_{\vect{s}\in\mathcal{L}_{\text{stab}}}\!\!\!\bigg(\int_{\mathcal{P}}\frac{d^{2}\vect{k}}{\pi}e^{-i\omega(\vect{k},\vect{s}+\vect{\ell}_{P})}\!\bigg)T(\vect{\ell}_{P})T(\vect{s}),
\end{equation}
where $\vect{\ell}_{X,Y,Z}=\vect{\alpha},\vect{\alpha}+\vect{\beta},\vect{\beta}$ respectively.
}}
\end{center}

It can be shown directly that $P^{}_{\text{m}}=P_{\text{m}}^{\dag}$ and $Y_{\text{m}}=iX_{\text{m}}Z_{\text{m}}$, as initially desired. Note also that \cref{eq:Pm_chi} takes the form of \cref{eq:O_chi} and therefore can be used to find the characteristic function of $P_{\text{m}}$:
\begin{multline}\label{eq:Pm_actual_chi}
\chi^{}_{P_{\text{m}}}\!(\vect{v})=2\!\!\!\sum_{\vect{s}\in\mathcal{L}_{\text{stab}}}\!\!\!\!\bigg(\!\int_{\mathcal{P}}\!\!d^{2}\vect{k}\,e^{i\omega(\vect{k},\vect{s}+\vect{\ell}_{P})}\!\bigg)\\
\times e^{i\omega(\vect{\ell}_{P},\vect{s})/2}\,\delta^{2}\big(\vect{v}-(\vect{s}+\vect{\ell}_{P})\big).
\end{multline}

However, the main use of \cref{eq:Pm_chi} is to perform the partial trace $\mathrm{tr}_{\mathcal{S}}$ numerically. Indeed, from \cref{eq:partial_trace_from_Pm}, one can calculate the partial trace of the state $\rho$ simply from the expectation values $\mathrm{tr}(\rho P_{\text{m}})$. If one is evaluating the partial trace of an individual state $\rho$, then it is most convenient to evaluate
\begin{multline}
    \mathrm{tr}(\rho P_{\text{m}})=\!\!\sum_{\vect{s}\in\mathcal{L}_{\text{stab}}}\!\!\!\bigg(\int_{\mathcal{P}}\frac{d^{2}\vect{k}}{\pi}e^{-i\omega(\vect{k},\vect{s}+\vect{\ell}_{P})}\!\bigg)\\
    \times e^{-i\omega(\vect{\ell}_{P},\vect{s})/2}\chi_{\rho}(\vect{\ell}_{P}+\vect{s}),
\end{multline}
where $\chi_{\rho}(\vect{v})$ is the characteristic function of $\rho$. Alternatively, if the calculations are being conducted on multiple states, then it is more convenient to obtain the operators \cref{eq:Pm_chi} themselves. For numerical Fock basis simulations, this can be done by truncating the sum over $\mathcal{L}_{\text{stab}}$, using a Fock basis representation of the displacement operators $T(\vect{v})$, and evaluating the integrals over $\mathcal{P}$ numerically if necessary. 

If the GKP code is rectangular, i.e.~$\vect{\alpha}=(\sqrt{\pi}a,0)$, $\vect{\beta}=(0,\sqrt{\pi}/a)$ and $\mathcal{P}=\big({-}\sqrt{\pi}a/2,\sqrt{\pi}a/2\big]\times\big({-}\sqrt{\pi}/(2a),\sqrt{\pi}/(2a)\big]$ for some $a>0$, the integrals over $\mathcal{P}$ can be evaluated analytically. In particular, we have
\begin{align}
\int_{-\pi/2}^{\pi/2}\!\!dx\,e^{-2ixn}&=\pi\delta_{n,0},&\int_{-\pi/2}^{\pi/2}\!\!dx\,e^{-ix(2n+1)}&=\frac{(-1)^{n}}{n+1/2},
\end{align}
for $n\in\mathbb{Z}$. Applying this to \cref{eq:Pm_chi} gives
\begin{subequations}\label{eq:sq_Pm}
\begin{align}
X_{\text{m}}&=\frac{1}{\pi}\sum_{n\in\mathbb{Z}}\frac{(-1)^{n}}{n+\frac{1}{2}}T\big((2n+1)\vect{\alpha}\big),\\
Y_{\text{m}}&=\frac{1}{\pi^{2}}\sum_{m,n\in\mathbb{Z}}\!\!\frac{T\big((2m+1)\vect{\alpha}+(2n+1)\vect{\beta}\big)}{(m+\frac{1}{2})(n+\frac{1}{2})},\label{eq:rect_Ym}\\
Z_{\text{m}}&=\frac{1}{\pi}\sum_{n\in\mathbb{Z}}\frac{(-1)^{n}}{n+\frac{1}{2}}T\big((2n+1)\vect{\beta}\big).
\end{align}
\end{subequations}
Note that one could also use $Y_{\text{m}}=iX_{\text{m}}Z_{\text{m}}$ instead of \cref{eq:rect_Ym}, although this is less numerically stable for simulations with truncated Fock space dimension.

One can also evaluate the integrals analytically for the hexagonal GKP code defined by
\begin{align}
    \vect{\alpha}_{\text{hex}}&=\sqrt{\pi}\bigg(\frac{\sqrt[4]{3}}{\sqrt{2}},-\frac{1}{\sqrt{2}\sqrt[4]{3}}\bigg),&\vect{\beta}_{\text{hex}}&=\sqrt{\pi}\bigg(0,\frac{\sqrt{2}}{\sqrt[4]{3}}\bigg),\tag{\ref{eq:hex_GKP}}
\end{align}
and with a hexagonal patch $\mathcal{P}$ given by the Voronoi cell of the logical lattice. Using Mathematica to evaluate the integrals we obtain
\begin{subequations}\label{eq:hex_Pm}
    \begin{align}
        X_{\text{m}}&=\frac{3}{\pi^{2}}\;\sum_{\mathclap{m,n\in\mathbb{Z}}}\;f(m,n)\,T\big((2m+1)\vect{\alpha}+2n\vect{\beta}\big),\\
        Y_{\text{m}}&=\frac{3}{\pi^{2}}\;\sum_{\mathclap{m,n\in\mathbb{Z}}}\;f(n,n-m)\,T\big((2m+1)\vect{\alpha}+(2n+1)\vect{\beta}\big),\\
        Z_{\text{m}}&=\frac{3}{\pi^{2}}\;\sum_{\mathclap{m,n\in\mathbb{Z}}}\;f(n,m)\,T\big(2m\vect{\alpha}+(2n+1)\vect{\beta}\big),\\
        &f(m,n)=\frac{(-1)^{m}\cos\big(\frac{\pi}{3}(m+n-1)\big)}{(m+n+\frac{1}{2})(m-2n+\frac{1}{2})},
    \end{align}
\end{subequations}
although in practice one only needs to calculate $X_{\text{m}}$ using this method since we have $Y_{\text{m}}=e^{i\pi a^\dag a/3}X_{\text{m}}e^{-i\pi a^{\dag}a/3}$ and $Z_{\text{m}}=e^{-i\pi a^\dag a/3}X_{\text{m}}e^{i\pi a^{\dag}a/3}$ in the hexagonal code.

Both \Cref{eq:sq_Pm,eq:hex_Pm} have recently been applied to characterize GKP state preparation in experiment~\cite{Matsos23} and a novel theoretical scheme~\cite{Kolesnikow23}, demonstrating that these expressions are practical and convenient to implement.

\section{Derivation of Measurement Error Estimate}\label{sec:measurement_error_estimate}
In this appendix, we derive the approximate expression of the measurement error
\begin{equation}
    M_{\text{error}}\approx\mathrm{erfc}\bigg(\frac{1}{2}\alpha_{1}\Big(\Delta^{2}+\frac{1-\eta}{\eta}\Big)^{-1/2}\bigg)\tag{\ref{eq:measurement_error_approx}}
\end{equation}
of a GKP logical Pauli $Z$ measurement using a binned measurement with bin size $b$ and homodyne measurement efficiency $\eta$. We consider any GKP code that is defined by the vectors $\vect{\alpha}=[\alpha_{1},\alpha_{2}]^{T}$ and $\vect{\beta}=[0,\pi/\alpha_{1}]^{T}$, which is rotated such that a logical measurement of $\bar{Z}=e^{i\pi q/\alpha_{1}}$ can be done with a binned position measurement with $b\approx\alpha_{1}$.

To show \cref{eq:measurement_error_approx}, we make the approximation $P(0|1)\approx P(1|0)$ (which is valid in the $\Delta\rightarrow0$ limit). As such, we have $M_{\text{error}}\approx P(1|0)$, the probability of obtaining a $-1$ logical $Z$ measurement outcome given an initial state $\ket{\bar{0}_{\Delta}}$. Now, the position wavefunction of the approximate codestate $\ket{\bar{0}_{\Delta}}$ is given exactly by
\begin{multline}
    \prescript{}{q}{\braket{x|\bar{0}_{\Delta}}}=\mathcal{N}_{0}\sum_{s\in\mathbb{Z}}e^{2is^{2}\alpha_{1}\alpha_{2}}e^{-\frac{1}{2}(2s\alpha_{1})^{2}\mathrm{tanh}(\Delta^{2})}\\
    \times e^{-\frac{1}{2}\mathrm{coth}(\Delta^{2})[x-2s\alpha_{1}\mathrm{sech}(\Delta^{2})]^{2}}.\tag{\ref{eq:0_bar_Delta}}
\end{multline}
Recall that a position measurement with efficiency $\eta$ is described by the POVM elements
\begin{equation}
    W_{\eta}(X)=\mathcal{N}_{\eta}\!\int dx\,\ket{x}_{q}\!\bra{x}\exp\Big({-}\frac{\eta}{1-\eta}(x-X)^{2}\Big),\tag{\ref{eq:inefficient_POVM}}
\end{equation}
where $X\in\mathbb{R}$ is the recorded outcome of the measurement and $\mathcal{N}_{\eta}=\sqrt{\eta/(\pi(1-\eta))}$ is a normalisation constant such that $\int dX\, W_{\eta}(X)=I$. From \cref{eq:0_bar_Delta,eq:inefficient_POVM} we can write down the probability of recording a $-1$ measurement outcome given an initial state $\ket{\bar{0}_{\Delta}}$ as
\begin{align}
    P(1|0)&=\sum_{t\in\mathbb{Z}}\int_{(2t+\frac{1}{2})b}^{(2t+\frac{3}{2})b}dX~\tr\big(W_{\eta}(X)\ket{\bar{0}_{\Delta}}\!\bra{\bar{0}_{\Delta}}\big)\tag{\ref{eq:measurement_error}}\\
    &=\mathcal{N}_{\eta}\sum_{t\in\mathbb{Z}}\int_{(2t+\frac{1}{2})b}^{(2t+\frac{3}{2})b}\!\!\!\!dX\!\!\int_{\mathbb{R}}\!dx\; e^{-\frac{\eta}{1-\eta}(x-X)^{2}}\big|\!\prescript{}{q}{\!\braket{x|\bar{0}_{\Delta}}}\!\big|^{2}.\label{eq:P_10_initial}
\end{align}

Next, we approximate the modulus squared of the wavefunction \cref{eq:0_bar_Delta} as
\begin{multline}\label{eq:0_Delta_squared}
    \big|\!\prescript{}{q}{\braket{x|\bar{0}_{\Delta}}}\!\big|^{2}\approx\mathcal{N}_{0}^{2}\sum_{s\in\mathbb{Z}}e^{-(2s\alpha_{1})^{2}\mathrm{tanh}(\Delta^{2})}\\
    \times e^{-\mathrm{coth}(\Delta^{2})[x-2s\alpha_{1}\mathrm{sech}(\Delta^{2})]^{2}},
\end{multline}
which assumes the cross-terms originating from squaring the sum over $s$ in \cref{eq:0_bar_Delta} are negligible. We approximate the normalisation constant in \cref{eq:0_bar_Delta}, $\mathcal{N}_{0}$, using
\begin{align}
    1&=\int_{\mathrlap{\mathbb{R}}}\;dx\,\big|\!\prescript{}{q}{\!\braket{x|\bar{0}_{\Delta}}}\!\big|^{2}\nonumber\\
    &\approx\mathcal{N}_{0}^{2}\sum_{s\in\mathbb{Z}}e^{-(2s\alpha_{1})^{2}\mathrm{tanh}(\Delta^{2})}\big(\pi\mathrm{tanh}(\Delta^{2})\big)^{1/2}\\
    \Rightarrow\;&\mathcal{N}_{0}^{2}\approx\bigg(\!\sqrt{\pi\mathrm{tanh}(\Delta^{2})}\,\sum_{s\in\mathbb{Z}}e^{-(2s\alpha_{1})^{2}\mathrm{tanh}(\Delta^{2})}\bigg)^{-1}.\label{eq:N_0_estimate}
\end{align}
One could approximate the sum in \cref{eq:N_0_estimate} to directly estimate $\mathcal{N}_{0}$, but we will not need to do this for our calculations.

Substituting \cref{eq:0_Delta_squared} into \cref{eq:P_10_initial} and integrating over $x$ gives
\begin{equation}\label{eq:P_10_progress}
    P(1|0)\approx\mathcal{N}'_{\eta,\Delta}\sum_{s\in\mathbb{Z}}I_{s}e^{-(2s\alpha_{1})^{2}\mathrm{tanh}(\Delta^{2})},
\end{equation}
where we've defined
\begin{subequations}
\begin{align}
    I_{s}&=\sum_{t\in\mathbb{Z}}\int_{(2t+\frac{1}{2})b}^{(2t+\frac{3}{2})b}\!\!\!\!dX\,e^{-\left(X-2s\alpha_{1}\mathrm{sech}(\Delta^{2})\right)^{2}\!/(2\sigma_{\eta,\Delta}^{2})},\\
    \mathcal{N}'_{\eta,\Delta}&=\mathcal{N}_{0}^{2}\bigg(1+\frac{1-\eta}{\eta}\mathrm{coth}(\Delta^{2})\bigg)^{\!-1/2},\label{eq:N_prime}\\
    \sigma_{\eta,\Delta}&=\frac{1}{\sqrt{2}}\bigg(\mathrm{tanh}(\Delta^{2})+\frac{1-\eta}{\eta}\bigg)^{1/2}\label{eq:sigma_eta_Delta}.
\end{align}
\end{subequations}
Now, $I_{s}$ can be approximated by setting $b=\alpha_{1}$ and $\mathrm{sech}(\Delta^{2})\approx 1$ (valid when $\Delta$ is small), giving
\begin{equation}
    I_{s}\approx\sum_{t\in\mathbb{Z}}\int_{(2t+\frac{1}{2})\alpha_{1}}^{(2t+\frac{3}{2})\alpha_{1}}\!\!\!\!dX\,e^{-\left(X-2s\alpha_{1}\right)^{2}\!/(2\sigma_{\eta,\Delta}^{2})}.
\end{equation}
Substituting $X'=X-2s\alpha_{1}$ and $t'=t-s$ gives
\begin{equation}
    I_{s}\approx\sum_{t'\in\mathbb{Z}}\int_{(2t'+\frac{1}{2})\alpha_{1}}^{(2t'+\frac{3}{2})\alpha_{1}}\!\!\!\!dX'\,e^{-{X'}^{2}/(2\sigma_{\eta,\Delta}^{2})},
\end{equation}
which is independent $s$. Finally, we simplify the domain of integration, which is valid when $\Delta$ is small such that the exponent is quickly decaying, as
\begin{align}
    I_{s}&\approx2\int_{\alpha_{1}/2}^{\infty}\!\!dX'\,e^{-{X'}^{2}/(2\sigma_{\eta,\Delta}^{2})}\\
    &=\sqrt{2\pi}\sigma_{\eta,\Delta}\,\mathrm{erfc}\bigg(\frac{\alpha_{1}}{2\sqrt{2}\sigma_{\eta,\Delta}}\bigg).\label{eq:Is_estimate}
\end{align}
Substituting \cref{eq:Is_estimate} back into \cref{eq:P_10_progress} gives
\begin{align}
    P(1|0)&\approx\sqrt{2\pi}\mathcal{N}'_{\eta,\Delta}\sigma^{\vphantom{\prime}}_{\eta,\Delta}\mathrm{erfc}\bigg(\frac{\alpha_{1}}{2\sqrt{2}\sigma_{\eta,\Delta}}\bigg)\nonumber\\
    &\qquad\qquad\qquad\qquad\times\sum_{s\in\mathbb{Z}}e^{-(2s\alpha_{1})^{2}\mathrm{tanh}(\Delta^{2})}\\
    &=\mathrm{erfc}\bigg(\frac{\alpha_{1}}{2\sqrt{2}\sigma_{\eta,\Delta}}\bigg),\label{eq:second_last_step}
\end{align}
where we have used \cref{eq:N_0_estimate,eq:N_prime} to arrive at \cref{eq:second_last_step}. Substituting \cref{eq:sigma_eta_Delta} and expanding $\tanh(\Delta^{2})\approx\Delta^{2}$ to lowest order gives the desired result, \Cref{eq:measurement_error_approx}.

\section{Comparing logical measurement with and without error-correction}\label{sec:measurement_scheme_comparison}

\begin{figure*}
    \includegraphics[]{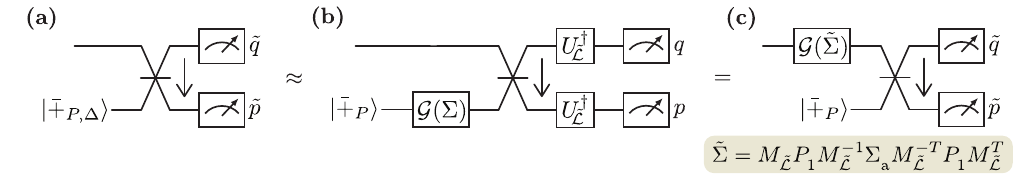}
    \caption{Error-corrected logical read-out with approximate codewords. (a) The ideal error-corrected logical read-out with the ideal codestates replaced with approximate ones. As described in the main text, we can apply methods similar to those in \cref{fig:tele_approx_circuit} to rewrite circuit (a) as circuit (c), which consists of an ideal error-corrected logical read-out circuit preceded by a Gaussian random displacement channel.}\label{fig:qec_measurement}
\end{figure*}

As discussed in the main text, we have previously pointed out in Appendix B in Ref.~\cite{Shaw24-1} that binned measurement operators do not always ideally read out the logical Pauli operators of the GKP code. Indeed, binned measurement operators are not ideal for $\bar{Y}$ read-out in the square GKP code, and for $\bar{X}$, $\bar{Y}$ and $\bar{Z}$ read-out in the hexagonal code. In theory, one can perform these measurements ideally by executing a round of ideal error correction alongside the logical Pauli read-out, and then use a decoder to infer the most likely logical outcome. However, this proposal neglects the effects of approximate codestates in the round of error correction. This appendix aims to show that such an error-corrected measurement scheme does not improve the measurement error probability in the presence of approximate codestates.

We will begin by describing how to perform a round of QEC alongside the logical read-out, and how we model the effect of approximate codestates. Then we will calculate the measurement error $M_{\text{error}}$ of this scheme, and compare it to the equivalent expression for binned measurement operators~\cref{eq:measurement_error_approx}. Much of the mathematical framework of these calculations has already been established in previous appendices, so we only provide a sketch of the derivations here. Note also that for simplicity, we assume ideal measurements ($\eta=1$) -- although we expect that incorporating measurement inefficiencies wouldn't significantly affect our results.

To measure the logical $\bar{P}$ of the code in an error-corrected way, we need to simultaneously measure $\bar{P}$ and \textit{one} of the stabilizer generators $S\neq\pm\bar{P}^{2}$. This is because the measurement outcome of a second stabilizer generator can always be inferred from the outcome of $\bar{P}^{2}$. Now, label the measurement-basis quadratures $\tilde{q}$ and $\tilde{p}$ such that $\bar{P}=e^{i\sqrt{\pi}\tilde{q}}$ and $S=e^{-2i\sqrt{\pi}\tilde{p}}$. This is defined such that for a $\bar{Z}$ measurement we have $\tilde{q}=\bar{q}$ and $\tilde{p}=\bar{p}$; while for $\bar{X}$ we have $\tilde{q}=-\bar{p}$ and $\tilde{p}=\bar{q}$, and for $\bar{Y}$ we have $\tilde{q}=\bar{q}-\bar{p}$ and $\tilde{p}=\bar{p}$.

To perform such an error-corrected logical measurement one can use the circuit given in \cref{fig:qec_measurement}(a), which uses a single ancilla mode, a 50-50 beamsplitter, and homodyne measurements in the $\tilde{q}$ and $\tilde{p}$ bases. The ancilla mode is initialized in the (approximate) GKP codestate $\ket{\bar{+}_{P,\Delta}}$, where $\ket{+_{P}}$ represents the $+1$-eigenstate of the Pauli operator $P$. However, to allow for the most general possible scheme, we allow for the approximate GKP codestate to have a biased noise distribution. Such a noise distribution could, in theory, be achieved by first preparing a GKP codestate with a different geometry and a symmetric noise distribution, and then applying a Gaussian unitary to prepare the desired state.

Now, to model the effect of the approximate codestate on the measurement, we can follow the methods described in \cref{sec:approx_Clifford_gates}. In particular, we can approximate \cref{fig:qec_measurement}(a) as \cref{fig:qec_measurement}(c), where $\mathcal{G}$ represents a Gaussian random displacement (GRD) channel [\cref{eq:GRD}] -- the derivation of this follows almost the same lines as in \cref{fig:tele_approx_circuit}. The only difference is that we allow the variance $\Sigma_{\text{a}}$ of the GRD channel on the ancilla mode to be non-symmetric, as discussed above. If the ancilla state is simply an approximate GKP codestate $\ket{\bar{+}_{P,\Delta}}=e^{-\Delta^{2}a^{\dag}a}\ket{\bar{+}_{P}}$, then $\Sigma_{\text{a}}=\mathrm{tanh}(\Delta^{2}/2)I$; but in general, we allow $\Sigma_{\text{a}}$ to be an arbitrary covariance matrix satisfying $\mathrm{det}(\Sigma_{\text{a}})=\tanh^{2}(\Delta^{2}/2)$. Taking this into account we end up with \cref{fig:qec_measurement}(c), which consists of an ideal QEC and logical read-out circuit preceded by a GRD channel with variance
\begin{equation}\label{eq:Sigma_tilde}
    \tilde{\Sigma}=M^{\vphantom{-T}}_{\tilde{\mathcal{L}}}P^{\vphantom{-T}}_{\vphantom{\tilde{1}}1}M_{\tilde{\mathcal{L}}}^{-1}\Sigma^{\vphantom{-T}}_{\vphantom{\tilde{1}}\text{a}} M_{\tilde{\mathcal{L}}}^{-T}P^{\vphantom{-T}}_{\vphantom{\tilde{1}}1}M_{\tilde{\mathcal{L}}}^{\vphantom{-1}T},
\end{equation}
where $M_{\tilde{\mathcal{L}}}$ is the matrix that transforms the quadrature operators $(q,p)\mapsto(\tilde{q},\tilde{p})$. Explicitly, we have
\begin{subequations}
\begin{align}
    M_{\tilde{\mathcal{L}}}&=[-\vect{\beta},\vect{\alpha}]/\sqrt{\pi},&P&=X;\\
    M_{\tilde{\mathcal{L}}}&=[\vect{\alpha},\vect{\alpha}+\vect{\beta}]/\sqrt{\pi},&P&=Y;\\
    M_{\tilde{\mathcal{L}}}&=[\vect{\alpha},\vect{\beta}]/\sqrt{\pi}=M_{\vphantom{\tilde{L}}\mathcal{L}},&P&=Z.
\end{align}
\end{subequations}
One can recover \cref{eq:Sigma_QEC} (in \cref{sec:approx_Clifford_gates}) from \cref{eq:Sigma_tilde} by substituting $M_{\tilde{\mathcal{L}}}\mapsto M_{\vphantom{\tilde{L}}\mathcal{L}}$ and $\Sigma_{\text{a}}\mapsto\mathrm{tanh}(\Delta^{2}/2)I$.

Next, we wish to estimate the measurement error $M_{\text{error}}=\big(P(0|1)+P(1|0)\big)/2$ associated with the circuit in \cref{fig:qec_measurement}. We will make a simplifying assumption that the input state on the data mode is simply an approximate GKP codestate $\ket{\bar{\psi}_{\Delta}}$ and can therefore be approximated by an ideal GKP codestate $\ket{\bar{\psi}}$ followed by a GRD channel $\mathcal{G}\big(\mathrm{tanh}(\Delta^{2}/2)I\big)$ following the methods of \cref{sec:error_estimates}. This is not necessarily the case, particularly if a logical gate has been applied immediately before the measurement, in which case the GRD channel will not be symmetrically distributed -- although we expect our results not to be significantly different in this case.

From \cref{fig:qec_measurement}(c), we can model the measurement error scheme as preparing first an ideal GKP codestate, then applying the noise channel $\mathcal{N}=\mathcal{G}(\Sigma)$ with $\Sigma=\mathrm{tanh}(\Delta^{2}/2)I+\tilde{\Sigma}$, performing a round of ideal QEC and then finally an ideal measurement of $\bar{P}$. As such we define the \textit{logical} noise channel
\begin{equation}
    \mathcal{N}_{\mathcal{L}}=\mathcal{J}[R^{\dag}]\circ\mathcal{C}_{\mathcal{P}}\circ\mathcal{N}\circ\mathcal{J}[R],\tag{\ref{eq:logical_noise_channel}}
\end{equation}
the same as in \cref{sec:approx_Clifford_gates}. Then, we can write the measurement error as
\begin{multline}
    M_{\text{error}}\approx\frac{1}{2}\Big(\!\bra{-_{P}}\mathcal{N}_{\mathcal{L}}(\ket{+_{P}}\!\bra{+_{P}})\ket{-_{P}}\\
    +\bra{+_{P}}\mathcal{N}_{\mathcal{L}}(\ket{-_{P}}\!\bra{-_{P}})\ket{+_{P}}\!\Big).
\end{multline}
Writing the logical noise channel in terms of its $\chi$-matrix elements \cref{eq:chi_mat_defn}, one can show that
\begin{equation}
    M_{\text{error}}=\sum_{P'\neq I,P} \chi_{P',P'},
\end{equation}
which is identical to~\cref{eq:fid_chi} except for the fact that sum over Pauli operators excludes the Pauli operator $P$ that is being measured. 

Because of this difference, we must revise a few definitions. We define the measurement region $\tilde{\mathcal{R}}$ as the set of vectors $\vect{v}$ whose closest vector the logical lattice $\mathcal{L}$ is logically equivalent either to the identity gate or to $\bar{P}$. Intuitively, if an error $T(\vect{v})$ with $\vect{v}\in\tilde{\mathcal{R}}$ is applied to an ideal GKP codestate, upon decoding this error will not flip the measurement outcome of $P$. Just as in \cref{sec:approx_Clifford_gates}, we define $\Sigma=2\mathrm{tanh}(\Delta^{2}/2)\Sigma_{0}$. Then, recall from \cref{sec:approx_Clifford_gates} that the effective distance $d_{\text{eff}}$ of the patch $\mathcal{P}$ is given by
\begin{equation}
    d_{\text{eff}}=2\min_{\vect{v}\in\partial\mathcal{P}}\sqrt{\vect{v}^{T}\Sigma_{0}^{-1}\vect{v}}.
\end{equation}
In words, $d_{\text{eff}}$ is twice the length of the shortest vector on the boundary of $\mathcal{P}$, under the metric defined by the inverse of the covariance matrix $\Sigma_{0}$. Meanwhile, the effective degeneracy $a_{\text{eff}}$ is half the number of vectors $\vect{v}\in\partial\mathcal{P}$ with a length $d_{\text{eff}}$ under the metric $\Sigma_{0}$. We now define the \textit{effective measurement} distance $\tilde{d}$ and degeneracy $\tilde{a}$, by adding to the definitions of $d_{\text{eff}}$ and $a_{\text{eff}}$ the requirement that the vectors $\vect{v}$ also lead to a logical error that flips the measurement outcome; i.e.~they do \textit{not} lead to a logical $\bar{I}$ or $\bar{P}$ upon decoding.

With these definitions we can follow the derivations in \cref{subsec:derivation,sec:approx_Clifford_gates} to show that
\begin{equation}
    M_{\text{error}}\approx\frac{1}{\sqrt{\mathrm{det}\Sigma}}\int_{\mathbb{R}^{2}\setminus\tilde{\mathcal{R}}}\frac{d^{2}\vect{v}}{2\pi}e^{-\vect{v}^{T}\Sigma^{-1}\vect{v}/2},
\end{equation}
where the only difference with \cref{eq:ent_infid_Sigma_int} is that the decoding patch $\mathcal{P}$ has been replaced by $\tilde{\mathcal{R}}$. Finishing the derivation gives
\begin{equation}\label{eq:M_error_qec}
    M_{\text{error}}\approx \tilde{a}\,\mathrm{erfc}\big(\tilde{d}/(2\sqrt{2}\Delta)\big),
\end{equation}
which is the same as \cref{eq:ent_fid_estimate_approx} except for the fact that the effective distance $d_{\text{eff}}$ and degeneracy $a_{\text{eff}}$ have been replaced by the effective \textit{measurement} distance $\tilde{d}$ and degeneracy $\tilde{a}$, respectively.

At this point it is useful to compare~\cref{eq:M_error_qec} to the equivalent expression for binned measurements (under the assumption of $\eta=1$), given by
\begin{equation}\label{eq:measurement_error_approx'}
    M_{\text{error}}\approx\mathrm{erfc}\big(b/(2\Delta)\big),\tag{\ref{eq:measurement_error_approx}$'$}
\end{equation}
where $b$ is the bin size of the measurement. The most important difference between \cref{eq:measurement_error_approx',eq:M_error_qec} is the additional $\sqrt{2}$ factor in \cref{eq:M_error_qec} which arises from the use of approximate codestates in the error-correction procedure. To overcome this, the effective measurement distance $\tilde{d}$ must be at least a factor of $\sqrt{2}$ larger than the bin size $b$.

We compared values of $\tilde{d}/b$ for all three measurement bases in the square GKP and hexagonal GKP codes. In the case where $\Sigma_{\text{a}}=\mathrm{tanh}(\Delta^{2}/2)I$, we found that in all cases $b=\tilde{d}$, and therefore the binned measurements have a lower measurement error. For each of these measurements, we can improve $\tilde{d}$ by using an ancilla with a highly asymmetric noise distribution. In the most general case, we allow $\Sigma_{\text{a}}$ to be an arbitrary positive-definite symmetric matrix with determinant $\mathrm{tanh}^{2}(\Delta^{2}/2)$. Maximizing the effective measurement distance $\tilde{d}$ using Mathematica yields $\tilde{d}=\sqrt{2}b$ in all cases, only breaking even with the binned measurement scheme. The binned measurement scheme is therefore preferable since it is significantly simpler to implement experimentally.

\section{Derivation of POVM of the Continuous Measurement Scheme}\label{sec:measurement_scheme_2_derivation}

In this appendix, we provide the details of the derivations of the POVM
\begin{multline}
    T_{g,\kappa,\eta,t}(X)=\mathcal{N}\!\int\! dx\,\ket{x}_{q_{1}}\!\!\bra{x}\int\! d\tilde{x}\,{\vphantom{\ket{x}}}_{q_{2}}\!\!\braket{\tilde{x}|\rho_{b}|\tilde{x}}_{q_{2}}\\
    \times \exp\bigg\{-\frac{1}{C}\Big[X-x-\frac{1}{2g\tau}\big(1-e^{-\kappa t/2}\big)^{2}\tilde{x}\Big]^{2}\bigg\},\tag{\ref{eq:measurement_scheme_2_general_POVM}}
\end{multline}
where $\mathcal{N}$ is a normalisation constant,
\begin{align}
    C&=\frac{\kappa\tau-\eta(1-e^{-\kappa t/2})^{4}}{4g^{2}\tau^{2}\eta},\tag{\ref{eq:c_expression}}\\
    X&=-\sqrt{\kappa/(8g^{2}\tau^{2}\eta)}\int_{0}^{t}dt'\,(1-e^{-\kappa t'/2})I(t'),\tag{\ref{eq:X_integral}}
\end{align}
and $\tau=t-(1-e^{-\kappa t/2})(3-e^{-\kappa t/2})/\kappa$.

We consider a GKP mode (described by ladder operators $a, a^{\dag}$ and position/momentum $q_{1}, p_{1}$) coupled via a Hamiltonian $H=-gq_{1}p_{2}$ to an ancilla readout mode (described by $b, b^{\dag}, q_{2}, p_{2}$). The ancilla mode is initialized in an arbitrary state $\rho_{b}$, which in the simplest case is the vacuum state $\ket{0}\!\bra{0}$. The ancilla mode undergoes homodyne detection at a rate $\kappa$ and efficiency $\eta$, and data is collected for a total measurement time $T$. We assume that the GKP mode experiences loss at a rate $\gamma\ll 1/T$ which we consider negligible for the following calculation. This is justified by the fact that typical $T_{1}$ times for microwave resonators are two to three orders of magnitude larger than measurement times that we are interested in -- roughly 1 $\mu$s -- to be comparable to measurement times currently achieved in transmon devices.

The goal of this appendix is to derive the POVM of the above measurement scheme in terms of the detected photocurrent record $\vect{I}(0,T)=\{I(t)\}_{t\in(0,T)}$ from time $0$ to time $T$. We follow the methods of Ref.~\cite{Warszawski20}, which we summarize here before presenting the details of the calculations.

We will begin by writing down the stochastic master equation (SME) that governs the dynamics of the normalized density matrix $\rho_{c}(t)$ conditioned on the detected photocurrent record $\vect{I}(0,t)$. Since the evolution of $\rho_{c}(t)$ depends on the measurement current, the SME will contain terms that depend on the Wiener increment $dW(t)=I(t)dt$ of the observed photocurrent. However, because $\rho_{c}(t)$ is normalized, the corresponding SME is necessarily non-linear. As such we instead consider a \textit{linearized} SME that describes the dynamics of an un-normalized density matrix $\bar{\rho}_{c}(t)\propto\rho_{c}(t)$. In this picture, the probability of obtaining a given measurement outcome $\vect{I}(0,T)$ is given by $\mathrm{Tr}\big(\bar{\rho}_{c}(T)\big)p_{\text{ost}}\big(\vect{I}(0,T)\big)$, where $p_{\text{ost}}$ is the chosen \textit{ostensible distribution} of the observed photocurrent record (see Ref.~\cite{Wiseman96}, Chapter 4 of Ref.~\cite{Wiseman09}, or Appendix A of Ref.~\cite{Warszawski20} for details). We can then solve the linearized SME by vectorizing the density matrix $\bar{\rho}_{c}(t)\mapsto\ket{\bar{\rho}_{c}(t)}$ and using Lie algebraic techniques to simplify the resultant expression.

We find that the state $\bar{\rho}_{c}(t)$ does not depend on the \textit{entire} photocurrent record $\vect{I}(0,t)$, but only on two integrals of the photocurrent labeled $R$ and $S$. The probability of a given measurement outcome $R,S$ can then be determined from the trace of $\bar{\rho}_{c}(T)$ and the ostensible distribution of $R$ and $S$, which can in turn be used to construct the POVM elements $W_{R,S}$. Finally, we will convert this two-mode POVM into a single-mode POVM by tracing out the ancilla mode given its initial state $\rho_{b}$ is known.

\subsection{Linearizing the SME}

We begin by writing down the stochastic master equation (SME) of the system
\begin{multline}\label{eq:nonlinear_master_eq}
    d{\rho}_{c}(t)=-i\big[H,{\rho}_{c}(t)\big]\,dt+\kappa\,\mathcal{D}[b]{\rho}_{c}(t)\,dt\\
    +\sqrt{\kappa\eta}\,\mathcal{H}[b]{\rho}_{c}(t)\,dW(t),
\end{multline}
where $H=-gq_{1}p_{2}$ is the system Hamiltonian, $\mathcal{D}[b]\rho=b\rho b^{\dag}-\frac{1}{2}\big\{b^{\dag}b,\rho\big\}$ is the Lindblad dissipator, $\mathcal{H}[b]\rho=b\rho+\rho b^{\dag}-\mathrm{Tr}(b\rho+\rho b^{\dag})$ describes the homodyne detection, and $dW(t)$ is the Wiener increment which is related to the relevant observed photocurrent $I(t)$ via $dW(t)=I(t)dt$. We use subscript $c$ to emphasize that $\rho_{c}(t)$ is the density operator conditioned on the record of the observed photocurrent $\vect{I}(0,t)$ from time $0$ to $t$.

\Cref{eq:nonlinear_master_eq} is defined such that $\mathrm{Tr}\big(\rho_{c}(t)\big)=1$ for all $t$; however, this normalization comes at the cost of the non-linearity in $\mathcal{H}[b]$. Furthermore, to calculate the measurement statistics, one would need to sample from the actual distribution of the photocurrent record in any simulation of \cref{eq:nonlinear_master_eq}. An alternative approach would be to instead consider the master equation for $\tilde{\rho}_{c}(t)$, which we define as the non-normalized density operator with norm $\mathrm{Tr}\big(\tilde{\rho}_{c}(t)\big)=p\big(\vect{I}(0,t)\big|\rho(0)\big)$. This way, calculating measurement statistics simply involves calculating the trace of the evolved density operator, with the measured photocurrent $\vect{I}(0,t)$ sampled uniformly. However, this has two problems: first, the SME is still non-linear; and second, sampling from $\vect{I}(0,t)$ uniformly results in the vast majority of states having negligibly small norm.

The solution is to choose an \textit{ostensible distribution} for $\vect{I}(0,t)$ that removes the non-linearity in \cref{eq:nonlinear_master_eq} while still resulting in states with a finite norm. In this picture, we track the evolution of a non-normalized density operator $\bar{\rho}_{c}(t)$, which is defined such that the probability of obtaining a measurement record $\vect{I}(0,t)$ is $p\big(\vect{I}(0,t)\big|\rho(0)\big)=\mathrm{Tr}\big(\bar{\rho}_{c}(t)\big)p_{\text{ost}}\big(\vect{I}(0,t)\big)$, where we call $p_{\text{ost}}\big(\vect{I}(0,t)\big)$ the ostensible distribution of $\vect{I}(0,t)$. If we choose the ostensible distribution of the photocurrent such that each $dW(t)$ obeys an independent mean-zero Gaussian distribution with variance $dt$, then the SME becomes
\begin{multline}\label{eq:linear_master_eq}
    d\bar{\rho}_{c}(t)=-i\big[{-gq_{1}p_{2}},\bar{\rho}_{c}(t)\big]dt+\mathcal{D}[\sqrt{\kappa}b]\bar{\rho}_{c}(t)dt\\
    +\sqrt{\kappa\eta}\big(b\bar{\rho}_{c}(t)+\bar{\rho}_{c}(t)b^{\dag}\big)dW(t),
\end{multline}
which is now linear. Moreover, to reproduce the measurement statistics of the system average we only need to sample $I(t)$ from the Gaussian ostensible distribution, since the additional contribution from the state of the system is taken into account by the norm of $\bar{\rho}_{c}(t)$.%Furthermore, we can the properties of $dW(t)$ given that $I(t)$ is Gaussian distributed, i.e. $dW(t)^{2}=dt$ (see Ref.~\cite{Gardiner85} for more details).

\subsection{Solving the linear SME}

To solve the linear SME~\cref{eq:linear_master_eq}, we vectorize the density operator by \textquotedblleft stacking\textquotedblright\ its rows. For an operator $\rho\in\mathcal{L}(\mathcal{H})$ acting on a Hilbert space $\mathcal{H}$ (which in our case contains two modes), we define the corresponding vectorized operator $\ket{\rho}\in\mathcal{H}\otimes\mathcal{H}^{*}$, where the resultant vector space now has two modes for each mode in $\mathcal{H}$, called the physical modes (with lowering operators $a, b$, etc.) and the fictitious modes ($\tilde{a}, \tilde{b},$ etc.).

For simplicity, we first consider $\mathcal{H}$ consisting of just a single bosonic mode. In particular, for an operator $\rho=\sum_{m,n\in\mathbb{Z}}\rho_{m,n}\ket{m}\!\bra{n}\in\mathcal{L}(\mathcal{H})$, we define the corresponding vectorized operator as $\ket{\rho}=\sum_{m,n\in\mathbb{Z}}\rho_{m,n}\ket{m}\!\ket{n}$ in the Fock basis. Alternatively, $\ket{\rho}=(\rho\otimes I)\ket{0}_{\beta}$, where $\ket{0}_{\beta}=\sum_{n\in\mathbb{Z}}\ket{n}\!\ket{n}=e^{a^{\dag}\tilde{a}^{\dag}}\ket{0}\!\ket{0}$ is the thermo-entangled ground state between the two modes. Equivalently, $\ket{0}_{\beta}$ is the simultaneous $0$-eigenstate of $\hat{\beta}=a-\tilde{a}^{\dag}$ and $\hat{\beta}^{\dag}=a^{\dag}-\tilde{a}$. For an operator $A=\sum_{n}\alpha_{n}a^{n}+\beta_{n}a^{\dag n}$ written in terms of ladder operators, we define the corresponding tilde'd operator as $\tilde{A}=\sum_{n}\alpha_{n}^{*}\tilde{a}^{n}+\beta_{n}^{*}\tilde{a}^{\dag n}$. With this definition, it can be shown that $(A\otimes I)\ket{0}_{\beta}=(I\otimes\tilde{A}^{\dag})\ket{0}_{\beta}$. These properties generalize straightforwardly to when $\mathcal{H}$ consists of two (or more) modes.

To vectorize the linear SME \cref{eq:linear_master_eq} we must obtain a stochastic differential equation in $d\ket{\bar{\rho}_{c}(t)}=(d\bar{\rho}_{c}(t)\otimes I)\ket{0}_{\beta}$. We can do this by replacing all the operators acting on the right of $\rho$ in \cref{eq:linear_master_eq} with operators acting on the fictitious modes, giving
\begin{multline}\label{eq:vectorised_SME}
    d\!\ket{\bar{\rho}_{c}(t)}=\Big(-i(H-\tilde{H})\,dt + \kappa\, D[b]\,dt\\
    +\sqrt{\kappa\eta}\,(b+\tilde{b})\,dW(t)\Big)\ket{\bar{\rho}_{c}(t)},
\end{multline}
where $D[b]=b\tilde{b}-\frac{1}{2}(b^{\dag}b+\tilde{b}^{\dag}\tilde{b})$. To first order in $dt$ this is equivalent to
\begin{multline}\label{eq:exponentiated_SME}
    \ket{\bar{\rho}_{c}(t+dt)}=\exp\Big(\sqrt{\kappa\eta}\,(b+\tilde{b})\,dW(t)\Big)\\
    \times\exp\Big(-i(H-\tilde{H})\,dt+\kappa\,D[b]\,dt\\
    -\frac{1}{2}\kappa\eta\,(b+\tilde{b})^{2}\,dt\Big)\ket{\bar{\rho}_{c}(t)},
\end{multline}
where we note that one can recover \cref{eq:vectorised_SME} from \cref{eq:exponentiated_SME} by Taylor expanding each exponential to second order and using $dW(t)^{2}=dt$.

\Cref{eq:exponentiated_SME} demonstrates the advantage of writing the SME in a vectorized form: now, the evolution operator is an exponential that is at most quadratic in the raising and lowering operators. In particular, the set of terms that are constant, linear or quadratic in $a,a^{\dag},\tilde{a},\tilde{a}^{\dag},b,b^{\dag},\tilde{b},\tilde{b}^{\dag}$ generate a closed Lie algebra $\mathfrak{l}$, allowing us to utilize Lie algebraic techniques to rearrange the expression. Writing the linear stochastic terms as $dL(t)=\sqrt{\kappa\eta}\,(b+\tilde{b})\,dW(t)$ and the quadratic terms as $Q=-i(H-\tilde{H})+\kappa\,D[b]-\frac{1}{2}\kappa\eta\,(b+\tilde{b})^{2}$ for simplicity, we can formally write the solution to \cref{eq:exponentiated_SME} as
\begin{equation}
    \ket{\bar{\rho}_{c}(t)}=\lim_{\delta t\rightarrow0}\prod_{j=1}^{J}\exp\big(dL(j\delta t)\big)\exp\big(Q\delta t\big)\ket{\rho(0)},
\end{equation}
where $t=J\delta t$. The first step to simplify this expression is to commute each linear term $dL(j\delta t)$ to the right of all the quadratic terms $Q$, modifying the linear term in the process. To commute the $j$th linear term past the $j$ quadratic terms to its right, we must calculate $dL'(t)$ such that $\exp\big(dL(j\delta t)\big)\exp\big(Qj\delta t\big)=\exp\big(Qj\delta t\big)\exp\big(dL'(j\delta t)\big)$. Using the commutation formula $e^{A}Be^{-A}=\sum_{n=0}^{\infty}\frac{1}{n!}\mathcal{C}_{A}^{n}[B]$, where $\mathcal{C}_{A}^{0}[B]=B$ and $\mathcal{C}_{A}^{n}[B]=[A,\mathcal{C}_{A}^{n-1}[B]]$ for $n\geq1$, we find that
\begin{multline}
\begin{aligned}
    dL'(j\delta t)&=e^{-Qj\delta t}dL(j\delta t)e^{Qj\delta t}\\
    &=\bigg(e^{-\kappa j\delta t/2}\sqrt{\kappa\eta}(b+\tilde{b})
\end{aligned}\\
    -g\sqrt{\frac{2\eta}{\kappa}}(1-e^{-\kappa j\delta t/2})(q_{1}+\tilde{q}_{1})\bigg)dW(j\delta t).
\end{multline}
Since $[dL'(t),dL'(t')]=0$ for all $t, t'$, we can combine all the linear terms into one exponential, which becomes an integral in the $\delta t\rightarrow 0$ limit, giving
\begin{subequations}\label{eq:unordered_solution}
\begin{align}
    \ket{\bar{\rho}_{R,S}(t)}&=\exp\big(Qt\big)\exp\big(L_{R,S}(t)\big)\ket{\rho(0)},\\
    L_{R,S}(t)&=\sqrt{\eta}R(b+\tilde{b}) -\frac{g\sqrt{2\eta}}{\kappa}S(q_{1}+\tilde{q}_{1}),\\
    R&=\sqrt{\kappa}\int_{0}^{t}\!dt'\,e^{-\kappa t'/2}\,I(t'),\\
    S&=\sqrt{\kappa}\int_{0}^{t}\!dt'\,(1-e^{-\kappa t'/2})\,I(t').
\end{align}
\end{subequations}
Note that we now write subscripts $R,S$ to indicate that the state $\bar{\rho}_{R,S}$ depends not on the \textit{entire} measurement record $\vect{I}(0,t)$, but only on the two relevant integrals of the photocurrent $R$ and $S$.

\subsection{Obtaining the POVM}

Now, we wish to find the POVM $\{W_{R,S}\}$ such that $\mathrm{Tr}(W_{R,S}\rho(t=0))$ is equal to the probability of obtaining the measurement outcomes $R$ and $S$ given an initial state $\rho(0)$, i.e.
\begin{subequations}
\begin{align}
\mathrm{Tr}\big(W_{R,S}\rho(0)\big)&=p\big(R,S\big|\rho(0)\big)\\
&=\big(\prescript{}{\beta}{\bra{0}}\!\prescript{}{\beta}{\bra{0}}\big)\ket{\bar{\rho}_{R,S}(t)}\,p_{\text{ost}}(R,S).
\end{align}
\end{subequations}
We will do this initially by instead evaluating $\Omega_{R,S}(x,x',\beta)={}_{q}\!\bra{x}{}_{c}\!\bra{\beta}W_{R,S}\ket{x'}_{q}\ket{\beta}_{c}/\pi$, which can be thought of as a mixed representation of $W_{R,S}$ in the position basis of the GKP mode $a$ and the Husimi Q-function representation in the ancilla mode $b$. From $\Omega_{R,S}$ one can reconstruct $W_{R,S}$ via the equation
\begin{multline}\label{eq:reconstruct_W}
W_{R,S}=\int_{\mathbb{R}}\!dx\int_{\mathbb{R}}\!dx'\int_{\mathbb{C}}\!d^{2}\beta\,\Big(\Omega_{R,S}(x,x',\beta)\\
\times\ket{x}_{q}{\vphantom{\ket{x}}}_{q}\!\bra{x'}\otimes \tilde{D}^{(+1)}(\beta)\Big).
\end{multline}
Here,
\begin{equation}
\tilde{D}^{(+1)}(\beta)=\int_{\mathbb{C}}\frac{d^{2}\alpha}{\pi}e^{\alpha b^{\dag}}e^{-\alpha^{*}b}e^{\beta\alpha^{*}-\beta^{*}\alpha}
\end{equation}
is the Fourier transform of the normal-ordered displacement operator, which reconstructs a given operator from its Q-function representation.

To calculate $\Omega_{R,S}$ we begin by evaluating the inner product $\mathrm{Tr}\big(W_{R,S}\ket{x}_{q}\ket{\beta}_{c}{\vphantom{\ket{x}}}_{q}\!\bra{x'}{\vphantom{\ket{x}}}_{c}\!\bra{\beta}\big)=\prescript{}{\beta}{\bra{0}}\!\prescript{}{\beta}{\bra{0}}e^{Qt}e^{L_{R,S}(t)}\ket{x}_{q}\!\ket{x'}_{q}\!\ket{\beta}_{c}\!\ket{\beta^{*}}_{c}$, where we adopt the convention of writing the two physical and ficticious modes in the order $a,\tilde{a},b,\tilde{b}$. Noting that ${}_{\beta}\!\bra{0}=\bra{0}\!\bra{0}e^{a\tilde{a}}$, our strategy to evaluate this inner product will be to partially normal-order the operators acting on the $b$ and $\tilde{b}$ modes to remove all $b^{\dag}$ operators to the left by annihilating them on the vacuum state. The remaining exponents written in terms of $q_{1},\tilde{q}_{1},b,\tilde{b}$ act trivially on the initial state $\ket{\rho(0)}=\ket{x}_{q}\!\ket{x'}_{q}\!\ket{\beta}_{c}\!\ket{\beta^{*}}_{c}$ and thus the inner product can be evaluated. To do this, we make use of a faithful 10-dimensional matrix representation of the Lie algebra $\mathfrak{l}$ given in Table 1 of Ref.~\cite{Warszawski20}. In particular, it can be shown that
\begin{multline}\label{eq:reordered_Q}
    e^{a\tilde{a}+b\tilde{b}}e^{Qt}=e^{a\tilde{a}}e^{c_{1}(b^{\dag}b+\tilde{b}^{\dag}\tilde{b})}e^{c_{2}(q_{1}b^{\dag}+\tilde{q}_{1}\tilde{b}^{\dag})}e^{b\tilde{b}}\\
    \times\exp\big(c_{3}(q_{1}b+\tilde{q}_{1}\tilde{b})+c_{4}(\tilde{q}_{1}b+q_{1}\tilde{b})\\+c_{5}(q_{1}^{2}+\tilde{q}_{1}^{2})+c_{6}q_{1}\tilde{q}_{1}+c_{7}(b+\tilde{b})^{2}\big),
\end{multline}
where
\begin{subequations}\label{eq:coefficients}
    \begin{align}
    c_{1}&=-\kappa t/2,\\
    c_{2}&=(\sqrt{2}g/\kappa)(1-e^{\kappa t/2}),\\
    c_{3}&=(\sqrt{2}g/\kappa)(1-e^{-\kappa t/2})(1+\eta-\eta e^{-\kappa t/2}),\\
    c_{4}&=(\sqrt{2}g/\kappa)(1-e^{-\kappa t/2})(\eta-1-\eta e^{-\kappa t/2}),\\
    c_{5}&=(g^{2}/\kappa^{2})\big(\,2(1-e^{-\kappa t/2})-\kappa t-\eta\kappa\tau\big),\\
    c_{6}&=(2g^{2}/\kappa^{2})\big({-2}(1-e^{-\kappa t/2})+\kappa t-\eta\kappa\tau\big),\\
    c_{7}&=(\eta/2)(e^{-\kappa t}-1),\\
    \tau&=t-(1-e^{-\kappa t/2})(3-e^{-\kappa t/2})/\kappa.
    \end{align}
\end{subequations}
Using the two equations
\begin{subequations}\label{eq:trace_evaluated}
\begin{gather}
\bra{0}\bra{0}\bra{0}\bra{0}e^{a\tilde{a}}e^{c_{1}(b^{\dag}b+\tilde{b}^{\dag}\tilde{b})}e^{c_{2}(q_{1}b^{\dag}+\tilde{q}_{1}\tilde{b}^{\dag})}e^{b\tilde{b}}=\prescript{}{\beta}{\bra{0}}\!\prescript{}{\beta}{\bra{0}},\\
\prescript{}{\beta}{\bra{0}}\!\prescript{}{\beta}{\bra{0}}\ket{x}_{q}\!\ket{x'}_{q}\!\ket{\beta}_{c}\!\ket{\beta^{*}}_{c}=\delta(x-x'),
\end{gather}
\end{subequations}
we can now evaluate the inner product
\begin{align}
    \prescript{}{\beta}{\bra{0}}\!\prescript{}{\beta}{\bra{0}}e^{Qt}&e^{L_{R,S}(t)}\ket{x}_{q}\!\ket{x'}_{q}\!\ket{\beta}_{c}\!\ket{\beta^{*}}_{c}\nonumber\\
    &=\delta(x-x')e^{f(x,x,\beta,\beta^{*})},
\end{align}
where $f(x,x,\beta,\beta^{*})$ represents the operator
\begin{align}
    f(q,\tilde{q},b,\tilde{b})&=L_{R,S}(t)+c_{3}(q_{1}b+\tilde{q}_{1}\tilde{b})+c_{4}(\tilde{q}_{1}b+q_{1}\tilde{b})\nonumber\\
    &+c_{5}(q_{1}^{2}+\tilde{q}_{1}^{2})+c_{6}q_{1}\tilde{q}_{1}+c_{7}(b+\tilde{b})^{2}
\end{align}
evaluated at $q=\tilde{q}=x$, $b=\beta$ and $\tilde{b}=\beta^{*}$. Substituting these values gives
\begin{multline}\label{eq:f_evaluated}
f(x,x,\beta,\beta^{*})=-\frac{4g^{2}\eta\tau}{\kappa}x^{2}\\
+\frac{4\sqrt{2}g\eta}{\kappa}\big(1-e^{-\kappa t/2}\big)^{2}x\mathrm{Re}(\beta)\\
-2\eta\big(1-e^{-\kappa t}\big)\mathrm{Re}(\beta)^{2}\\
+2\sqrt{\eta}R\mathrm{Re}(\beta)-\frac{2\sqrt{2}g\sqrt{\eta}}{\kappa}Sx.
\end{multline}

Now that we have calculated the contribution to $\Omega_{R,S}$ from the norm of $\bar{\rho}_{R,S}(t)$, we must now determine the ostensible distribution $p_{\text{ost}}(R,S)$, given that each $dW(t)$ is a Wiener increment, i.e.~an independent Gaussian random variable with mean 0 and variance $dt$. Since this is a Gaussian distribution (as $R$ and $S$ are integrals of Gaussian distributed random variables), we only need to calculate the mean and covariances to determine the distribution. Using $E[dW(t)]=0$ it is straightforward to show that $E[R]=E[S]=0$, and from $E[dW(t)dW(t')]=\delta(t-t')dt dt'$, we obtain
\begin{equation}\label{eq:R_variance}
\begin{aligned}
E[R^{2}]&=\kappa\!\int_{0}^{t}\!\int_{0}^{t}\!e^{-\kappa(t'+t'')/2}E[dW(t')dW(t'')]\\
&=\kappa\!\int_{0}^{t}\!dt'\,e^{-\kappa t'}=1-e^{-\kappa t}.
\end{aligned}
\end{equation}
Similarly,
\begin{align}
E[S^{2}]&=\kappa\!\int_{0}^{t}\!dt'\,(1-e^{-\kappa t'/2})^{2}=\kappa\tau,\label{eq:S_variance}\\
E[RS]&=\kappa\!\int_{0}^{t}\!dt'\,(e^{-\kappa t'/2}-e^{-\kappa t'})=(1-e^{-\kappa t/2})^{2}.\label{eq:RS_covariance}
\end{align}
Defining $\Sigma=\begin{bmatrix}E[R^{2}]&E[RS]\\E[RS]&E[S^{2}]\end{bmatrix}$ as the covariance matrix of the ostensible distribution, we thus obtain
\begin{subequations}
\begin{multline}\label{eq:p_ost_RS}
p_{\text{ost}}(R,S)=\frac{1}{2\pi\sqrt{\det{\Sigma}}}\exp\bigg(\!-\frac{1}{2}\big(R^{2}\Sigma^{-1}_{11}\\
+2RS\Sigma^{-1}_{12}+S^{2}\Sigma^{-1}_{22}\big)\bigg)
\end{multline}
\vspace{-0.6 cm}
\begin{multline}\label{eq:p_ost_evaluated}
=\frac{1}{2\pi\sigma}\exp\bigg(\!-\frac{1}{2}\Big(\frac{\kappa\tau R^{2}}{\sigma^{2}}+\frac{2RS}{4-\kappa t\coth(\kappa t/4)}\\
+\frac{S^{2}}{\kappa t-4\tanh(\kappa t/4)}\Big)\bigg),
\end{multline}
\end{subequations}
where we define $\sigma^{2}=\det\Sigma=\kappa t(1-e^{-\kappa t})-4(1-e^{-\kappa t/2})^{2}$ for convenience.

Using \cref{eq:reconstruct_W,eq:trace_evaluated,eq:f_evaluated,eq:p_ost_evaluated}, we can now write the POVM elements as
\begin{multline}\label{eq:two-mode_POVM}
W_{R,S}=p_{\text{ost}}(R,S)\!\int_{\mathbb{R}}\!dx\!\int_{\mathbb{C}}\frac{d^{2}\beta}{\pi}\Big(e^{f(x,x,\beta,\beta^{*})}\\
\times\ket{x}_{q}{\vphantom{\ket{x}}}_{q}\!\bra{x}\otimes\tilde{D}^{(+1)}(\beta)\Big).
\end{multline}
Next, we wish to turn this two-mode POVM into a single-mode POVM $\{V_{R,S,\rho_{b}}\}$ by tracing out the ancilla mode (initialized in the state $\rho_{b}$) via the equation $
V_{R,S,\rho_{b}}=\mathrm{Tr}_{b}\big(W_{R,S}\rho_{b}\big)$. Given in the position representation $\rho_{b}(y,y')={}_{q}\!\bra{y}\rho_{b}\ket{y'}_{q}$, we can formally write
\begin{multline}\label{eq:position_P_function}
\mathrm{Tr}\big(\tilde{D}^{(+1)}(\beta)\rho_{b}\big)=\int_{\mathbb{R}}\!dx\int_{\mathbb{C}}\frac{d^{2}\alpha}{\pi}\Big(\rho_{b}(x,x-\sqrt{2}\alpha_{R})\\
\times e^{|\alpha|^{2}/2-i\alpha_{R}\alpha_{I}+\beta\alpha^{*}-\alpha\beta^{*}+i\sqrt{2}\alpha_{I}x}\Big),
\end{multline}
where $\alpha_{R}=\mathrm{Re}(\alpha)$ and $\alpha_{I}=\mathrm{Im}(\alpha)$. Note however that in practice we cannot evaluate the integral \cref{eq:position_P_function} on its own due to the divergent $e^{|\alpha|^{2}/2}$ term. However, substituting \cref{eq:position_P_function} into \cref{eq:two-mode_POVM} and performing the resultant integrals gives our desired result
\begin{align}
    V_{S,\rho_{b}}&=\int_{\mathbb{R}}\!dR\,\mathrm{Tr}_{b}(W_{R,S}\rho_{b})\nonumber\\
    &=\mathcal{N}\!\int\! dx\,\ket{x}_{q_{1}}\!\!\bra{x}\int\! d\tilde{x}\,{\vphantom{\ket{x}}}_{q_{2}}\!\!\braket{\tilde{x}|\rho_{b}|\tilde{x}}_{q_{2}}\nonumber\\
    &\times \exp\bigg\{{-}\frac{1}{C}\Big[X-x-\frac{1}{2g\tau}\big(1-e^{-\kappa t/2}\big)^{2}\tilde{x}\Big]^{2}\bigg\},\tag{\ref{eq:measurement_scheme_2_general_POVM}}
\end{align}
where
\begin{subequations}
\begin{align}
    \mathcal{N}&=\Big(2\pi\big(\kappa\tau-\eta(1-e^{-\kappa t/2})^{4}\big)\Big)^{-1/2},\\
    X&=-S/\sqrt{8g^{2}\tau^{2}\eta},\\
    C&=\frac{\kappa\tau-\eta(1-e^{-\kappa t/2})^{4}}{4g^{2}\tau^{2}\eta}.\tag{\ref{eq:c_expression}}
\end{align}
\end{subequations}

\subsection{The conditional post-state}\label{subsec:post_state}

As a final exercise, we consider the situation where the ancilla state has been set to the vacuum state as in the main text, i.e.~$\rho_{b}=\ket{0}\!\bra{0}$. Here, we wish to find the state of the first mode of the system conditioned on the measurement outcome, i.e.~$\mathrm{tr}_{b}\big(W_{R,S}\,\rho(0)\otimes\ket{0}\!\bra{0}\big)$. In vectorized form, this is given by
\begin{equation}
    {\vphantom{\ket{x}}}_{\beta,b}\bra{0}e^{Qt}e^{L_{R,S}(t)}\ket{\rho(0)}\otimes\ket{0}\!\ket{0},
\end{equation}
up to a normalization, where ${\vphantom{\ket{x}}}_{\beta,b}\!\bra{0}$ is the maximally entangled state across the $b$ and $\tilde{b}$ modes. Using \cref{eq:unordered_solution,eq:reordered_Q,eq:coefficients} we can simplify this expression to
\begin{multline}
    \exp\bigg(\!{-}\frac{g^{2}}{\kappa}\Big(t-2(1-e^{-\kappa t/2})/\kappa\Big)(q_{1}-\tilde{q}_{1})^{2}\!\bigg)\\
    \times\exp\bigg(\!{-}\frac{g^{2}\eta\tau}{\kappa}(q_{1}+\tilde{q}_{1}-2X)^{2}\!\bigg)\ket{\rho(0)},
\end{multline}
again up to a normalization factor. Intuitively, the first exponent suppresses off-diagonal terms in the position basis expansion of $\rho(0)$, while the second exponent suppresses terms whose position is far away from $X$. As $t$ becomes large, $\tau\approx t-3/\kappa$, so the constant of decay in the second exponent is smaller than that of the first exponent by a factor of roughly $\eta$. Importantly, as $t\rightarrow\infty$, $\ket{\rho(0)}$ is projected into a position eigenstate $\ket{X}_{q}{\vphantom{\ket{x}}}_{q}\!\bra{X}$, and for finite times, the state will consist of some finite mixture of finitely squeezed states. It may be easier to prepare a new GKP codestate for subsequent computations from this final state than if the mode were reset to the vacuum.
\end{document}